\documentclass[a4paper,12pt]{article}
\usepackage[english]{babel}
\usepackage{amsmath}
\usepackage{amssymb}
\usepackage{bbm}
\usepackage{xcolor}
\usepackage{cite}
\usepackage{multirow}
\usepackage{xcolor,hyperref}
\usepackage{graphicx}
\usepackage{enumitem}
\usepackage{subfigure}
\usepackage{enumitem}
\usepackage{ulem}
\usepackage[titletoc,toc,title]{appendix}
\usepackage[font=footnotesize,labelfont=bf]{caption}
\usepackage[paperwidth=210mm,paperheight=297mm,
lmargin=2cm,rmargin=2cm,
tmargin=2cm,bmargin=2.5cm,
headsep=0.5cm,headheight=0.5cm,
footskip=1cm]{geometry}
\Roman{section}
\newcommand{\naw}[1]{\left(#1\right)}
\newcommand{\ket}[1]{\left|#1\right>}
\newcommand{\bra}[1]{\left<#1\right|}

\newcommand{\com}[1]{\left[#1\right]}

\DeclareMathOperator{\arccosh}{arccosh}

\title{Quantum Cournot model based on general entanglement operator}
\author{Katarzyna Bolonek-Laso\'n\footnote{katarzyna.bolonek@uni.lodz.pl}  }
\date{%
	\small{	Department of Statistical Methods, Faculty of Economics and Sociology\\ University of Lodz, 41/43 Rewolucji 1905 St., 90-214 Lodz, Poland}}
\begin{document}
\maketitle

\begin{abstract}
The properties of the Cournot model based on the most general entanglement operator containing quadratic expressions which is symmetric with respect to the exchange of players are considered. The degree of entanglement of games dependent on one and two squeezing parameters and their payoff values in Nash equilibrium are compared. The analysis showed that the relationship between the degree of entanglement of the  initial state of the game and the payoff values in Nash equilibrium is ambiguous. The phase values included in the entanglement operator have a strong influence on the final outcome of the game. In a quantum duopoly based on the initial state of a game that depends on one squeezing parameter, the maximum possible payoff in Nash equilibrium cannot be reached when the value of the phase parameter is greater than zero, in contrast to a game that depends on two parameters.\end{abstract}

\section{Introduction}
Since the publication of Eisert, Wilkens and Lewenstein's papers \cite{EisertWL}, \cite{EisertW} on the quantization procedure for the classical game in 1999 and 2000, the topic of quantum games has been the subject of extensive research \cite{Meyer}$\div$\cite{bolonek2}.  The main advantage of quantum games over their classical counterparts is the potential for higher payoffs when the initial state of the quantum game is maximally entangled; on the other hand, for the vanishing squeezing parameter the game reduces to classical form. 

The question arises what is the relevance of quantum games in real economic market situations. It seems that the valuable answer can be derived from the remarks made already at the beginning of the subject \cite{EisertWL},\cite{Enk}. As noticed by Eisert et al., any quantum game can be played by purely classical means; a quantum game can be replaced by a classical one if one allows for considerable extension of the classical game we have started with \cite{Enk}. The extended classical game provides the proper payoffs, however at the expense of time and less economical use of the resources. This reasoning can be reversed: given a complicated classical game that requires large resources one can ask whether the quantum game can be found which reproduces the relevant payoffs but is more economical way as far as resources are concerned. This is also valid question in real market economy. For example, it can happen that the valuable predictions are not possible in real time (which is sometimes dramatically important) if one doesn't refer to the quantum game which reproduces the payoffs of the classical extended one. From this point of view the quantum game theory is the particular case of quantum computations. It should be noted that finding the quantum game embracing a complicated classical one may be a difficult task. Therefore, it seems that the reverse strategy is more promising: one selects the most important simple classical games, quantize them and, finally, looks for the classical games derived from the set of quantum ones. 
	The authors of \cite{EisertWL} refer also to different context. They speculate, on the basis of Dawkin's famous book ''\textit{The selfish Gene}'' \cite{Dawkins},  that ''games of survival are being played already on the molecular level where quantum mechanics dictates the rules'' \cite{EisertWL}. Finally, the extremely exotic possibility is that the human consciousness has the quantum origin \cite{Penrose}.

The quantization scheme proposed by Eisert et al.~dealt with games possessing a discrete set of strategies. However, in the classical game theory, we are not always dealing with a discrete set of strategies. For games that have applications in economics, an optimal solution belonging to a continuous set is often sought. For games with a continuous set of strategies, a quantization scheme was introduced by Li, Du and Massar \cite{Li}, who presented their proposal using a simple competition model such as the Cournot duopoly as an example.

This initiated a series of papers showing that in other models of market competition known in economics, such as Stackelberg's duopoly model, Bertrand's duopoly model, or the model with more than two participants \cite{varian}, higher payoffs in Nash equilibrium can be obtained by using quantum strategies instead of classical ones \cite{LoKiang0,LoKiang3,khan,rahaman,FracSladk,LoYeung}.
The main difference between the Cournot, Bertrand and Stackelberg duopoly models is as follows. In the Cournot model, assuming constant production costs, producers try to determine the volume of production in order to obtain the highest possible profit, whereas in the Bertrand model the producer's decision variable is the price of the product rather than the quantity produced. In the third model, the Stackelberg model, the firm decides whether it is more profitable to be the price leader or to follow another leader.
In addition to the basic versions of the above models that we can find in economics textbooks \cite{varian}, modifications of these models have also been analyzed. The Cournot duopoly model has also been considered for nonlinear price functions, quadratic cost functions \cite{wang}, \cite{sekiguchi}, \cite{wang1} or when the knowledge of the game is not the same for both players \cite{frackiewiczbilski}. 
The cases with incomplete information were considered in the Bertrand \cite{QinChen} and Stackelberg duopoly \cite{LoKiang1}, \cite{frackiewicz2}. 
Bertrand's model was also considered under the assumption that products offered by oligopolists differ from each other \cite{LoKiang2}.

In 2005, Qin et al.~\cite{Qin2}, using the Bertrand model as an example, proposed quantization scheme for asymmetric games with a continuous set of strategies, that is, when, for example, players/companies operate under different conditions or they have a different set of information. This scheme is based on an entanglement operator that depends on two squeezing parameters and has also been used in the analysis of the asymmetric Stackelberg model \cite{WangXia,Zhong}. 

Cournot's quantum model was also considered in asymmetric form, that is using an asymmetric entangled state \cite{LiQin}. The authors of this model modified the model of Li et al. \cite{Li} using additionally two single-mode electromagnetic fields. As a result, the entanglement operator depended on three parameters. 

Since quantum game models offer the possibility of higher payoffs than their classical counterparts when the initial state of the game is entangled, it seems natural to ask whether, while preserving the properties of the classical game, it is possible to obtain higher payoffs using more than one squeezing parameter. In discussing the properties of the game, we primarily refer to its symmetry. That is, we assume that the quantum game is symmetric with respect to the exchange of players, provided that the classical counterpart is also symmetric. In the present paper we consider such a model, which means that the game entanglement operator is symmetric with respect to the exchange of players but has a more general form than the one proposed by  Li et al., and depends on two entanglement parameters. To check the dependence of the payoff on the degree of entanglement of the initial state of the game we compute the entanglement entropy.  Our study aims to compare the properties of a quantum game based on an initial state that depends on a single squeezing parameter with those of a game that depends on two squeezing parameters, while maintaining the symmetry of the game. 

The paper is organized as follows. In Section \ref{sec1}, we describe the classical Cournot model. The quantization scheme of games with a continuous strategic space proposed by Li et al. is presented in Section \ref{sec2}. The form of  general entanglement operator and the Cournot quantum duopoly model based on it are presented in Section \ref{sec3}. The section \ref{sec4} is devoted  to a comparison of the payoffs of players in duopoly models based on one and two squeezing parameters, as well as an analysis of the degree of entanglement of the initial states of the models under consideration. The final section contains some conclusions.
Some detailed calculations are included in the appendices.

	\section{The  Cournot's  duopoly model}\label{sec1}
	The eternal question posed by the entrepreneurs is what the optimal level of production should be to achieve maximum profit. Obviously, the answer depends on the prevailing market conditions. The analysis of business behavior is frequently conducted under the conditions of perfect competition or monopoly. However, these situations are far from reality, and thus oligopoly models, which better reflect the real market, are often considered. 
	
	One of the most well-known oligopoly models is the Cournot duopoly, which is  based on the following assumptions:
	\begin{itemize}
	\item[1)] there are only two producers of a given good,
	\item[2)] the product of both producers is homogeneous,
	\item[3)] the cost of production is the same in both companies and the marginal cost is constant,
	\item[4)] the product price is set by buyers, while sellers adjust the production quantity according to the already established price,
	\item[5)] each producer simultaneously estimates the demand for their product and sets the quantity of their production assuming that the production volume of the competitor will not change.
\end{itemize}
We can describe this simple model as follows. Let
\begin{equation}
	Q=q_1+q_2
\end{equation} denotes the total quantity of production, where $q_1$, $q_2$ are the production quantities of company 1 and 2, respectively, while
\begin{equation}
	P(Q)=\left\{\begin{array}{lll}
	p-Q	& \text{for} & p>Q  \\
	0	& \text{for} & p<Q
	\end{array}\right.
	\end{equation} is the market-clearing price, where $p>0$.    
The total cost to company $i$ of producing the quantity $q_i$ is
\begin{equation}
	C_i(q_i)=cq_i
\end{equation}
which means that the marginal cost equals $c$, where we assume $c<p$.
The payoffs can be written as
\begin{equation}
	\pi_i(q_1,q_2)=q_i(P(Q)-c)=q_i(p-c-q_1-q_2)=q_i(k-q_1-q_2),\label{a4}
\end{equation}
with $k=p-c$, $i=1,2$. In the context of game theory, a duopoly can be defined as a two-person game in which each player/company aims to identify the optimal strategy, within the set of all possible strategies $q_i$, that will yield the greatest profit, under the assumption that the other player strategy is fixed.  Thus, solving the optimization problem $\max\limits_{q_i}\pi_i(q_i,q_j^*)$ we find the pair $(q_1^*,q_2^*)$ 
\begin{equation}
	q_1^*=q_2^*=\frac{k}{3}\label{a5}
\end{equation}
which is the Nash equilibrium. Substituting the solution \eqref{a5} into the payoffs function \eqref{a4} we get
\begin{equation}
\pi_1(q_1^*,q_2^*)=	\pi_2(q_1^*,q_2^*)=\frac{k^2}{9}
\end{equation} 
It is important to note that, in the Cournot model, there is no cooperation between companies, and they make their decisions simultaneously.

\section{Quantum version of the Cournot's duopoly}\label{sec2}

The quantization scheme of the games with continuous strategic space was proposed by H. Li, J. Du and S. Massar \cite{Li}. The transition from the classical game to its quantum counterpart involves the preparation of a Hilbert space for each player and the assignment of specific quantum states to the potential outcomes of each classical strategy. Given that the set of strategies in the classical duopoly is continuous, it is necessary to utilize a Hilbert space of a continuous-variable quantum system. In this context, the authors have selected two single-mode electromagnetic fields with a continuous set of eigenstates.
 The initial state 
 \begin{equation}
 \hat{J}(\beta)\ket{0}_1\otimes\ket{0}_2\label{abb7}
 \end{equation}
 is the starting point of the game; here $\ket{0}_i$ for $i=1,2$ are the single-mode vacuum states of two electromagnetic fields. $\hat{J}(\beta)$ is the entangling operator which must satisfy two conditions:
 \begin{enumerate}[label={(\roman*)}]
 \item  symmetry with respect to the exchange of
 players/companies (in accordance with the symmetry of the classical game);
 \item 	the classical game must be faithfully represented by its quantum counterpart; this will be satisfied if $\hat{J}(\beta)=\hat{J}(\beta)^\dagger=\hat{I}$ for some particular $\beta$ (in this case $\beta=0$). 
 	\end{enumerate}
  The operator $\hat{J}(\beta)$ has form
 \begin{equation}
 	\hat{J}(\beta)=e^{-\beta(\hat{a}_1^\dagger \hat{a}_2^\dagger-\hat{a}_1\hat{a}_2)},\label{ab8}
 \end{equation}
where $\beta\geq 0$; $a_i^\dagger$ ($a_i$) is the creation (annihilation) operator of $i$-th company ''electromagnetic field''.  In fact, eq.~\eqref{ab8} describes the two-mode squeezing operator where $\beta$ is called the squeezing parameter \cite {gerry}. Thus the action of operator $\hat{J}(\beta)$ on the state $\left|00\right>$ creates the two-mode squeezed vacuum state (eq.~\eqref{abb7}).
The quantum strategies of each player are represented by unitary operators 
\begin{equation}
	\hat{D}_i(x_i)=e^{x_i(\hat{a}_i^\dagger-\hat{a}_i)},\qquad x_i\in\left[0,\infty\right),\quad i=1,2.
\end{equation}
 Hence the final state is given by
 \begin{equation}
\ket{\psi_f}=\hat{J}(\beta)^\dagger(\hat{D}_1(x_1)\otimes\hat{D}_2(x_2))\hat{J}(\beta)\ket{00},
\end{equation}
where $\ket{00}=\ket{0}_1\otimes\ket{0}_2$.
The quantities of companies are determined by the final measurements $q_i=\bra{\psi_f}\hat{X}_i\ket{\psi_f}$.
The functions $\pi_i^Q(x_1,x_2)$ determined according eq.~\eqref{a4} reach maximal value for $	x_i^*=\frac{k\cosh\beta}{1+2e^{2\beta}}$; thus the payoffs corresponding to the Nash equilibrium are
\begin{equation}
	\pi_1^Q(x_1^*,x_2^*)=\pi_2^Q(x_1^*,x_2^*)=\frac{k^2 e^\beta\cosh\beta}{(3\cosh\beta+\sinh\beta)^2}.\label{ab14}
\end{equation}
The profits depend on squeezing parameter; when $\beta=0$ the quantum game acquires the classical form described in Sec.~\ref{sec1}.
However, for $\beta\rightarrow \infty$ the payoffs in the Nash equilibrium converge to $k^2/8$. Therefore, in the case of a maximally entangled quantum game, the players are able to achieve higher profits than they would in a  classical game.

\section{More general form of entangling operator}\label{sec3}
From the previous section we see that the optimal outcomes of the quantum game depend on the value of the squeezing parameter $\beta$; the higher its value, the more profits players can make. This observation gives rise to the question of whether the enhanced profits could be attained in the quantum Cournot duopoly if the entanglement operator were to possess a more general form and were to depend, for instance, on two squeezing parameters.

Let us suppose that the entanglement operator has the following form:
\begin{equation}
	\hat{J}_1(\delta,\xi)=e^{\delta^*(\hat{a}_1^2+\hat{a}_2^2)-\delta(\hat{a}_1^{\dagger\,2}+\hat{a}_2^{\dagger\,2})+\xi^*\hat{a}_1\hat{a}_2-\xi\hat{a}_1^\dagger\hat{a}_2^\dagger},\label{a15}
\end{equation}  
where $\delta=\alpha e^{i\phi}$, $\xi=\beta e^{i\theta}$;  $\theta,\phi\in[0,2\pi)$, $\alpha,\beta\in\left[0,\infty\right)$ are the squeezing parameters.
The operator $\hat{J}_1(\delta,\xi)$ reduces to $\hat{J}(\beta)$ described by eq.~\eqref{ab8} for $\alpha=0$ and $\theta=0$, thus the parameter $\beta$ is equivalent to the squeezing parameter of the two-mode squeezing operator \cite{gerry}. On the other hand, for $\beta=0$ the $\hat{J}_1(\delta,\xi)$ reduces to two single-mode squeezing operators with  $\alpha\geq 0$ as squeezing parameter\cite{gerry}. Note, however, that  the operator $\hat{J}_1(\delta,\xi)$ is not equal to the product of the two-mode and two single-mode squeezing operators because the term $\delta^*(\hat{a}_1^2+\hat{a}_2^2)-\delta(\hat{a}_1^{\dagger\,2}+\hat{a}_2^{\dagger\,2})$ do not commute with  $\xi^*\hat{a}_1\hat{a}_2-\xi\hat{a}_1^\dagger\hat{a}_2^\dagger$.
Equation \eqref{a15} describes the most general form of the operator containing the quadratic terms in creation/annihilation operators,
 which is symmetric with respect to the exchange of players. An entanglement operator that also contains quadratic expressions but does not satisfy the symmetry condition for player substitution was presented in reference \cite{LiQin}. A comparison of the  operator $\hat{J}_1(\delta,\xi)$ with the operator discussed in reference \cite{LiQin} is included in the Appendix \ref{AA1}.

Following the scheme of Li et al. \cite{Li} we have to find
\begin{equation}
	\tilde{q}_i=\bra{\psi_{f_1}}\hat{X}_i\ket{\psi_{f_1}},\quad i=1,2,\label{b16}
\end{equation} 
where the final state is now of the form
\begin{equation}
\ket{\psi_{f_1}}=\hat{J}_1^\dagger(\delta,\xi)(\hat{D}_1(x_1)\otimes\hat{D}_2(x_2))\hat{J}_1(\delta,\xi)\ket{00}.
\end{equation}
To do this we need to determine the expressions 
\begin{equation}
	\begin{split}
	&\hat{J}_1(\delta,\xi)\hat{a}_i\hat{J}_1^\dagger(\delta,\xi)\equiv	e^{\hat{A}}\hat{a}_ie^{-\hat{A}},\\
&\hat{J}_1(\delta,\xi)\hat{a}_i^\dagger\hat{J}_1^\dagger(\delta,\xi)\equiv	e^{\hat{A}}\hat{a}_i^\dagger e^{-\hat{A}}
	\end{split}\label{a17}
\end{equation}
for $i=1,2$, where 
\begin{equation}
	\hat{A}=\delta^*(\hat{a}_1^2+\hat{a}_2^2)-\delta(\hat{a}_1^{\dagger\,2}+\hat{a}_2^{\dagger\,2})+\xi^*\hat{a}_1\hat{a}_2-\xi\hat{a}_1^\dagger\hat{a}_2^\dagger.
\end{equation}
Let us write
\begin{equation}
	\hat{a}_i(t)\equiv e^{t\hat{A}}\hat{a}_ie^{-t\hat{A}}.
\end{equation}
Differentiating the above equation with respect to $t$, we get
\begin{equation}
\dot{\hat{a}}_i(t)\equiv e^{t\hat{A}}\hat{A}\hat{a}_ie^{-t\hat{A}}-e^{t\hat{A}}\hat{a}_i\hat{A}e^{-t\hat{A}}=[\hat{A},\hat{a}_i(t)].\label{a20}
\end{equation}
Using eq.~\eqref{a20} we find
\begin{equation}
	\begin{split}
		& \dot{\hat{a}}_1(t)=\xi\hat{a}_2^\dagger(t)+2\delta\hat{a}_1
^\dagger(t)\\
&\dot{\hat{a}}_2(t)=\xi\hat{a}_1^\dagger(t)+2\delta\hat{a}_2
^\dagger(t)\\
	& \dot{\hat{a}}_1^\dagger(t)=\xi^*\hat{a}_2(t)+2\delta^*\hat{a}_1(t)\\
		& \dot{\hat{a}}_2^\dagger(t)=\xi^*\hat{a}_1(t)+2\delta^*\hat{a}_2(t)
	\end{split}
\end{equation}
or, in the matrix form
\begin{equation}
	\left(\begin{array}{c}
	\dot{\hat{a}}_1(t)\\
	\dot{\hat{a}}_2(t)\\
	\dot{\hat{a}}_1^\dagger(t)\\
	\dot{\hat{a}}_2^\dagger(t)
	\end{array}\right)=\left(\begin{array}{cccc}
0 & 0 & 2\delta & \xi\\
0 & 0 & \xi & 2\delta\\
2\delta^* & \xi^* & 0 & 0\\
\xi^* & 2\delta^* & 0 & 0
\end{array}\right)\left(\begin{array}{c}
\hat{a}_1(t)\\
\hat{a}_2(t)\\
\hat{a}_1^\dagger(t)\\
\hat{a}_2^\dagger(t)
\end{array}\right).\label{a22}
\end{equation}
The solution of eq.~\eqref{a22} is
\begin{equation}
	\mathbf{a}(t)=e^{tM}\mathbf{a}(0)
\end{equation}
where
\begin{displaymath}
	\mathbf{a}(t)=\left(\begin{array}{c}
		\hat{a}_1(t)\\
		\hat{a}_2(t)\\
		\hat{a}_1^\dagger(t)\\
		\hat{a}_2^\dagger(t)
	\end{array}\right),\qquad M=\left(\begin{array}{cccc}
	0 & 0 & 2\delta & \xi\\
	0 & 0 & \xi & 2\delta\\
	2\delta^* & \xi^* & 0 & 0\\
	\xi^* & 2\delta^* & 0 & 0
\end{array}\right);
\end{displaymath}
for $t=1$ we have 
\begin{equation}
	\mathbf{a}(1)=e^{M}\mathbf{a}(0).
	\end{equation}
Accordingly, the explicit form of eq.~\eqref{a17} can be obtained by  computing $e^M$, which assumes the form (see Appendix \ref{AA2} for detailed calculations)
{\footnotesize 
\begin{equation}
	e^M=\left(\begin{array}{cccc}
	\frac{1}{2}(\cosh a+\cosh b)	& 	\frac{1}{2}(-\cosh a+\cosh b) & 	\frac{1}{2}(ad\sinh a+bf\sinh b) & \frac{1}{2}(-ad\sinh a+bf\sinh b) \\
		\frac{1}{2}(-\cosh a+\cosh b)	& 	\frac{1}{2}(\cosh a+\cosh b) & 	\frac{1}{2}(-a d\sinh a+bf\sinh b) & \frac{1}{2}(a d\sinh a+bf\sinh b) \\
	\frac{1}{2}\naw{\frac{\sinh a}{ad}+\frac{\sinh b}{bf}}	& \frac{1}{2}\naw{-\frac{\sinh a}{ad}+\frac{\sinh b}{bf}} & 	\frac{1}{2}(\cosh a+\cosh b)	& 	\frac{1}{2}(-\cosh a+\cosh b)  \\
		\frac{1}{2}\naw{-\frac{\sinh a}{ad}+\frac{\sinh b}{bf}}	& \frac{1}{2}\naw{\frac{\sinh a}{ad}+\frac{\sinh b}{bf}} & 	\frac{1}{2}(-\cosh a+\cosh b)	& 	\frac{1}{2}(\cosh a+\cosh b) 
	\end{array}
	\right)\label{a26}
\end{equation}}
where
\begin{equation}
\begin{split}
	& a=\sqrt{4\alpha^2+\beta^2-4\alpha\beta\cos(\theta-\phi)},\\
	& b=\sqrt{4\alpha^2+\beta^2+4\alpha\beta\cos(\theta-\phi)},\\
	& d=\frac{e^{i(\theta+\phi)}}{2e^{i\theta}\alpha-e^{i\phi}\beta},\\
&  f=\frac{e^{i(\theta+\phi)}}{2e^{i\theta}\alpha+e^{i\phi}\beta}.	
\end{split}	\label{ab27}
	\end{equation}
Eqs.~\eqref{a17} take the form
\begin{equation}
	\begin{split}
\hat{J}_1(\delta,\xi)\hat{a}_1\hat{J}_1^\dagger(\delta,\xi)=&	\frac{1}{2} ((\cosh a+\cosh b)\hat{a}_1-(\cosh a-\cosh b)\hat{a}_2\\
&-(ad\sinh a-bf\sinh b)\hat{a}_1^\dagger+(ad\sinh a+bf\sinh b)\hat{a}_2^\dagger),\\
\hat{J}_1(\delta,\xi)\hat{a}_2\hat{J}_1^\dagger(\delta,\xi)=&	\frac{1}{2} ((\cosh a+\cosh b)\hat{a}_2-(\cosh a-\cosh b)\hat{a}_1\\
&-(ad\sinh a-bf\sinh b)\hat{a}_2^\dagger+(ad\sinh a+bf\sinh b)\hat{a}_1^\dagger),\\
\hat{J}_1(\delta,\xi)\hat{a}_1^\dagger\hat{J}_1^\dagger(\delta,\xi)=&	\frac{1}{2abdf} (abdf(\cosh a+\cosh b)\hat{a}_1^\dagger-(\cosh a-\cosh b)\hat{a}_2^\dagger\\
&+(bf\sinh a+ad\sinh b)\hat{a}_1-(bf\sinh a-ad\sinh b)\hat{a}_2),\\
\hat{J}_1(\delta,\xi)\hat{a}_2^\dagger\hat{J}_1^\dagger(\delta,\xi)=&	\frac{1}{2abdf} ((\cosh a+\cosh b)\hat{a}_2^\dagger-abdf(\cosh a-\cosh b)\hat{a}_1^\dagger\\
&+(bf\sinh a+ad\sinh b)\hat{a}_2-(bf\sinh a-ad\sinh b)\hat{a}_1). 
\end{split}
\end{equation}
The final measurement defined by the eq.~\eqref{b16} yields
\begin{equation}
	\begin{split}
		\tilde{q}_1=&\frac{1}{4}\left[\naw{2\cosh a+2\cosh b+\naw{\frac{1}{ad}+ad}\sinh a+\naw{\frac{1}{bf}+bf}\sinh b }\tilde{x}_1\right.\\
		&\left.-\naw{2\cosh a-2\cosh b+\naw{\frac{1}{ad}+ad}\sinh a-\naw{\frac{1}{bf}+bf}\sinh b }\tilde{x}_2\right],\\
			\tilde{q}_2=&\frac{1}{4}\left[\naw{2\cosh a+2\cosh b+\naw{\frac{1}{ad}+ad}\sinh a+\naw{\frac{1}{bf}+bf}\sinh b }\tilde{x}_2\right.\\
		&\left.-\naw{2\cosh a-2\cosh b+\naw{\frac{1}{ad}+ad}\sinh a-\naw{\frac{1}{bf}+bf}\sinh b }\tilde{x}_1\right].
	\end{split}\label{ac29}
\end{equation}
In the next step, we determine, according to the eq.~\eqref{a4}, the quantum profit functions $\tilde{u}_{1,2}^Q(\tilde{x}_1,\tilde{x}_2)$; their explicit form is given in Appendix \ref{AA3}. From the conditions
\begin{align}
		&\frac{\partial \tilde{u}_1^Q(\tilde{x}_1,\tilde{x}_2)}{\partial \tilde{x}_1}=0,\label{ab30}\\	
		&\frac{\partial \tilde{u}_2^Q(\tilde{x}_1,\tilde{x}_2)}{\partial \tilde{x}_2}=0.\label{ab31}	
\end{align}
 we get
\begin{small}\begin{equation}
	\tilde{x}_1^*=\tilde{x}_2^*=\frac{k(2(\cosh a+\cosh b)+(\frac{1}{ad}+ad)\sinh a+(\frac{1}{bf}+bf)\sinh b)}{(2\cosh b+(\frac{1}{bf}+bf)\sinh b)(2\cosh a+4\cosh b+(\frac{1}{ad} +ad)\sinh a+2(\frac{1}{bf}+bf)\sinh b)}
\end{equation}\end{small}
provided
\begin{displaymath}
	\begin{split}
&2ad\cosh a+\sinh a+a^2d^2\sinh a\neq 0,\\
& 2bf\cosh b+\sinh b+b^2f^2\sinh b\neq 0.
\end{split}
	\end{displaymath}
The payoffs corresponding to the Nash equilibrium read
\begin{small}
\begin{equation}
	\begin{split}
	\tilde{u}_1^Q(\tilde{x}_1^*,\tilde{x}_2^*)&=\tilde{u}_2^Q(\tilde{x}_1^*,\tilde{x}_2^*)=\\
	&=\frac{k^2(2\cosh b+(\frac{1}{bf}+bf)\sinh b)(2\cosh a+2\cosh b+(\frac{1}{ad}+ad)\sinh a+(\frac{1}{bf}+bf)\sinh b)}{2(2\cosh a+4\cosh b+(\frac{1}{ad}+ad)\sinh a+2(\frac{1}{bf}+bf)\sinh b)^2}
	\end{split}\label{ab33}
\end{equation}\end{small}
with $a,\,b,\,d,\,f$ defined by eq.~\eqref{ab27}. The payoffs depend on four parameters: two phase parameters $\theta,\,\phi\in[0,2\pi]$ and two squeezing parameters $\alpha,\,\beta\in\left[0,\infty\right)$.

 In the case of $\alpha=0$, the entanglement operator $\hat{J}_1(\delta,\xi)$ \eqref{a15} reduces to an operator that depends on one squeezing parameter and one phase parameter
\begin{equation}
	\hat{J}(\xi)=e^{-\xi\hat{a}_1^\dagger\hat{a}_2^\dagger+\xi^*\hat{a}_1\hat{a}_2}, \quad \xi=\beta e^{i\theta}\label{ac34}
\end{equation}
and the payoffs read
\begin{equation}
	u_1^Q(x_1^*,x_2^*)=u_2^Q(x_1^*,x_2^*)=\frac{k^2\cosh \beta(\cosh \beta+\cos\theta\sinh\beta)}{(3\cosh\beta+\cos\theta\sinh\beta)^2}.\label{ab35}
\end{equation}
If we additionally put $\theta=0$, we obtain exactly the duopoly model described by Li et al.

Both functions $u_{1,2}^Q(x_1^*,x_2^*)$ and $\tilde{u}_{1,2}^Q(\tilde{x}_1^*,\tilde{x}_2^*)$ do not exceed the value $\frac{k^2}{8}$. The first one approaches its maximum value for $\theta=0$ and $\beta\rightarrow\infty$. For the second function, it can be shown analytically that its maximum value is $\frac{k^2}{8}$. Just note that $\tilde{u}_{1,2}^Q(\tilde{x}_1^*,\tilde{x}_2^*)$ can be written in the form
\begin{equation}
	\tilde{u}_{1,2}^Q(\tilde{x}_1^*,\tilde{x}_2^*)=\frac{k^2z_1(z_1+z_2)}{2(z_2+2z_1)^2}=\frac{k^2}{2}(z-z^2)
\end{equation}
where 
\begin{displaymath}
	\begin{split}
	& z_1=2\cosh b+(\frac{1}{bf}+bf)\sinh b,\\
	& z_2=2\cosh a+(\frac{1}{ad}+ad)\sinh a,\\
	& z=\frac{z_1}{2z_1+z_2}.
		\end{split}
\end{displaymath}
Thus, for $z=\frac{1}{2}$, the function 	$\tilde{u}_{1,2}^Q(\tilde{x}_1^*,\tilde{x}_2^*)$ reaches its highest value equal to $\frac{k^2}{8}$.

\begin{figure}[h]
	\begin{center}
		\subfigure[the payoff $u_{1,2}^Q(x_1^*,x_2^*)$ as a function of the squeezing parameter $\beta$ and phase parameter $\theta$.]{\label{subfig1a}\includegraphics[scale=0.35]{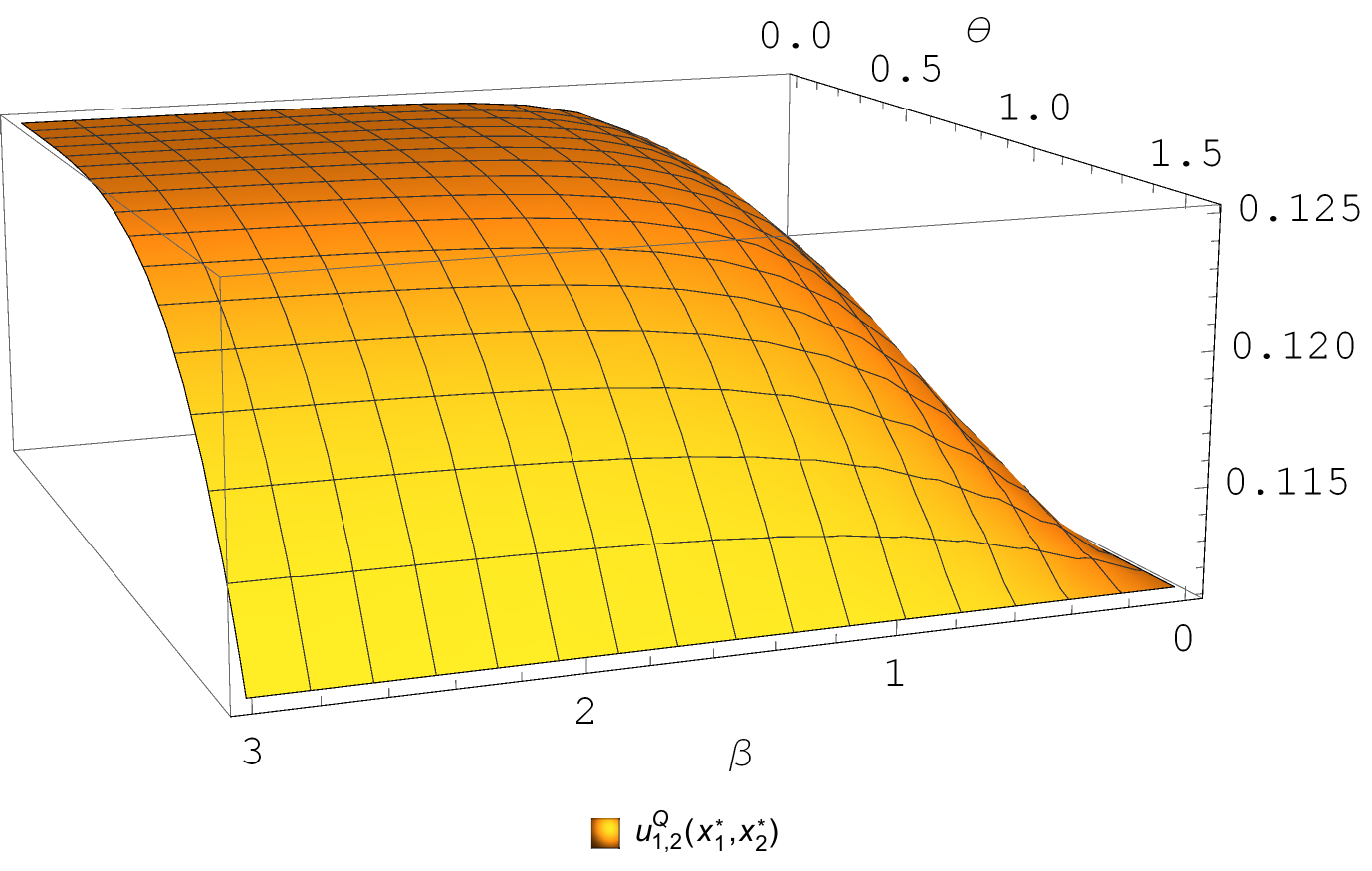}}\qquad
	\subfigure[The payoff $u_{1,2}^Q(x_1^*,x_2^*)$  for selected values of the phase parameter $\theta$.  ]{\label{subfig1b}\includegraphics[scale=0.45]{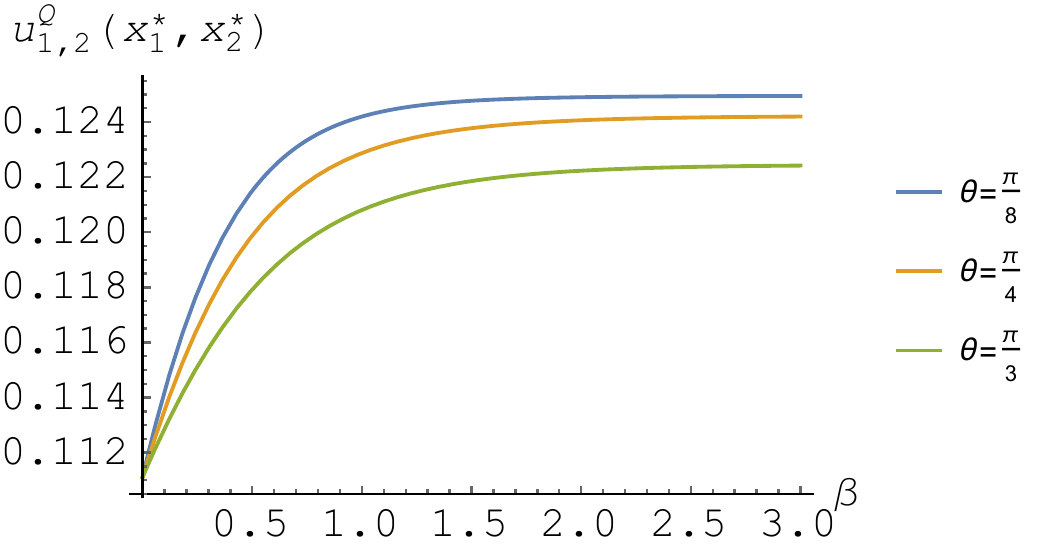}}\end{center}
	\caption{The payoff at quantum Nash equilibrium described by eq.~\eqref{ab35} for $k=1$.} \label{F1}	
\end{figure}

Let us examine the payoff function defined by the eq.~\eqref{ab35} in greater detail. For higher values of phase parameter its values increase at a slower rate with an increase in the $\beta$ parameter (see Fig.~\ref{F1}). The function $u_{1,2}^Q(x_1^*,x_2^*)$   goes to $\frac{k^2(1+\cos\theta)}{(3+\cos\theta)^2}$ when $\beta\rightarrow\infty$. Hence, in the limit, it attains its minimal value for $\theta=\frac{\pi}{2}$, which is equal to the payoff in the classic Cournot model, and its maximal value for $\theta=0$, which is equal to $\frac{k^2}{8}$. As illustrated in Fig.~\ref{F1}, for the $\theta\in(0,\frac{\pi}{2})$ the  players cannot reach the maximum value of payoff in the model depending on only one squeezing parameter $\beta$.

\begin{figure}[h!]
	\begin{center}
		\subfigure[$\phi=\theta=\frac{\pi}{4}$]{\label{subfig2a}\includegraphics[width=0.3\textwidth]{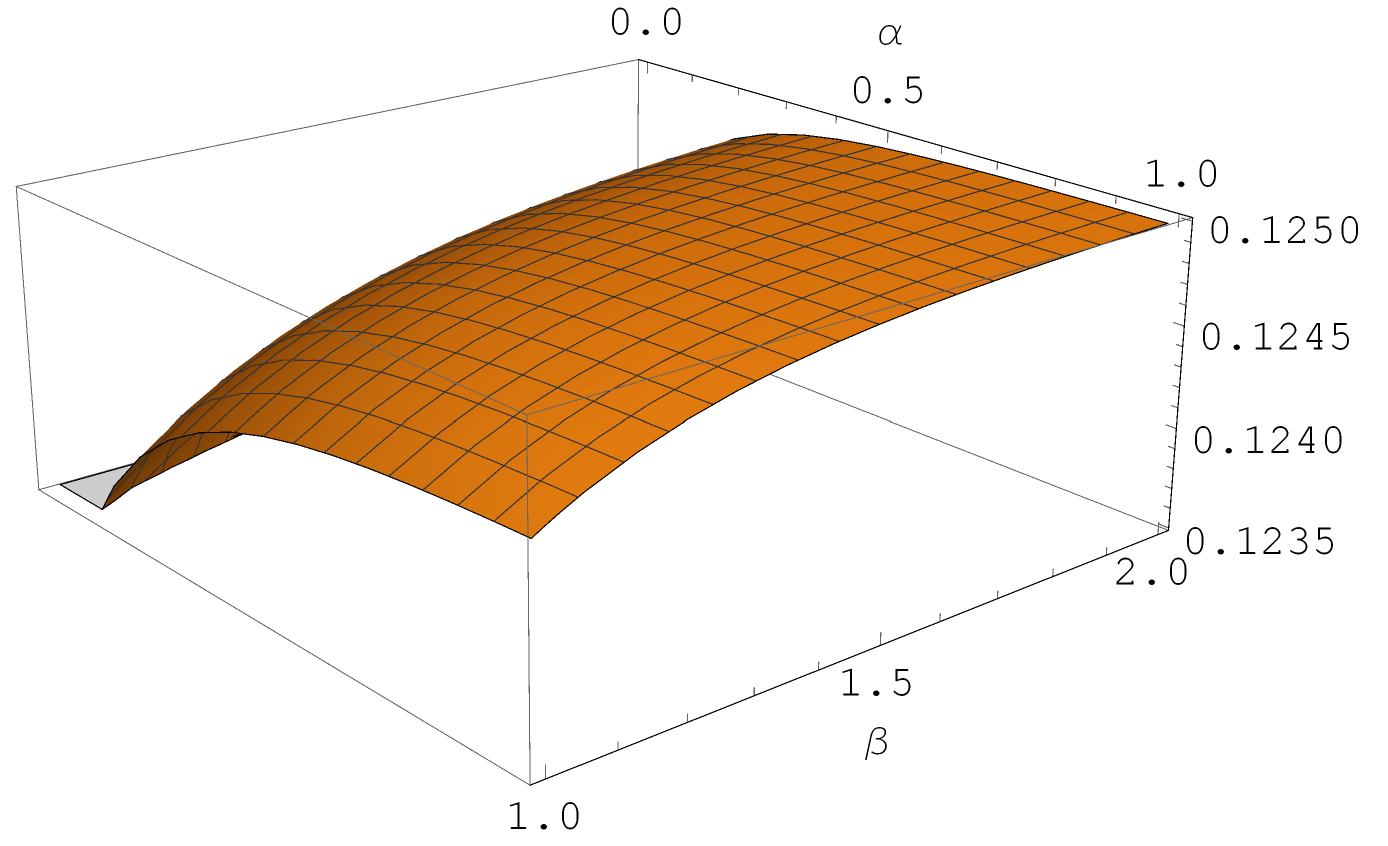}}\,\,
	\subfigure[$\phi=\frac{\pi}{3}$, $\theta=\frac{3\pi}{4}$]{\label{subfig2b}\includegraphics[width=0.3\textwidth]{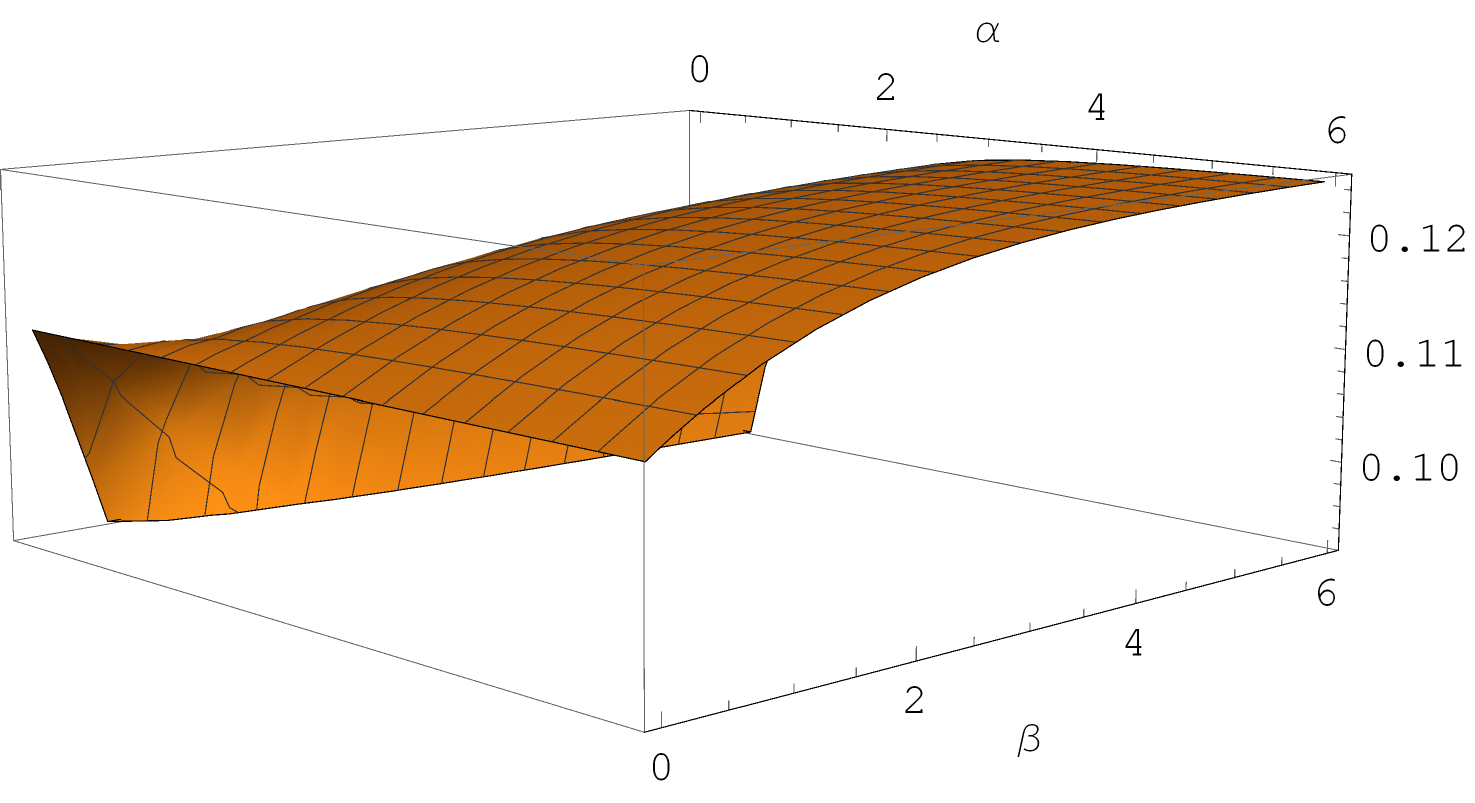}}
	\subfigure[$\phi=\frac{3\pi}{4}$, $\theta=\frac{\pi}{3}$]{\label{subfig2c}\includegraphics[width=0.3\textwidth]{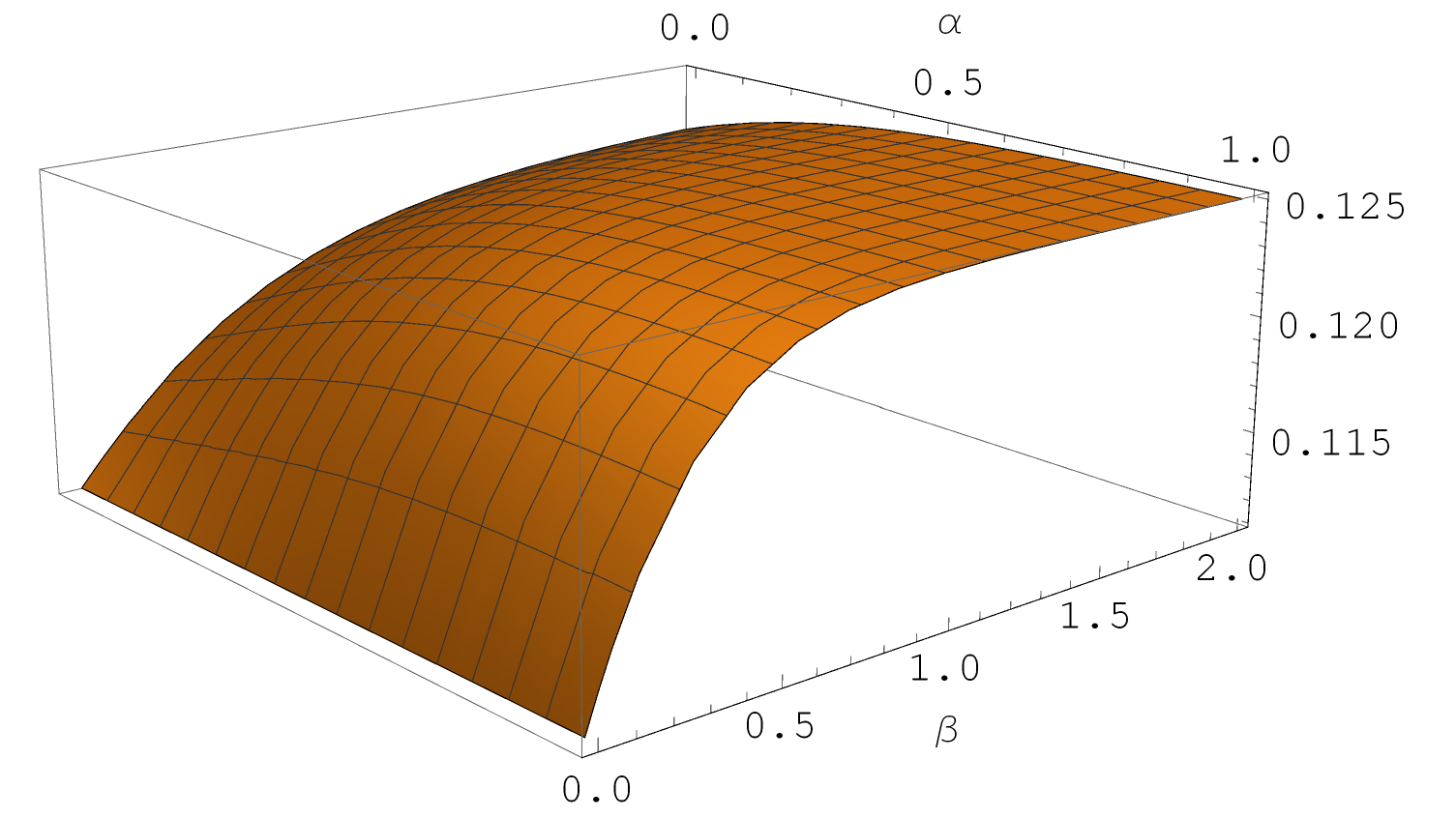}}
\end{center}
	\caption{The payoffs at quantum Nash equilibrium described by eq.~\eqref{ab33} as a function of the squeezing parameters $\alpha$, $\beta$; $k=1$.} 	\label{F2}
\end{figure}

The payoff function described by eq.~\eqref{ab33} depends on four parameter. Obviously  $\tilde{u}_{1,2}^Q(\tilde{x}_1^*,\tilde{x}_2^*)$ reduces to	$u_{1,2}^Q(x_1^*,x_2^*)$, $i=1,2$, for $\delta=0$. The graphs of this function for selected values of phase parameters are given on Figure \ref{F2}. 
In contrast, when the entanglement operator depends on two squeezing parameters, the payoff function $\tilde{u}_{1,2}^Q(\tilde{x}_1^*,\tilde{x}_2^*)$ can reach its maximum value even if $\theta$ and $\phi$ are different from zero, see Figure \ref{F2}. 
 In particular, when $\theta=\phi=\frac{\pi}{4}$ (Fig.~\ref{subfig2a}) and $\beta\rightarrow\infty$ the payoff goes to 0.1242 for $\alpha=0$ and to 0.125 for $\alpha\rightarrow\infty$. As  can be seen in Fig.~\ref{subfig2a} the function $\tilde{u}_{1,2}^Q(\tilde{x}_1^*,\tilde{x}_2^*)$ reaches its maximum value very quickly: for $\alpha=1$ and $\beta=2$ the payoff is already 0.1249. When the values of the phase parameters are different, as in Figs.~\ref{subfig2b} and \ref{subfig2c}, reaching the maximum possible payoff sometimes requires higher values of the squeezing parameters. With $\alpha=0$ and $\beta\rightarrow\infty$ the payoff (Fig.~\ref{subfig2b}) is only 0.0557; to get 0.1249 the parameter $\alpha$ needs to be about 4.5. A comparison of Figs.~\ref{subfig2b} and \ref{subfig2c} suggests that the phase parameter $\theta$ has a greater effect on reducing the payoff value with fixed $\alpha$ and $\beta$ than the parameter $\phi$.
In the case of the function described in Fig.~\ref{subfig2c}, substituting $\alpha=0$ and going with $\beta$ to infinity, we obtain the payoff equal to 0.1224, whereas a value equal to 0.1249 is already obtained, for example, with $\alpha=0.7,\beta=2$. 

\begin{figure}[h!]
	\begin{center}
		\includegraphics[scale=0.4]{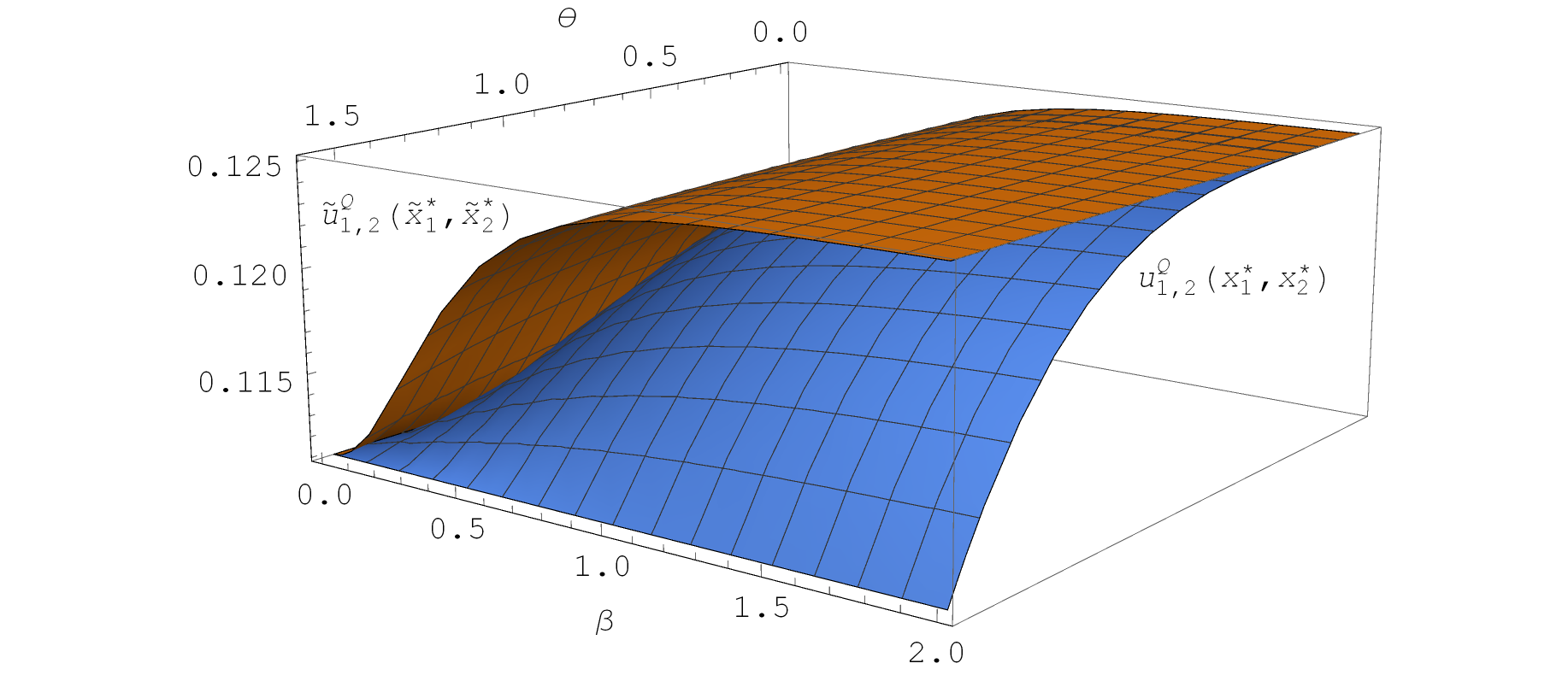}\end{center}
	\caption{The payoffs at quantum Nash equilibrium described by eqs.~\eqref{ab35} (the blue surface) and \eqref{ab33} (the orange surface) for $\phi=\theta$ and $\alpha=\beta $; $k=1$.} \label{F3}	
\end{figure}

If we put $\theta=\phi$ and $\alpha=\beta$ in the equation \eqref{ab33}, the result is a function that depends on only two parameters, just like the function ${u}_{1,2}^Q({x}_1^*,{x}_2^*)$ described by eq.~\eqref{ab35}. The graphs of both functions are shown in Fig.~\ref{F3}. 
We can see that the phase parameter, which has a significant effect on the payoff ${u}_{1,2}^Q({x}_1^*,{x}_2^*)$, which reaches a minimum value for $\theta=\frac{\pi}{2}$ despite the increasing value of the parameter $\beta$, does not affect to the same extent the function $\tilde{u}_{1,2}^Q(\tilde{x}_1^*,\tilde{x}_2^*)$ which reaches a value of about 0.1244 for $\theta=\frac{\pi}{2}$ and $\beta =1$.

\section{Degree of entanglement versus payoff of the quantum game}\label{sec4}

Both for the games with discrete as well as continuous sets of strategies, the advantage of quantum games as compared to their classical counterparts is due to the entanglement of the initial state of the game. For example, as one can infer from Figure \ref{F3}, the largest payoff values in a Nash equilibrium are obtained with maximally entangled states. As the entanglement parameter approaches zero the quantum game coincides with its classical counterpart.

Let us compare the degrees of entanglement of the games based on $\hat{J}(\xi)$ and $\hat{J}_1(\delta,\xi)$ (defined by eqs.~\eqref{ac34} and \eqref{a15}) and their payoff values in a Nash equilibrium. To determine the degree of entanglement of a game, it is necessary to use some measure of entanglement.  For Gaussian states, which we are dealing with, many measures of entanglement can be found  in the literature \cite{plenio},\cite{adesso}. 

One of the simplest measures is the entropy of entanglement which, for a pure state, can be expressed in terms of the covariance matrix. In the case of a bipartite Gaussian pure state, the entanglement entropy is of the form \cite{Tommaso}
\begin{equation}
	S=\sum_{i=1}^{n_{sub}}\com{\naw{\mu_i+\frac{1}{2}}\log_2\naw{\mu_i+\frac{1}{2}}-\naw{\mu_i-\frac{1}{2}}\log_2\naw{\mu_i-\frac{1}{2}}}\label{k38}
\end{equation}
where $\mu_i$ are the symplectic eigenvalues of the correlation matrix $\sigma$ of any of the subsystems. It remains to explain what symplectic eigenvalues are. 

Let us define the column vector
\begin{equation}
	\hat{r}=\left(\begin{array}{c}
		\hat{X}_1\\
			\hat{P}_1\\
				\hat{X}_2\\
				\hat{P}_2
\end{array}\right)		
\end{equation}
where $\hat{X}_k$, $\hat{P}_k$ , $k=1,2$, are position- and momentum operators with the canonical commutator $[\hat{X}_k,\hat{P}_l]=i\delta_{kl}$. The vector $\hat{r}$ obeys the following commutation relations
\begin{equation}
	[\hat{r}_k,\hat{r}_l]=i\Omega_{kl}
\end{equation}
where 
\begin{equation}
	\Omega=\bigoplus\limits_{k=1}^{n}\Bigg(\begin{matrix}
		0 & 1 \\
		-1 & 0\end{matrix}\Bigg)=\mathbbm{1}_n\otimes \Bigg(\begin{matrix}
		0 & 1 \\
		-1 & 0\end{matrix}\Bigg)\label{k42a}	
\end{equation}
is a symplectic matrix, $\mathbbm{1}_n$ being the $n\times n$ identity matrix.
The inverse of the matrix $\Omega$ is
\begin{equation}
	\Omega^{-1}=\Omega^T=-\Omega. \label{k42}
\end{equation}
For Gaussian states we can define Gaussian unitary operation as a transformation that maps a given Gaussian state onto another Gaussian state. Such a transformation can be assigned a unique symplectic transformation, $S\in Sp(2n,\mathbbm{R})$, which preserves the commutation relations
\begin{equation}
	i\Omega=[\hat{r}'_k,\hat{r}'_l]=S[\hat{r}_k,\hat{r}_l]S^T,
\end{equation} 
thus $\Omega=S\Omega S^T$.\\
The covariance matrix of n-mode system defined by
\begin{equation}
	\sigma_{kl}=\frac{1}{2}\left<\{\hat{r}_k,\hat{r}_l\}\right>-\left<\hat{r}_k\right>\left<\hat{r}_l\right>,\label{k44}
\end{equation}
where $\{\cdot,\cdot\}$ denotes an anti-commutator, transforms under the action of a symplectic transformation as follows
\begin{equation}
	\sigma'=S\sigma S^T.
\end{equation}
According to Williamson's theorem \cite{williamson}, any real symmetric positive-define matrix of order $2n\times 2n$ can be diagonalized by an appropriate symplectic transformation $S_w\in Sp(2n,\mathbbm{R})$:
\begin{equation}
	S_w\sigma S_w^T=\sigma_w
\end{equation}
where 
\begin{equation}
	\sigma_w=diag(\mu_1,\mu_2,\ldots,\mu_n,\mu_1,\mu_2,\ldots,\mu_n).
\end{equation}
 All the $\mu_i$ are real and the $n$ distinct eigenvalues of $\sigma_w$ are the symplectic eigenvalues of the covariance matrix $\sigma$. 
 
 In order to find the values of the $\mu_i$, let us consider, for any given covariance matrix $\sigma$, a new matrix $K$  such that
 \begin{equation}
 	\sigma=-K\Omega.\label{k48}
 \end{equation} 
By multiplying both sides of the equation \eqref{k48} from the right by $\Omega^{-1}$ and using the equation \eqref{k42}, we obtain
\begin{equation}
	K=\sigma\Omega.
\end{equation}
If two covariance matrices, denoted by $\sigma$ and $\sigma'$, are related by a symplectic transformation, then the corresponding $K$ and $K'$ matrices are related by a similarity transformation
\begin{equation}
	K'=\sigma'\Omega=S\sigma S^T\Omega=S\sigma \Omega S^{-1}=SKS^{-1}
\end{equation}
which means that the matrices K and K' have the same eigenvalues. It can be checked that the eigenvalues of the matrix $K'=\sigma_w\Omega$ are equal to $\{\pm i\mu_i\}$. Thus, in order to find the symplectic eigenvalues of the covariance matrix $\sigma$, it is sufficient to determine the eigenvalues of the matrix $K=\sigma\Omega$.

Now we can find how the entanglement entropy of initial state depends on the entanglement operator.
\begin{enumerate}[label={\Alph*.}]
\item \textbf{ The degree of entanglement of the initial state $\ket{\Psi_0}=\hat{J}(\xi)\ket{00}$} \\
First, we determine the covariance matrix of one of the subsystems (as the covariance matrices for both subsystems are identical). According to the eq.~\eqref{k44} it reads
\begin{equation}
	\sigma_1=\begin{pmatrix}
	\frac{1}{2}\cosh(2\beta) & 0\\
	0 & \frac{1}{2}\cosh(2\beta)
	\end{pmatrix}
\end{equation}
from which we can directly determine the symplectic eigenvalues.
Thus, the entanglement entropy, eq.~\eqref{k38} has form
\begin{equation}
	S_1=\frac{\cosh(2\beta)\ln(\coth\beta)+\ln(\cosh\beta\sinh\beta)}{\ln 2}.\label{ac52}
\end{equation}
As we can see from Figure \ref{F4} the values of the entanglement measure $S_1$ increase with the $\beta$.
\begin{figure}[!h]
	\begin{center}
	\includegraphics[scale=0.4]{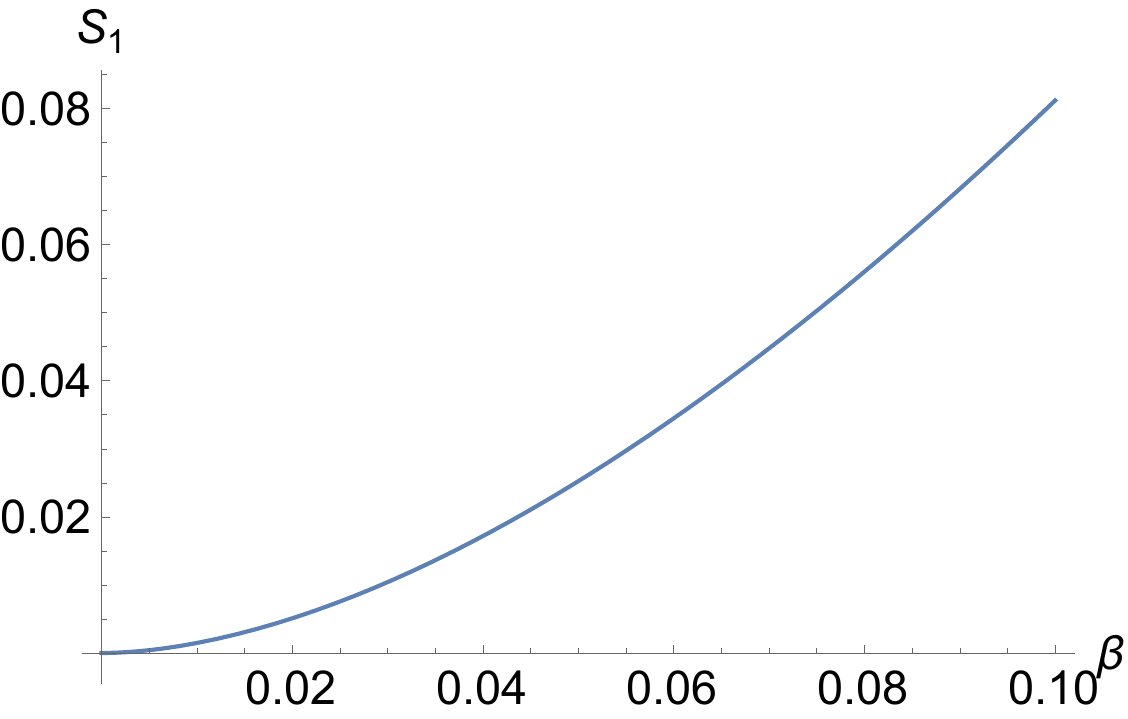}\end{center}
\caption{The entanglement entropy of state $\ket{\Psi_0}=\hat{J}(\xi)\ket{00}$}\label{F4}
\end{figure} 

\item \textbf{The degree of entanglement of the initial state $\big|\tilde{\Psi}_0\big>=\hat{J_1}(\delta,\xi)\ket{00}$} \\
In the case of an initial state based on the entanglement operator $\hat{J}_1(\delta,\xi)$, the covariance matrix $\tilde{\sigma}_1$ of the first subsystem has a more complicated form (again the covariance matrix of the second subsystem is identical to that of the first). Its matrix elements are as follows
\begin{equation}
	\begin{split}
	& \tilde{\sigma}_{11}=\frac{1}{8}(2(\cosh(2a)+\cosh(2b))-(\frac{1}{ad}+ad)\sinh(2a)-(\frac{1}{bf}+bf)\sinh(2b)),\\
	&\tilde{\sigma}_{12}=\tilde{\sigma}_{21}=\frac{1}{8}((ad-\frac{1}{ad})\sinh(2a)+(bf-\frac{1}{bf})\sinh(2b)),\\
	& \tilde{\sigma}_{22}=\frac{1}{8}(2\cosh(2a)+2\cosh(2b)+(ad+\frac{1}{ad})\sinh(2a)+(bf+\frac{1}{bf})\sinh(2b))
\end{split}\end{equation}
where $a,\, b,\, d,\,f$ are defined by eq.~\eqref{ab27}.
According to the procedure described above, we multiply the covariance matrix of the subsystem $\tilde{\sigma}_1$ by the symplectic matrix $\Omega$, defined by eq.~\eqref{k42a}, and then we look for the eigenvalues of the resulting matrix. This procedure yields the following symplectic eigenvalues
\begin{small}
\begin{equation}
	\begin{split}
\tilde{\mu}_1&=\tilde{\mu}_2=\frac{1}{4}\sqrt{2+2\cosh(2a)\cosh(2b)-(\frac{ad}{bf}+\frac{bf}{ad})\sinh(2a)\sinh(2b)}\\
&=\frac{1}{4}\bigg(2+2\cosh(2\sqrt{4\alpha^2+\beta^2-4\alpha\beta\cos(\theta-\phi)})
\cosh(2\sqrt{4\alpha^2+\beta^2+4\alpha\beta\cos(\theta-\phi)})\\
& -\frac{2(4\alpha^2-\beta^2)\sinh(2\sqrt{4\alpha^2+\beta^2-4\alpha\beta\cos(\theta-\phi)})
	\sinh(2\sqrt{4\alpha^2+\beta^2+4\alpha\beta\cos(\theta-\phi)})}{\sqrt{16\alpha^4+\beta^4-8\alpha^2\beta^2\cos(2\theta-2\phi)}}\bigg)^{1/2}\end{split}\label{ad54}
\end{equation}\end{small}
 and the entanglement entropy
 \begin{equation}
 	\tilde{S}_1= \frac{1}{\ln 4}\naw{(1+2\tilde{\mu}_1)\ln\bigg(\tilde{\mu}_1+\frac{1}{2}\bigg)+(1-2\tilde{\mu}_1)\ln\bigg(\tilde{\mu}_1-\frac{1}{2}\bigg)}.
 \end{equation}
 The eigenvalues \eqref{ad54} are periodic functions of $\theta$ and $\phi$ with period $\pi$ and depend on the difference between the phase parameters. Therefore, the entropy function $\tilde{S}_1$ actually depends on three parameters $\alpha,\beta>0$, and $0\leq \theta-\phi< \pi$.

Let us see how the function $\tilde{S}_1$ behaves in a few special cases:
\begin{itemize}
	\item [a)] for $\phi=\theta$ and $\alpha,\beta>0$ we obtain $\tilde{S}_1=S_1$, where $S_1$ is described by eq.~\eqref{ac52};
	\item[b)] for $\alpha=\beta$ and $0\leq\theta-\phi< \pi$, the symplectic eigenvalue and the entanglement entropy increase with the entanglement parameter $\beta$; the fastest grow corresponds to  $\theta-\phi=\frac{\pi}{2}$, see Figure \ref{F5}.
	
\begin{figure}[!h]
	\begin{center}
			\subfigure[]{\label{fig5a}\includegraphics[width=0.4\textwidth]{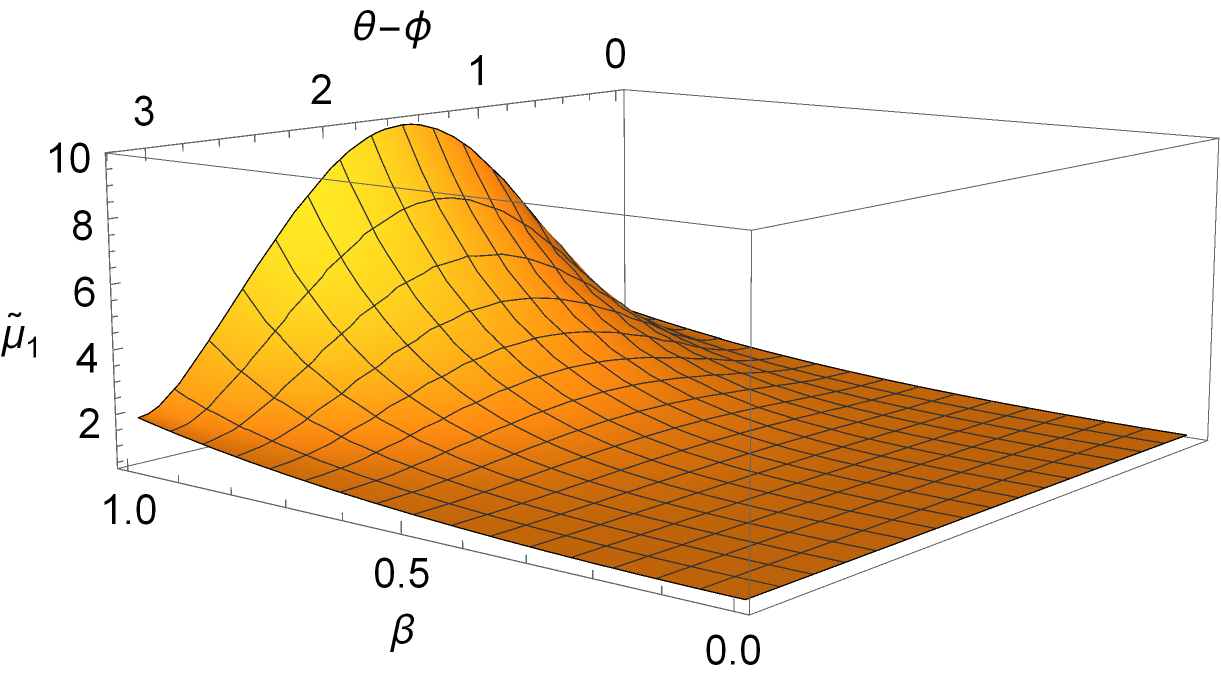}}\quad
		\subfigure[]{\label{fig5b}\includegraphics[width=0.4\textwidth]{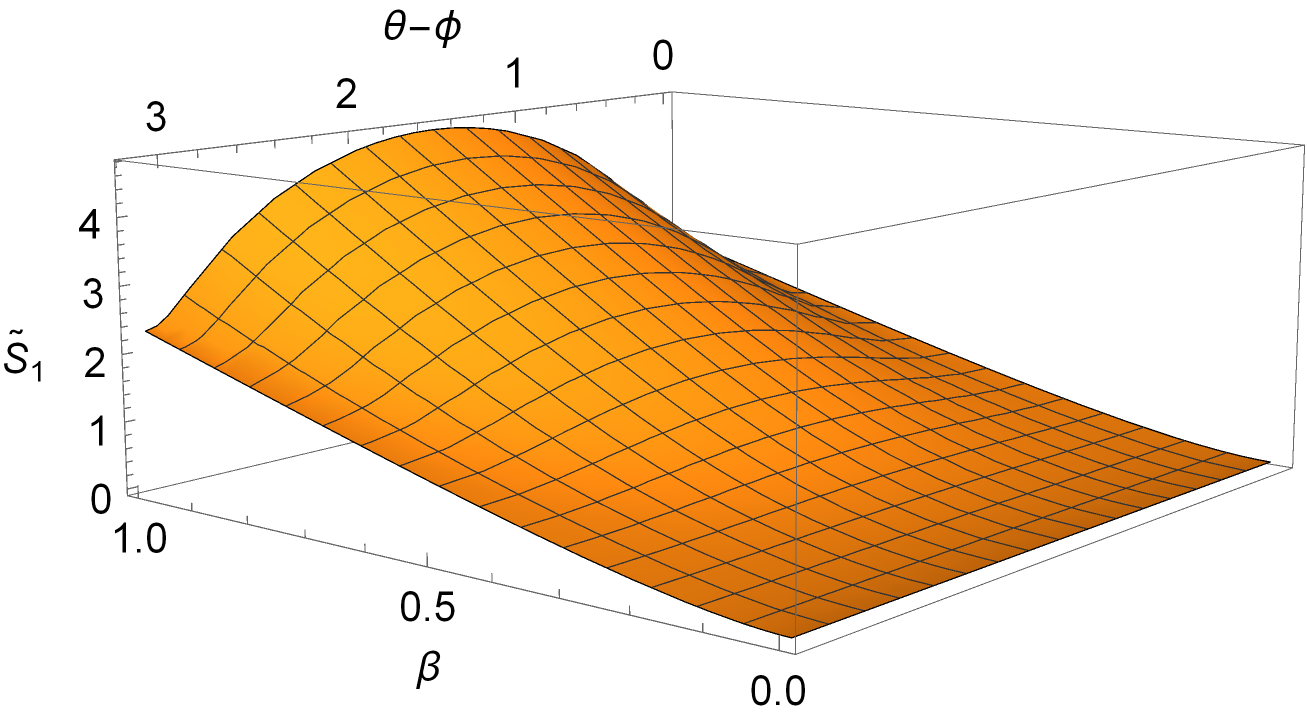}}\end{center}
	\caption{The symplectic eigenvalue $\tilde{\mu}_1$ (a) of the covariance matrix of subsystem and the entanglement entropy $\tilde{S}_1$ (b) for $\alpha=\beta$ and $0\leq\theta-\phi<\pi$.}\label{F5}
\end{figure} 
\item[c)] for $\alpha,\beta>0$ and $\theta-\phi=\{\frac{\pi}{6},\frac{\pi}{4},\frac{\pi}{3}\}$, it can be seen that as $\alpha$ and $\beta$ increase, the entanglement entropy value also increases, see Figure \ref{F6}.
\begin{figure}[h!]
	\begin{center}
		\subfigure[$\theta-\phi=\frac{\pi}{6}$]{\label{fig6b}\includegraphics[width=0.3\textwidth]{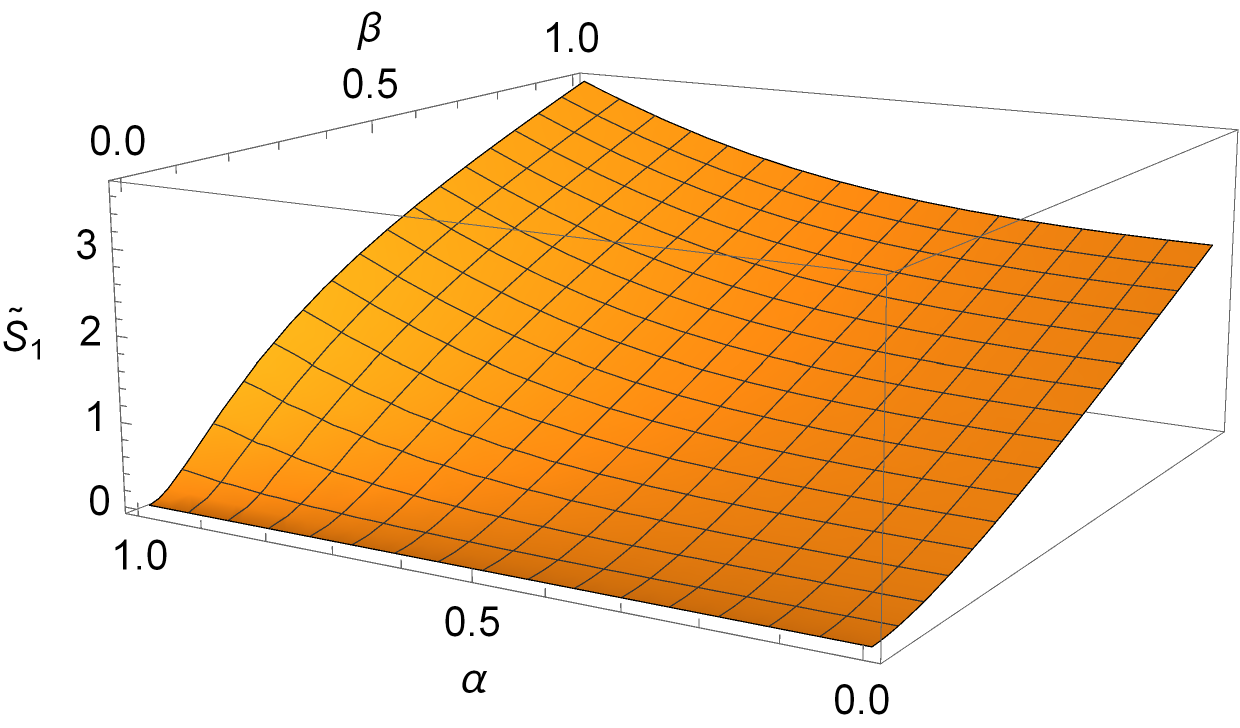}}\quad
		\subfigure[$\theta-\phi=\frac{\pi}{4}$]{\label{fig6a}\includegraphics[width=0.3\textwidth]{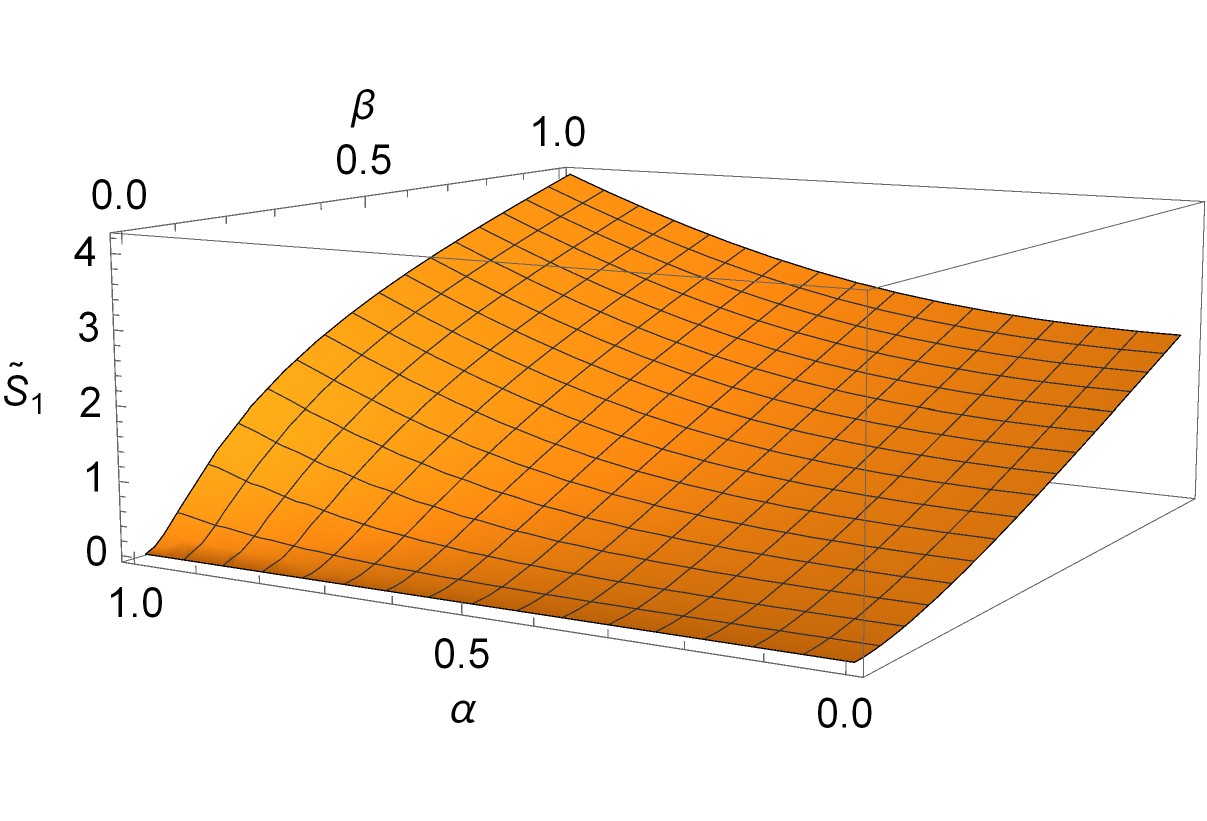}}\quad
		\subfigure[$\theta-\phi=\frac{\pi}{3}$]{\label{fig6c}\includegraphics[width=0.3\textwidth]{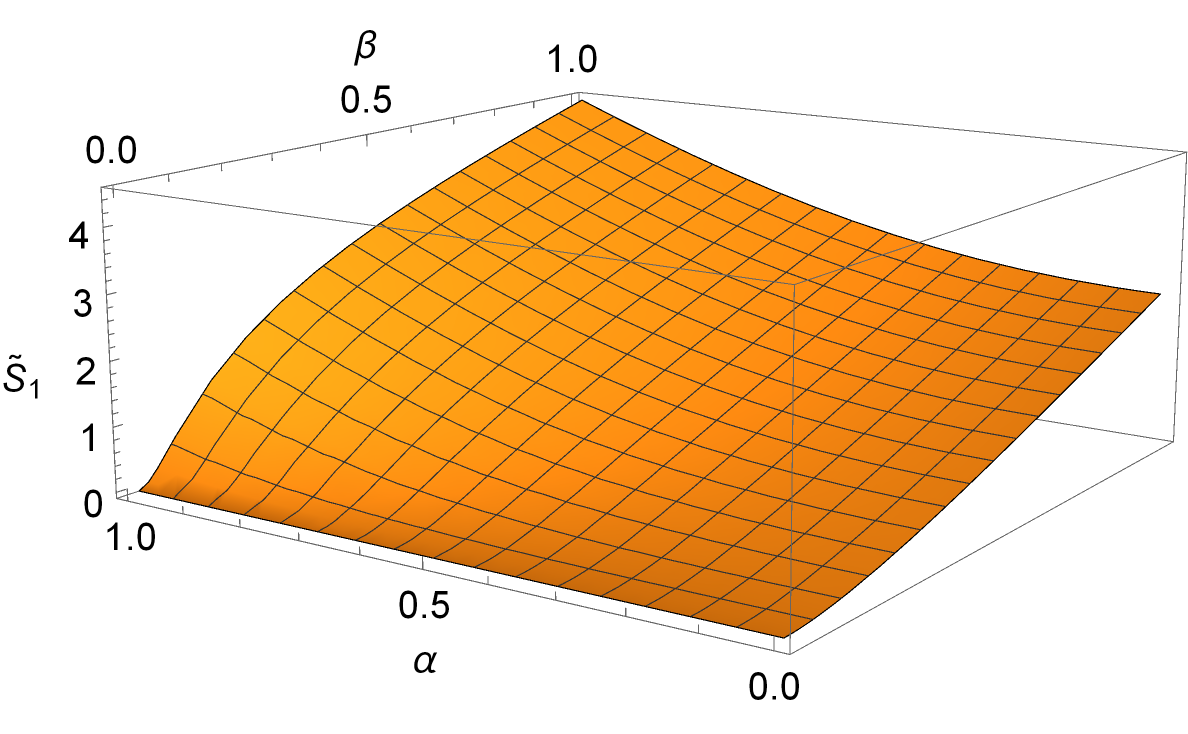}}	
	\end{center}
	\caption{The entanglement entropy $\tilde{S}_1$ for determined value of $\theta-\phi$.} \label{F6}	
\end{figure} 
\end{itemize} 
\end{enumerate}

Now, let us try to analyze how the payoffs of the players in the Cournot model depend on the degree of entanglement of the initial state of the game.
\newpage

\textbf{Case I.} 

\begin{figure}[h!]
	\begin{center}
	\subfigure[$\phi=\theta=\frac{\pi}{6}$]{\label{fig:edge-7a}\includegraphics[width=0.45\textwidth]{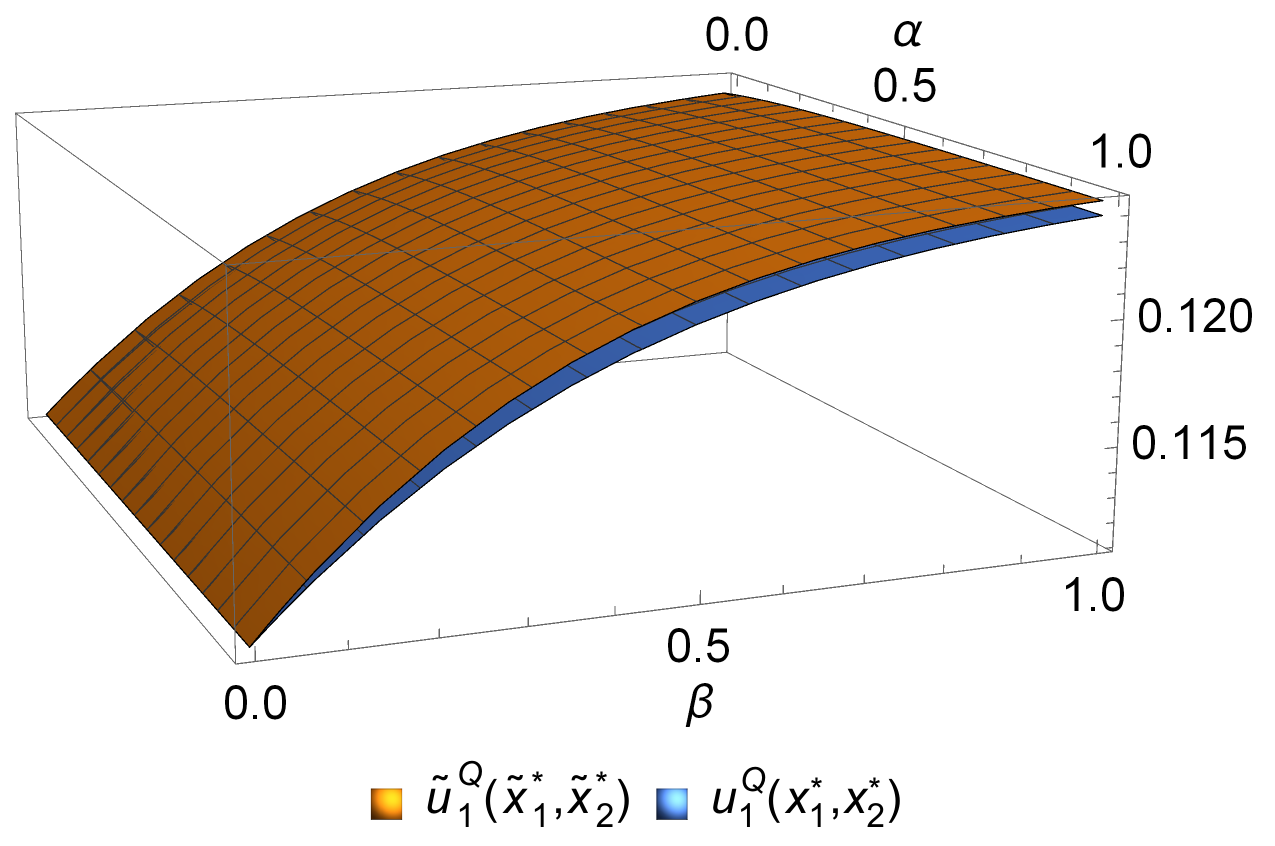}}\qquad
		\subfigure[the payoffs from subpoint (a) for selected $\alpha$]{\label{fig:edge-7b}\includegraphics[width=0.45\textwidth]{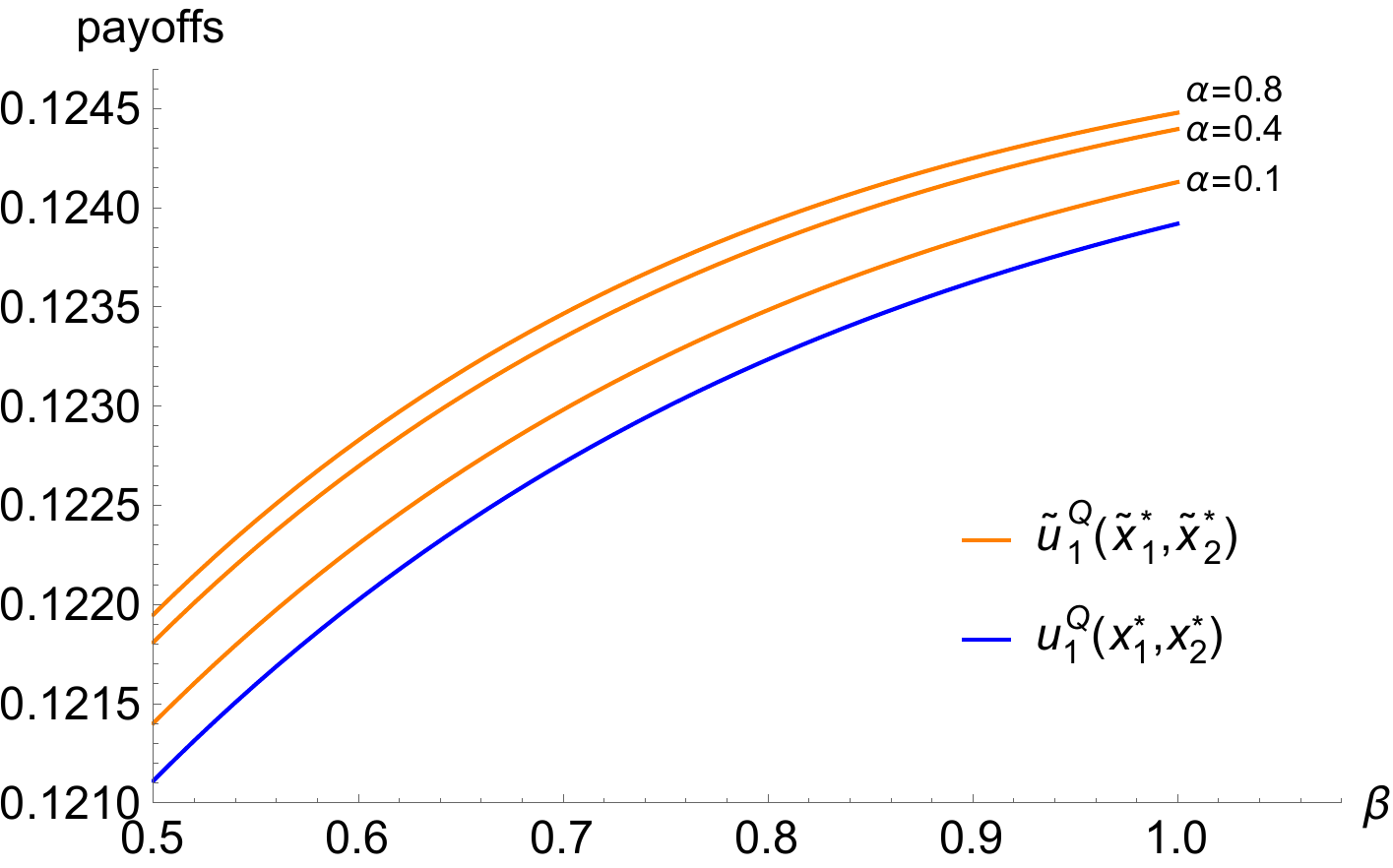}}\\
	\subfigure[$\phi=\theta=\frac{\pi}{3}$]{\label{fig:edge-7c}\includegraphics[width=0.45\textwidth]{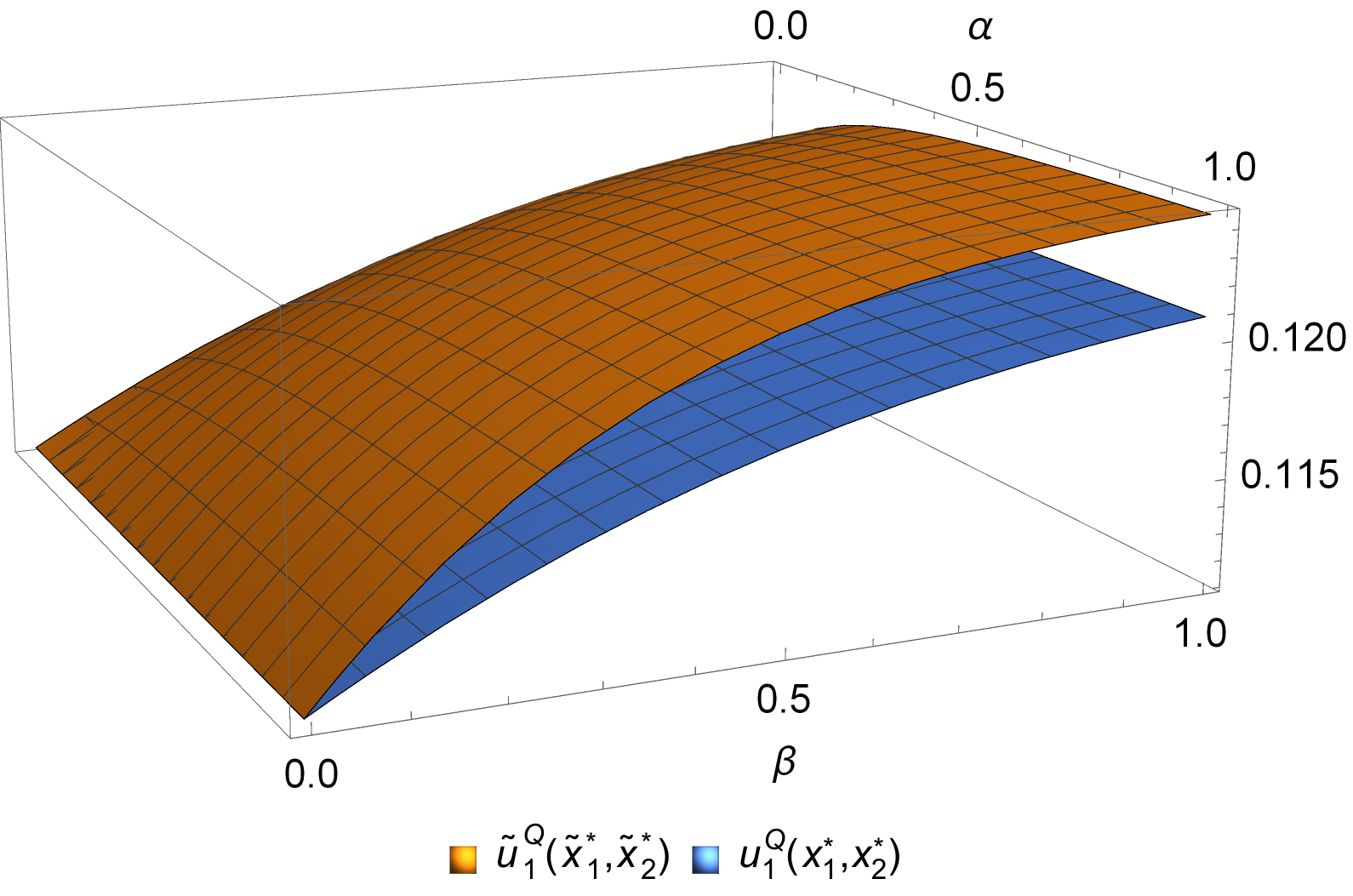}}\qquad
		\subfigure[the payoffs from subpoint (c) for selected $\alpha$]{\label{fig:edge-7d}\includegraphics[width=0.45\textwidth]{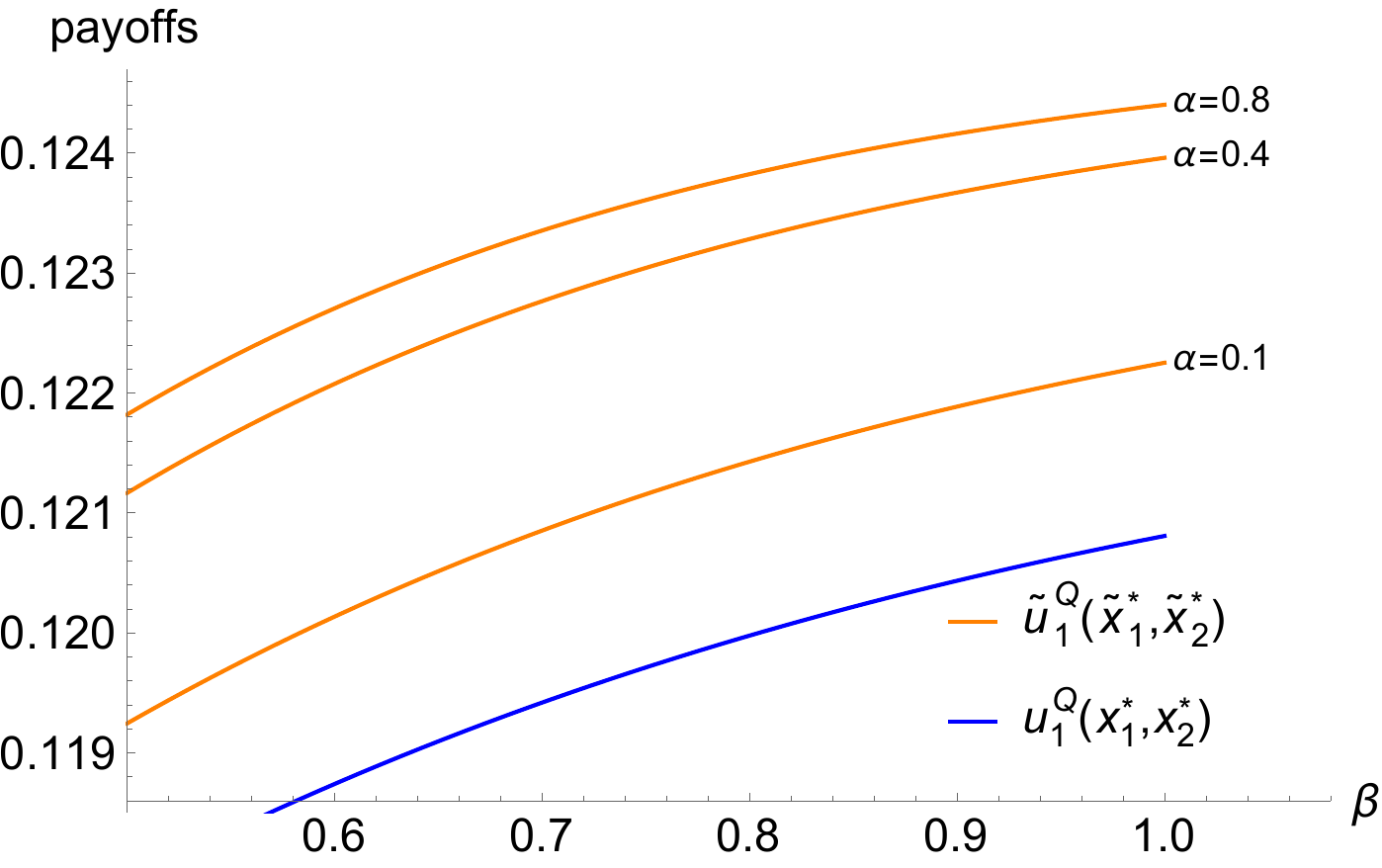}}\\
	\subfigure[$\phi=\theta=\frac{7\pi}{5}$]{\label{fig:edge-7e}\includegraphics[width=0.45\textwidth]{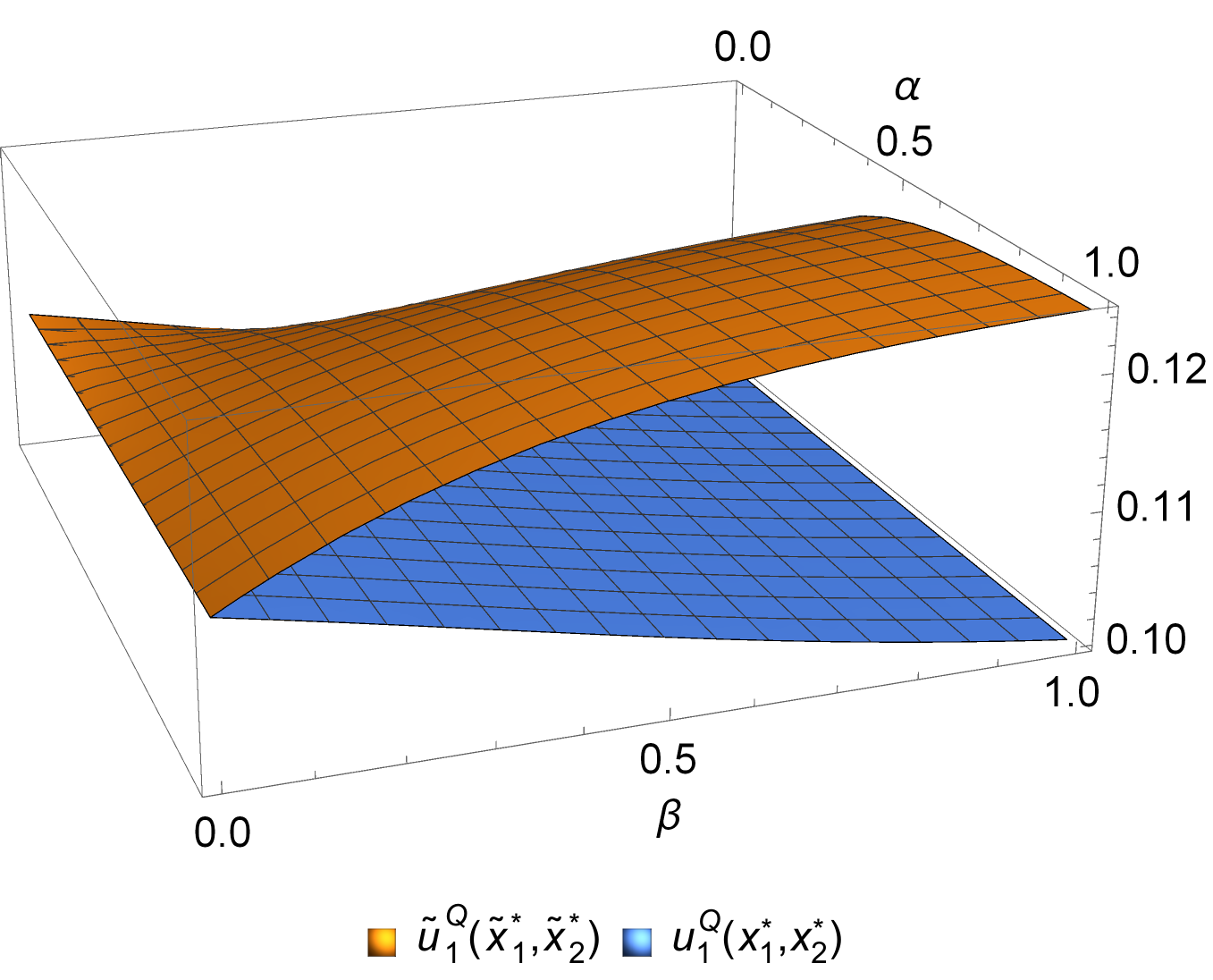}}\qquad
		\subfigure[the payoffs from subpoint (e) for selected $\alpha$]{\label{fig:edge-7f}\includegraphics[width=0.45\textwidth]{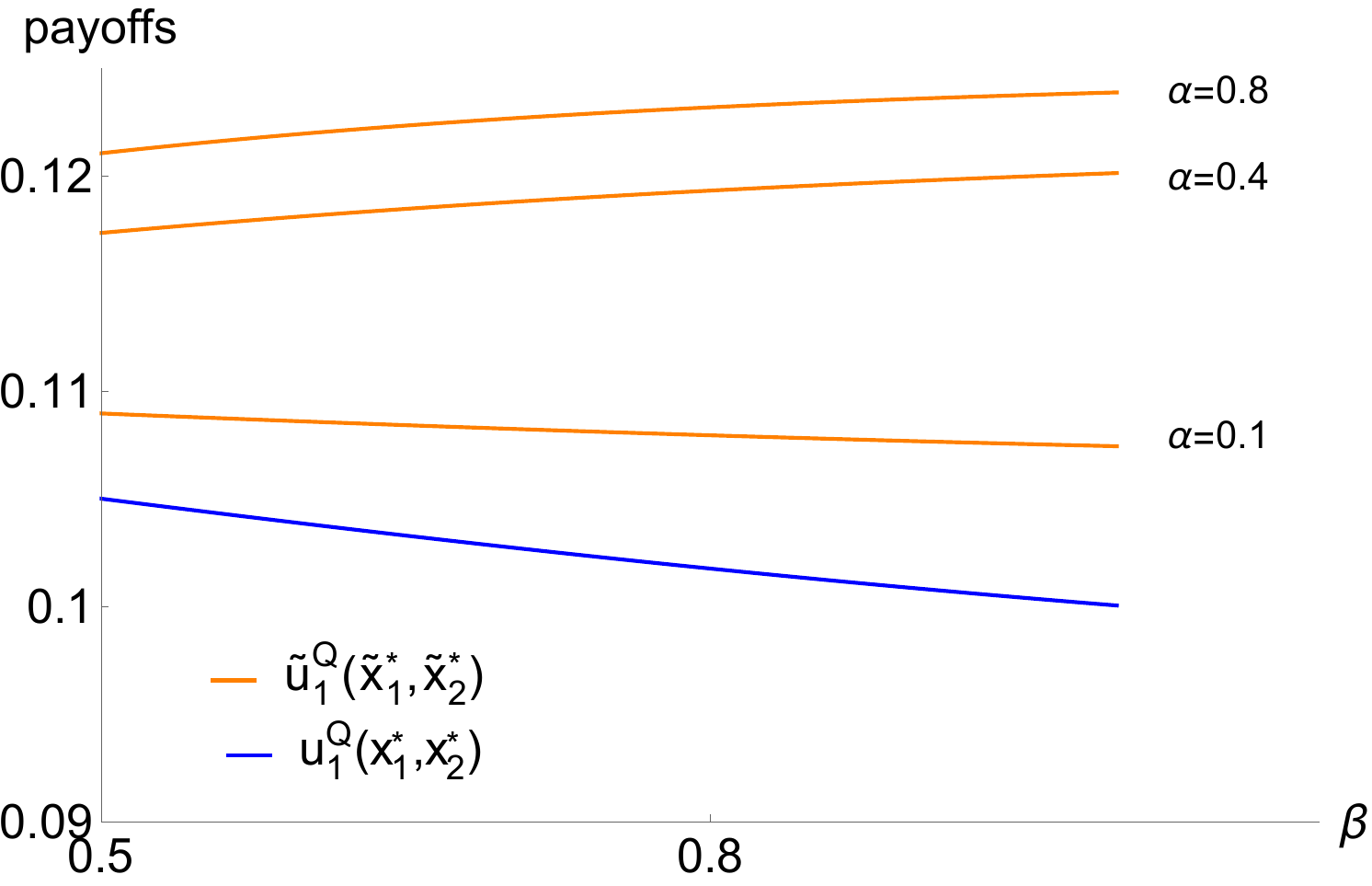}}\\
				
	\end{center}
	\caption {The payoffs in Nash equilibrium of games based on the entangled operators $\hat{J}(\xi)$ and $\hat{J}_1(\delta,\xi)$ for $\theta=\phi$, $k=1$.}\label{F7}	
\end{figure}

Let $\phi=\theta$; then, according to the measures $S_1$ and $\tilde{S}_1$, the entanglement degrees of the initial states $\ket{\Psi_0}$, $|\tilde{\Psi}_0\big>$ are the same, the entanglement measure depending on only one squeezing parameter $\beta$. However, the payoff functions based on these states are distinct and dependent on entanglement and phase parameters. Assuming specific values of the phase parameters (see Figure \ref{F7}), it can be observed that the payoff  $\tilde{u}_1^Q(\tilde{x}_1^*,\tilde{x}_2^*)$ (eq.~\eqref{ab33}) is  always higher than the payoff $u_1^Q(x_1^*,x_2^*)$ (eq.~\eqref{ab35}).

Furthermore, it can be observed that the difference between the payoffs of the games under consideration depend on the values of the parameters $\phi,\,\theta$. The payoff  $u_1^Q(x_1^*,x_2^*)$ is
\begin{itemize}  
	\item[-] an increasing function of the parameter $\beta$ for  $\theta\in\big[0,\frac{\pi}{2}\big)\cup\big(\frac{3\pi}{2},2\pi\big]$, 
	\item[-]  a constant function for $\theta=\{\frac{\pi}{2},\frac{3\pi}{2}\}$, 
	\item[-] a decreasing function as $\beta$ increases for  $\theta\in\big(\frac{\pi}{2},\frac{3\pi}{2}\big) $.
\end{itemize}
The effect of the parameter $\theta$ on $\tilde{u}_1^Q(\tilde{x}_1^*,\tilde{x}_2^*)$ can also be studied. For relatively small values of the parameter $\alpha$, the function $\tilde{u}_1^Q(\tilde{x}_1^*,\tilde{x}_2^*)$ behaves similarly to $u_1^Q(x_1^*,x_2^*)$, i.e. for $\theta\in\big(\frac{\pi}{2},\frac{3\pi}{2}\big)$ the payoff decreases as $\beta$ increases. However, as the value of the parameter $\alpha$ increases, the properties of the function $\tilde{u}_1^Q(\tilde{x}_1^*,\tilde{x}_2^*)$ change and it transforms into a function which increases with the parameter $\beta$ (see Fig. \ref{fig:edge-7e} and \ref{fig:edge-7f}).   

\begin{figure}[h!]
	\begin{center}
		\subfigure[$\alpha=0.3$, $\beta=0.5$]{\label{fig:edge-8a}\includegraphics[width=0.3\textwidth]{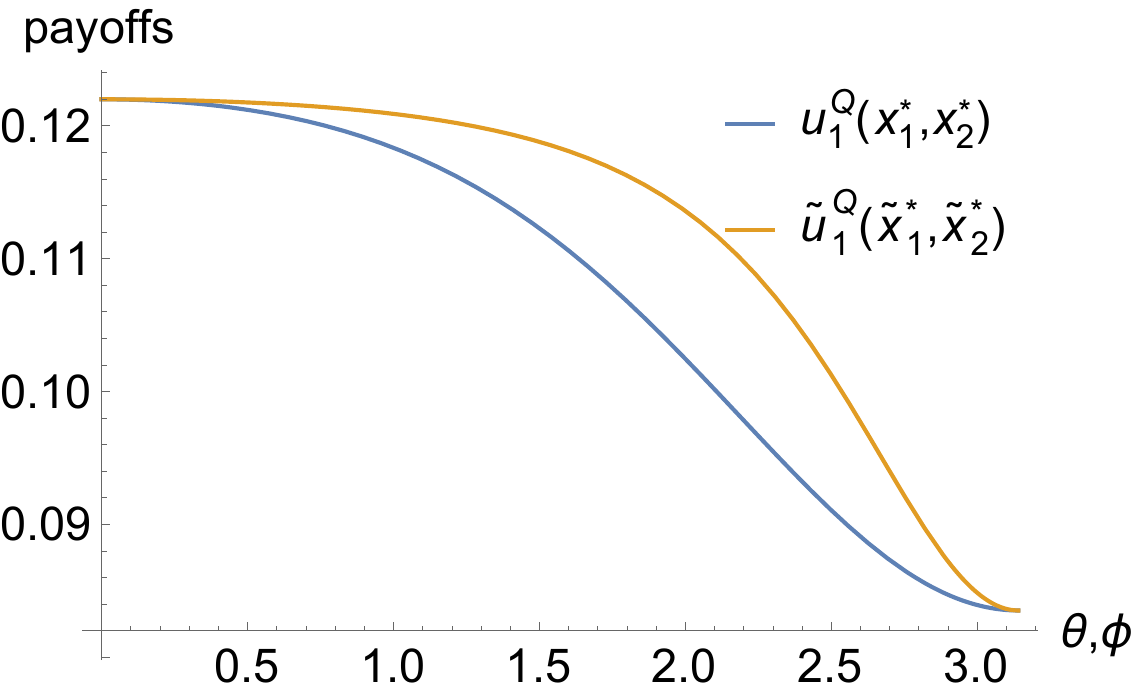}}\qquad
		\subfigure[$\alpha=0.1$, $\beta=1$]{\label{fig:edge-8b}\includegraphics[width=0.3\textwidth]{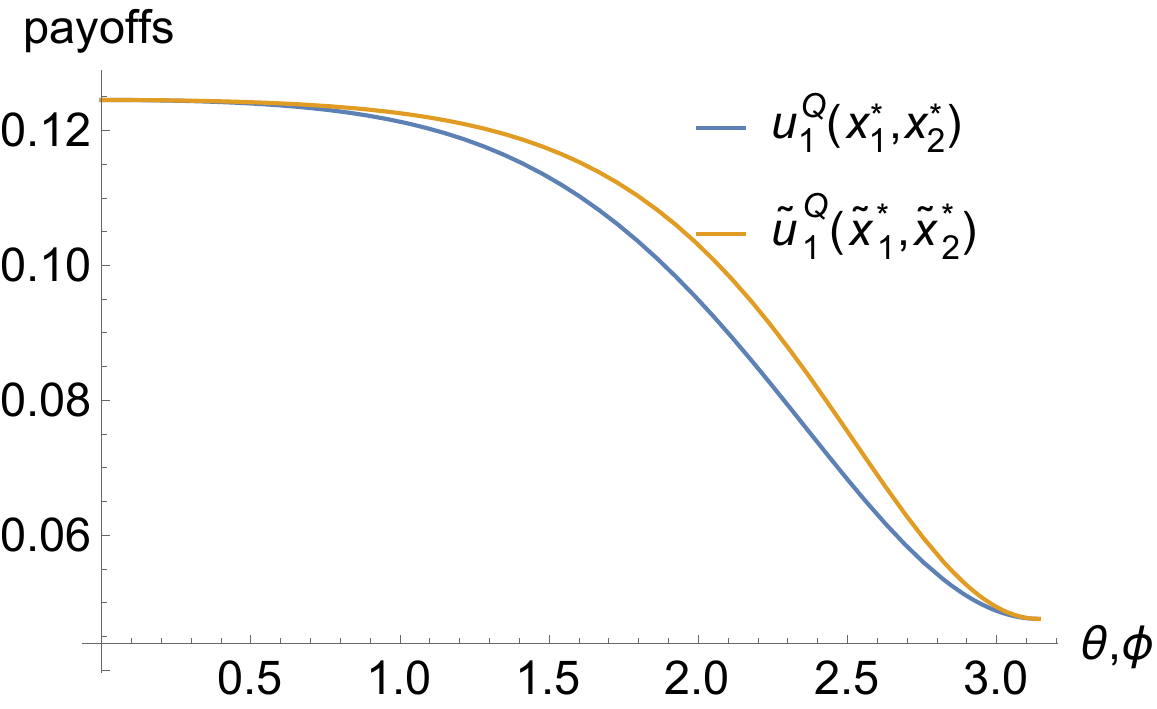}}
		\subfigure[$\alpha=1$, $\beta=0.2$]{\label{fig:edge-8c}\includegraphics[width=0.3\textwidth]{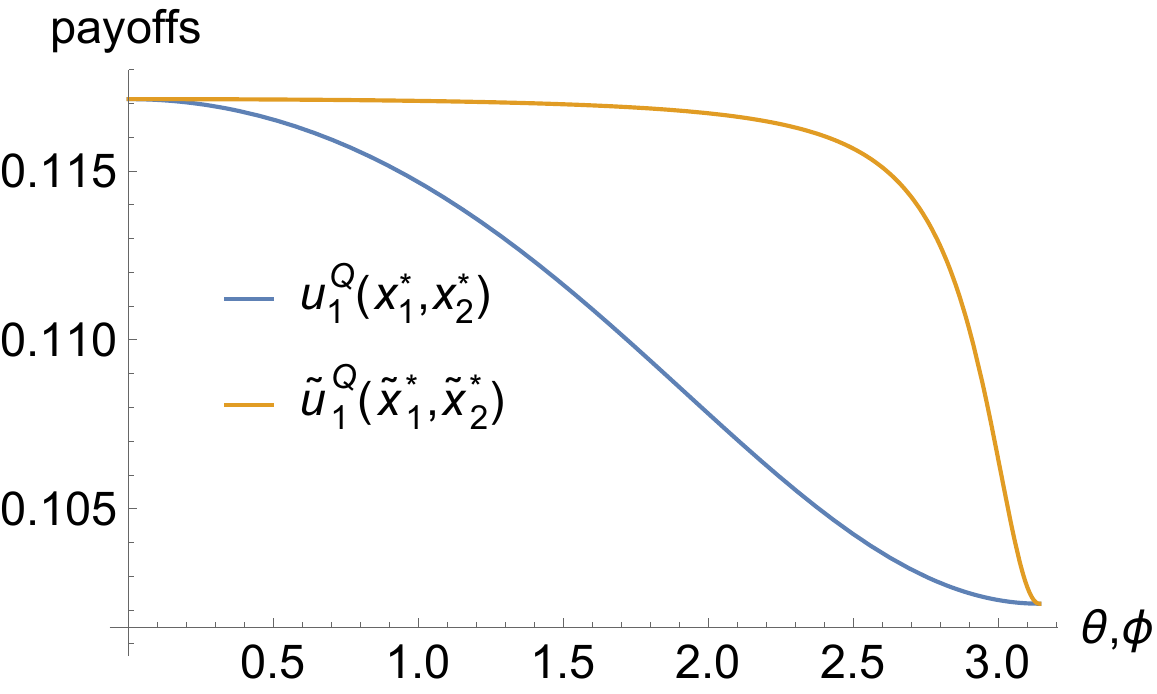}}\qquad
\end{center}
\caption{The payoffs $\tilde{u}_1^Q(\tilde{x}_1^*,\tilde{x}_2^*)$ and $u_1^Q(x_1^*,x_2^*)$ for selected values of parameters $\alpha$, $\beta$ and  $\theta=\phi$.} \label{F8}.	
\end{figure}

The dependence of payoff functions on the phase parameters for selected values of the entanglement parameters is illustrated on Figure \ref{F8}. It can be observed that as the value of the parameter $\alpha$ increases, the  the difference between the payoffs $\tilde{u}_1^Q(\tilde{x}_1^*,\tilde{x}_2^*)$ and $u_1^Q(x_1^*,x_2^*)$  also increases.

\vspace{1cm}

\textbf{Case II.} 

Now, let us see how the payoff functions behave when  $\alpha=\beta$  and $\theta,\phi\in\big[0,2\pi)$. In this case, the entanglement entropy $S_1$ has different form than $\tilde{S}_1$. The payoff function ${u}_1^Q({x}_1^*,{x}_2^*)$ depends on two variables, whereas the function $\tilde{u}_1(\tilde{x}_1^*,\tilde{x}_2^*)$ depends on three. Therefore, let us analyze some cases with specific values of one of the phase parameters. 

\begin{figure}[h!]
	\begin{center}
		\subfigure[$\phi=\frac{\pi}{3}$, ]{\label{fig9a}\includegraphics[width=0.3\textwidth]{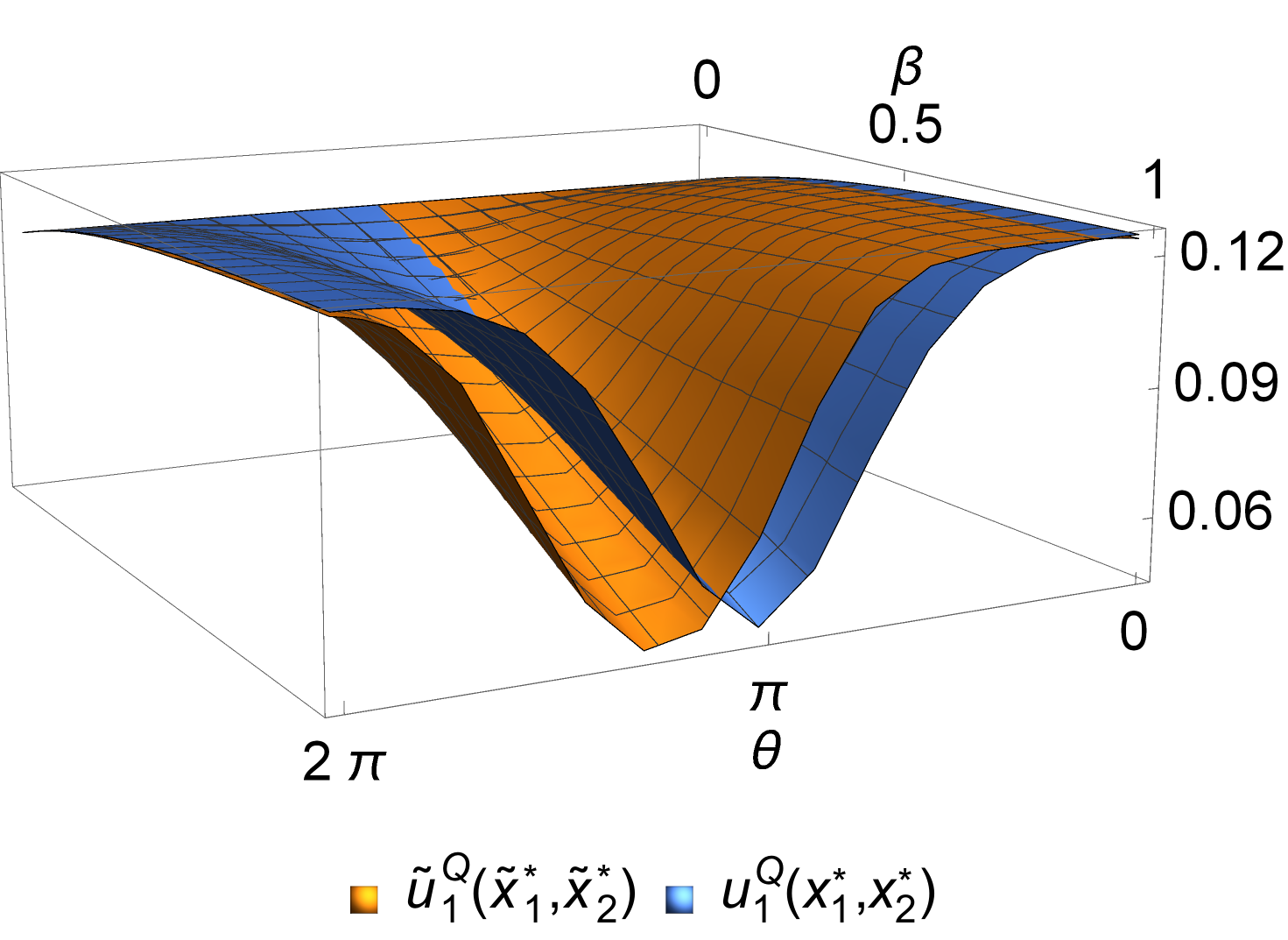}}\qquad
		\subfigure[$\phi=\pi$]{\label{fig9b}\includegraphics[width=0.3\textwidth]{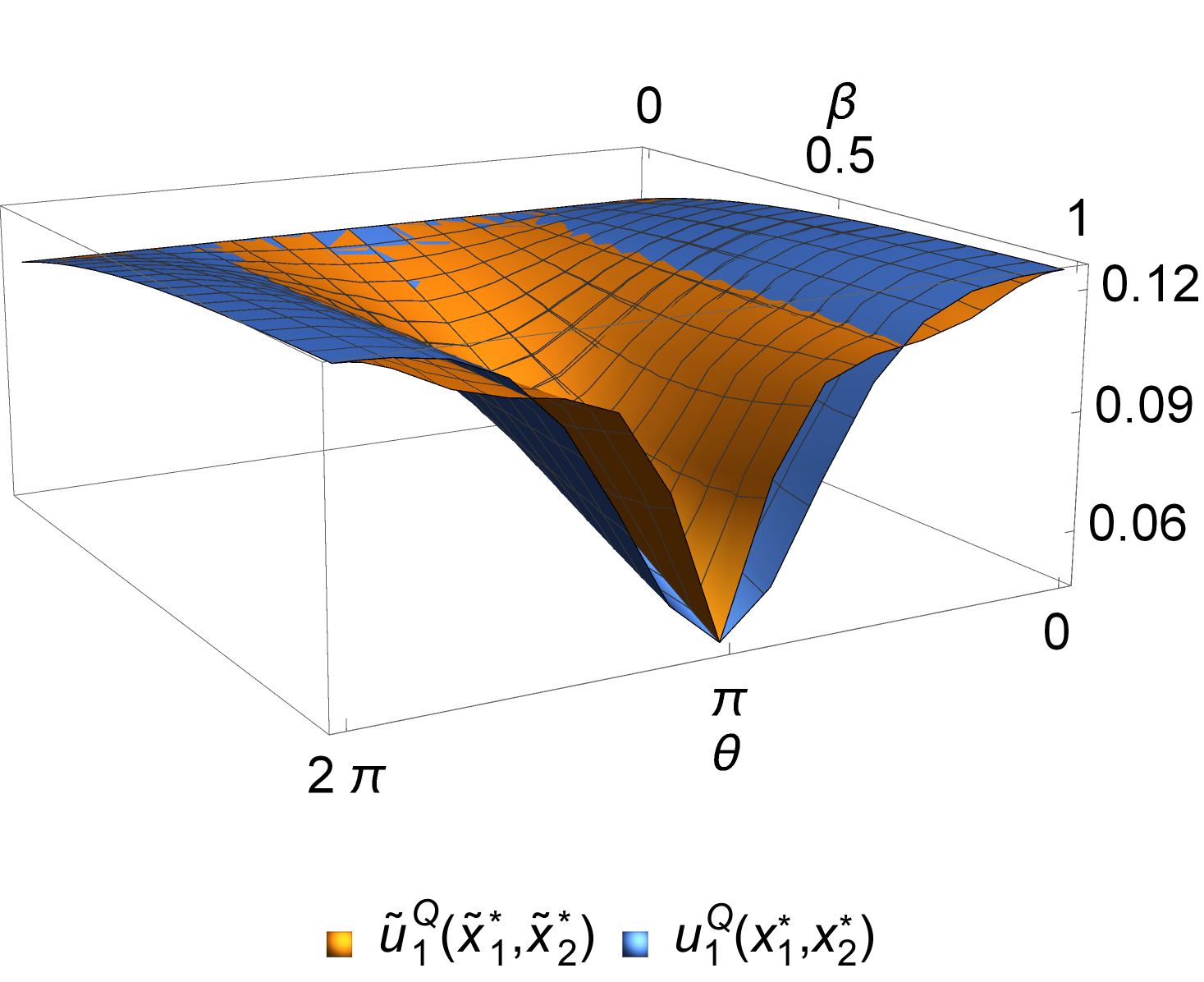}}\qquad
		\subfigure[$\phi=\frac{7\pi}{5}$]{\label{fig9c}\includegraphics[width=0.3\textwidth]{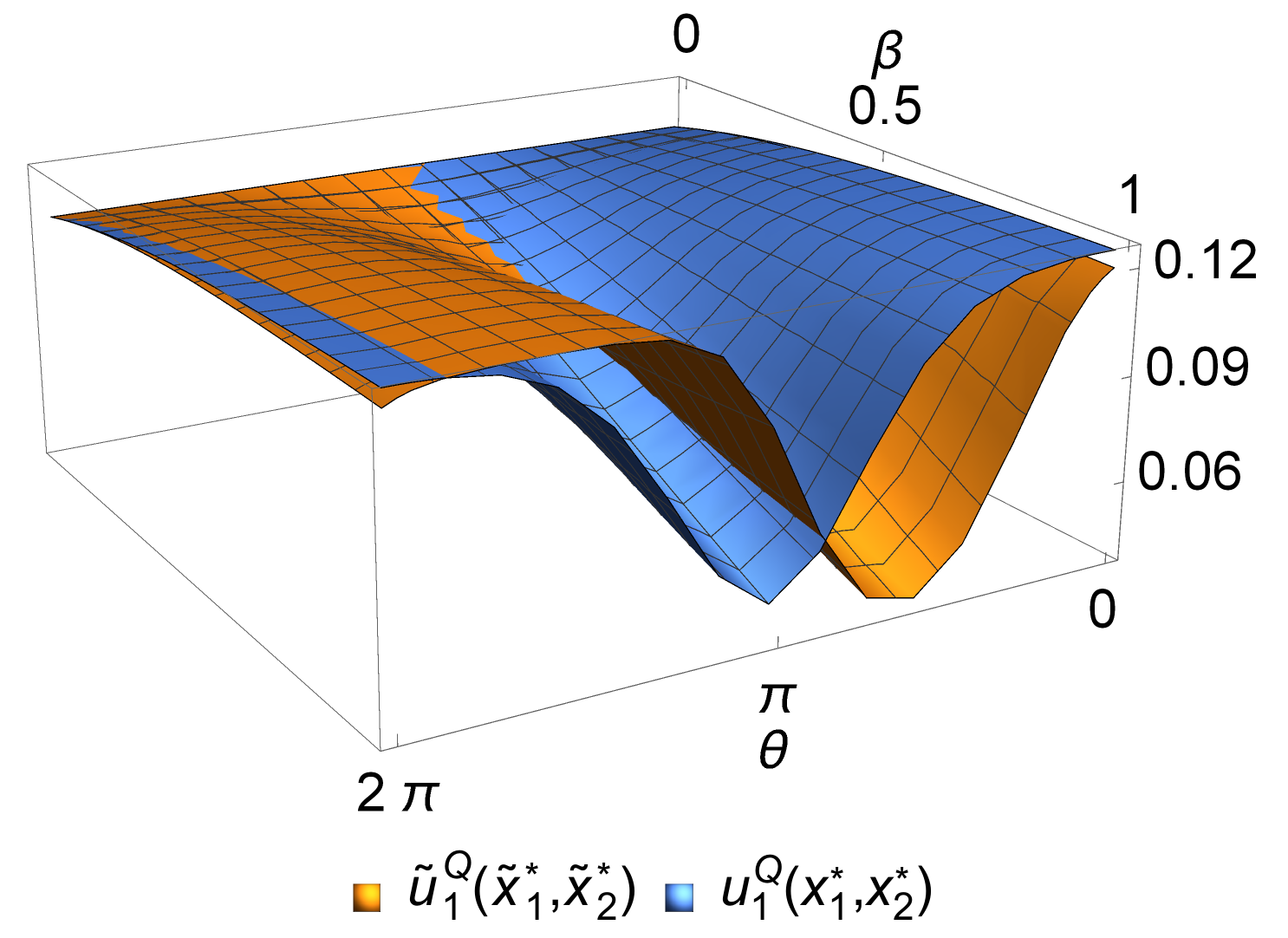}}\\
		
	\end{center}
	\caption{The payoffs $\tilde{u}_1^Q(\tilde{x}_1^*,\tilde{x}_2^*)$ and $u_1^Q(x_1^*,x_2^*)$ for $\alpha=\beta$ and  selected values of parameter $\phi$.} \label{F9}	
\end{figure}

\begin{figure}[h!]
	\begin{center}
		\includegraphics[width=0.6\textwidth]{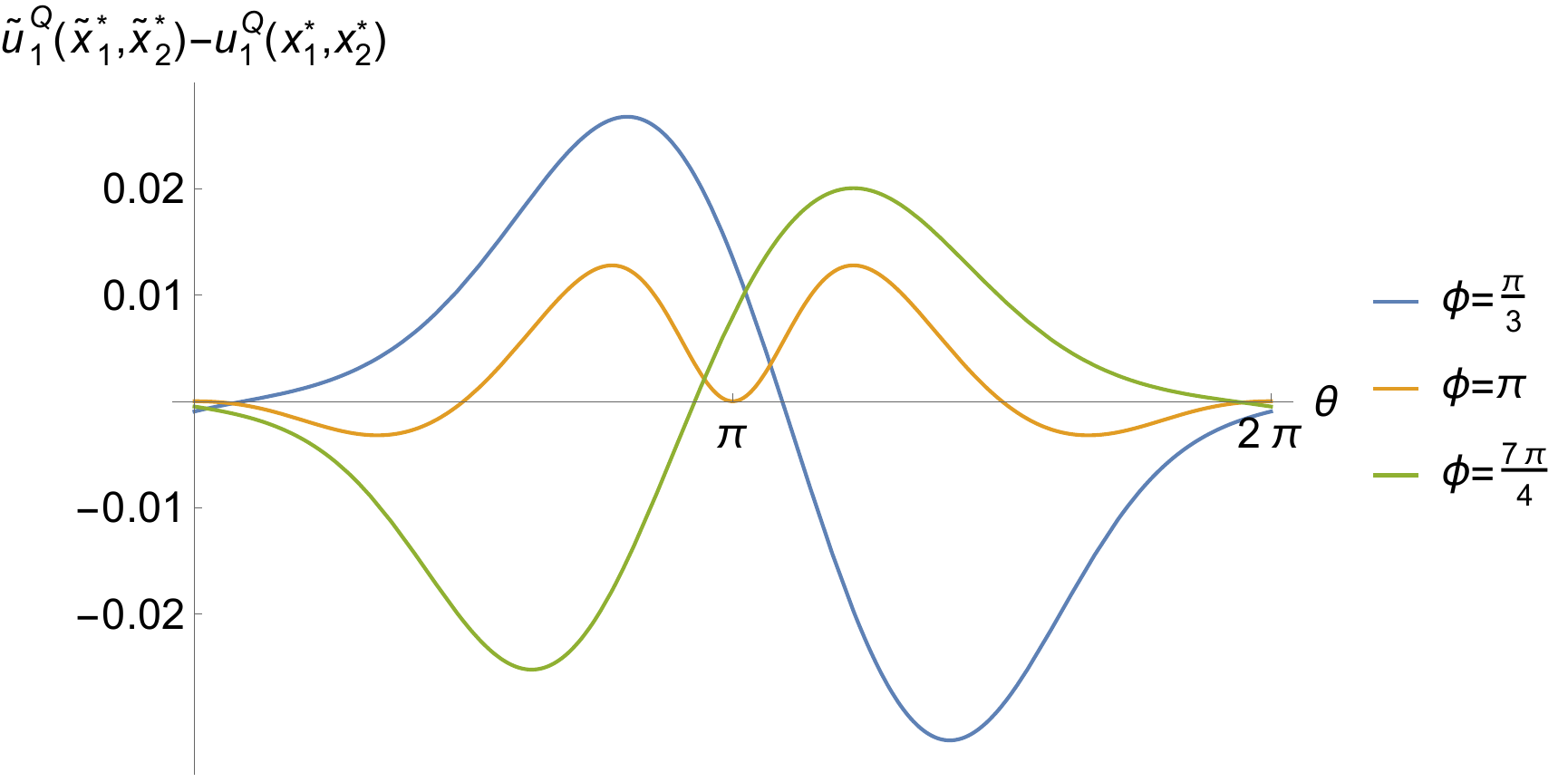}
	\end{center}
	\caption{The difference between payoffs  $\tilde{u}_1^Q(\tilde{x}_1^*,\tilde{x}_2^*)$ and $u_1^Q(x_1^*,x_2^*)$ for selected values of $\phi$ and $\alpha=\beta=0.8$.}\label{F10}
\end{figure}

First, let us analyze the properties of $\tilde{u}_1(\tilde{x}_1^*,\tilde{x}_2^*)$ for specific values of the parameter $\phi$. In contrast to the case I, the values of $\tilde{u}_1(\tilde{x}_1^*,\tilde{x}_2^*)$ are not always greater 
 than those of ${u}_1^Q({x}_1^*,{x}_2^*)$ (Fig.~\ref{F9}). By analyzing the difference between the payoffs obtained in games based on the entanglement operator  $\hat{J}(\xi)$ and $\hat{J}_1(\delta,\xi)$, it can be seen that the phase parameters can be chosen in such a way that ${u}_1^Q({x}_1^*,{x}_2^*)$ gives significantly more favorable results as compared to $\tilde{u}_1(\tilde{x}_1^*,\tilde{x}_2^*)$, see Fig.~\ref{F10}.

The $\theta$ and $\phi$ parameters enter function $\tilde{u}_1^Q(\tilde{x}_1^*,\tilde{x}_2^*)-u_1^Q(x_1^*,x_2^*)$ only as linear arguments of the cosine function thus this function is a periodic one with respect to both $\theta$ and $\phi$ parameters with period $2\pi$.	
By fixing a specific value of the parameter $\beta$, it is possible to numerically identify the values of the phase parameters that result in the maximum or minimum value of the difference between the functions $\tilde{u}_1^Q(\tilde{x}_1^*,\tilde{x}_2^*)$ and $u_1^Q(x_1^*,x_2^*)$; the results are shown in Table \ref{tab1}.  As one can see, a set of parameter values can be selected that yield more favorable results in the game based on $\hat{J}(\xi)$.

\begin{figure}[h!]
	\begin{center}
		\subfigure[$\phi=\frac{\pi}{3}$, $\alpha=\beta$ ]{\label{fig:edge-a10}\includegraphics[width=0.3\textwidth]{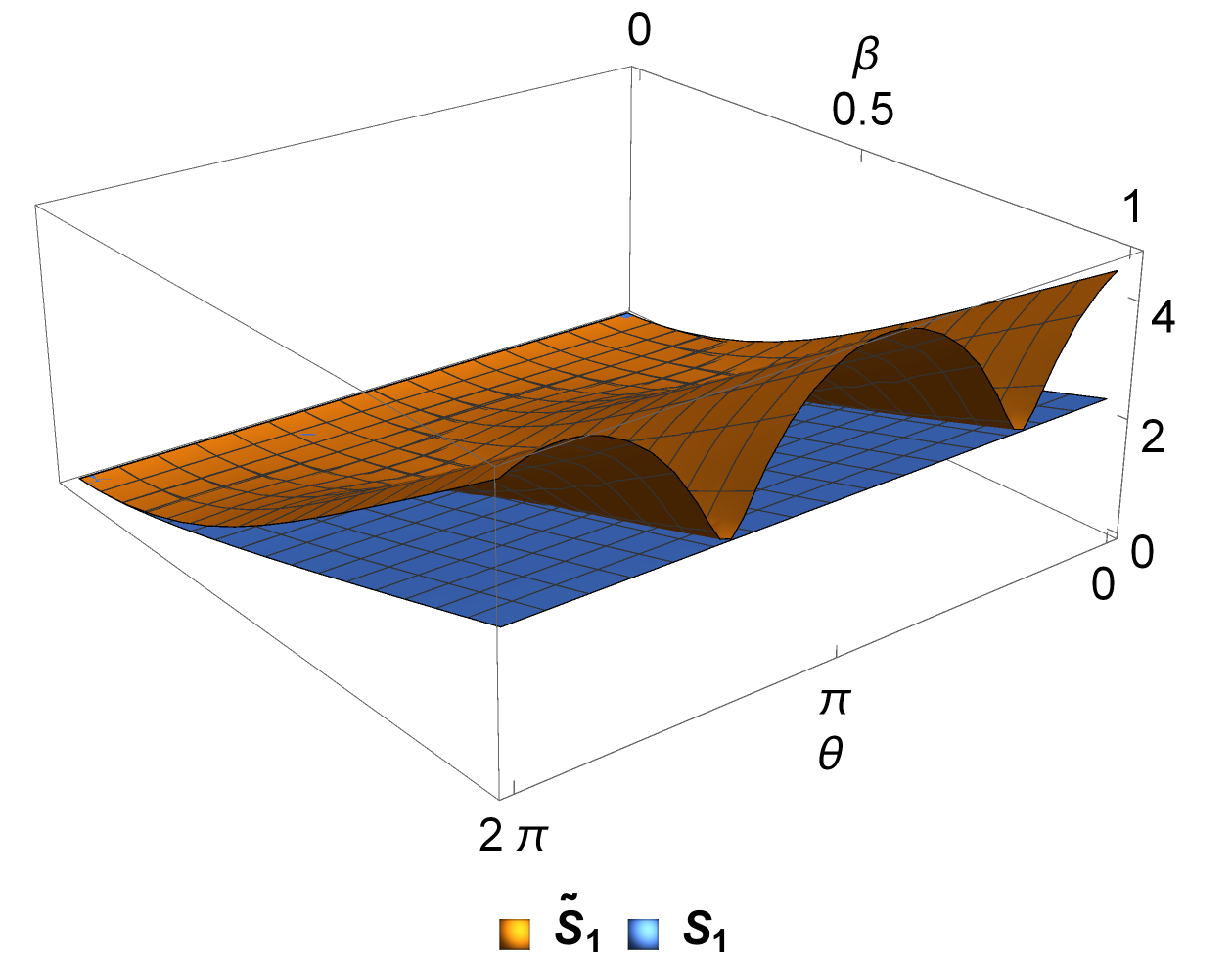}}\qquad
		\subfigure[$\alpha=\beta=0.7$]{\label{fig:edge-10b}\includegraphics[width=0.5\textwidth]{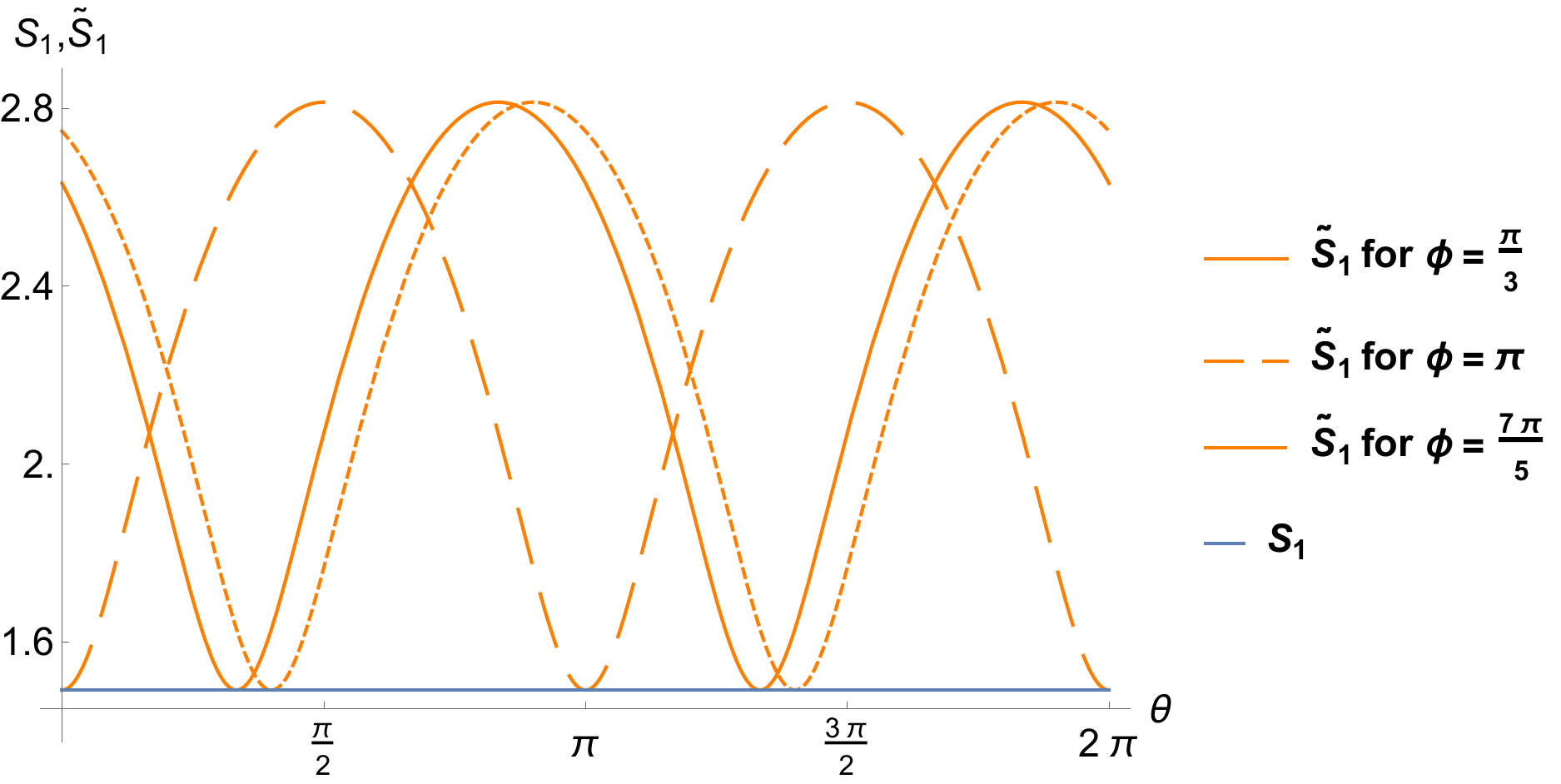}}
		
	\end{center}
	\caption{The entanglement entropies $\tilde{S}_1$ and $S_1$ for selected values of phase parameter $\phi$.} \label{F11}	
\end{figure}

 If we compare the degree of entanglement of the two games, we find that the entanglement entropy $\tilde{S}_1$ is always greater than or equal to $S_1$ (Fig.~\ref{F11}). Even in the cases when $u_1^Q(x_1^*,x_2^*)$ payoffs are higher than $\tilde{u}_1^Q(\tilde{x}_1^*,\tilde{x}_2^*)$, the selected measure of entanglement indicates a higher degree of entanglement for the model based on $\hat{J}_1(\delta,\xi)$ as compared with that based on $\hat{J}(\xi)$ (Table \ref{tab1}).

\begin{table}[h!]
\begin{footnotesize}	\begin{center}
	\begin{tabular}{cl|c|c|c|c|}
		\cline{3-6}
		&	& \multicolumn{4}{c|}{fixed $\alpha$ and $\beta$, with $\alpha=\beta$} \\
				\cline{3-6}
	&	 & 0.2 & 0.5 & 1 & 5\\
		 \hline
\multicolumn{2}{|c|} {$\max(\tilde{u}_1^Q(\tilde{x}_1^*,\tilde{x}_2^*)-u_1^Q(x_1^*,x_2^*))$ } & 0.0030 & 0.0195 & 0.0648 & 0.1250   \\
\multicolumn{1}{|r}{\multirow{2}{*}{estimated parameters $\Bigg\{$}}&{$\theta$} & 2.0525 & 2.4987 & 2.8611 & 3.1416 \\
\multicolumn{1}{|c}{} &$\phi$  & 1.8324 & 2.1704 & 2.5492 & 3.3488\\
\multicolumn{1}{|l}{\multirow{2}{*}{entropy for the estimated parameter $\theta$ and $\phi$ $\Bigg\{$}}&{$\tilde{S}_1$} & 0.2491 & 1.0499 & 3.1337 & 24.2017\\
\multicolumn{1}{|c}{}&{$S_1$} & 0.2471 & 0.9514 & 2.3369 & 13.8696\\
\hline 
\multicolumn{2}{|c|} {$\min(\tilde{u}_1^Q(\tilde{x}_1^*,\tilde{x}_2^*)-u_1^Q(x_1^*,x_2^*))$} &-0.0003 & -0.0235 & -0.0832 & -0.1250   \\
\multicolumn{1}{|r}{\multirow{2}{*}{estimated parameters $\Bigg\{$}}&{$\theta$} & 6.2832 & 1.7687 & 1.5048 & 0.1151 \\
\multicolumn{1}{|c}{} &{$\phi$}  & 4.5132 & 4.0706 & 3.7487 & 3.1994\\
\multicolumn{1}{|l}{\multirow{2}{*}{entropy for the estimated parameter $\theta$ and $\phi$ $\Bigg\{$}}&{$\tilde{S}_1$} & 0.2860 & 1.3707 & 4.3336 & 23.5971\\
\multicolumn{1}{|c}{}&{$S_1$} & 0.2471 & 0.9514 & 2.3369 & 13.8696\\
	\hline 
	\end{tabular}\end{center}\end{footnotesize}
\caption{Values of phase parameters estimated numerically, for which the difference between payoffs at fixed $\alpha,\beta$ ($\alpha=\beta$) values is the largest or smallest.}\label{tab1}
\end{table}

\begin{figure}[h!]
	\begin{center}
		\subfigure[$\theta=\frac{\pi}{3}$, ]{\label{fig12a}\includegraphics[width=0.3\textwidth]{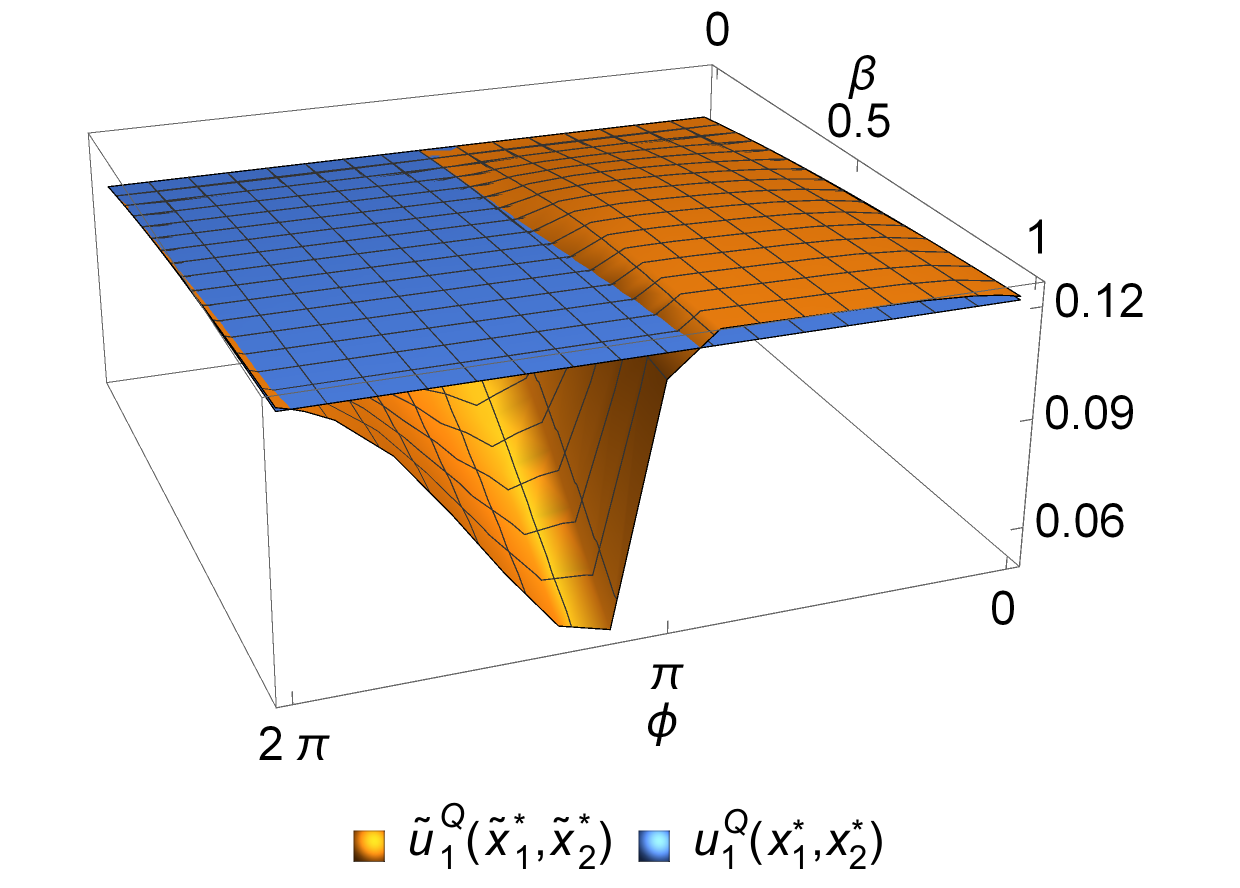}}\qquad
		\subfigure[$\theta=\pi$]{\label{fig12b}\includegraphics[width=0.3\textwidth]{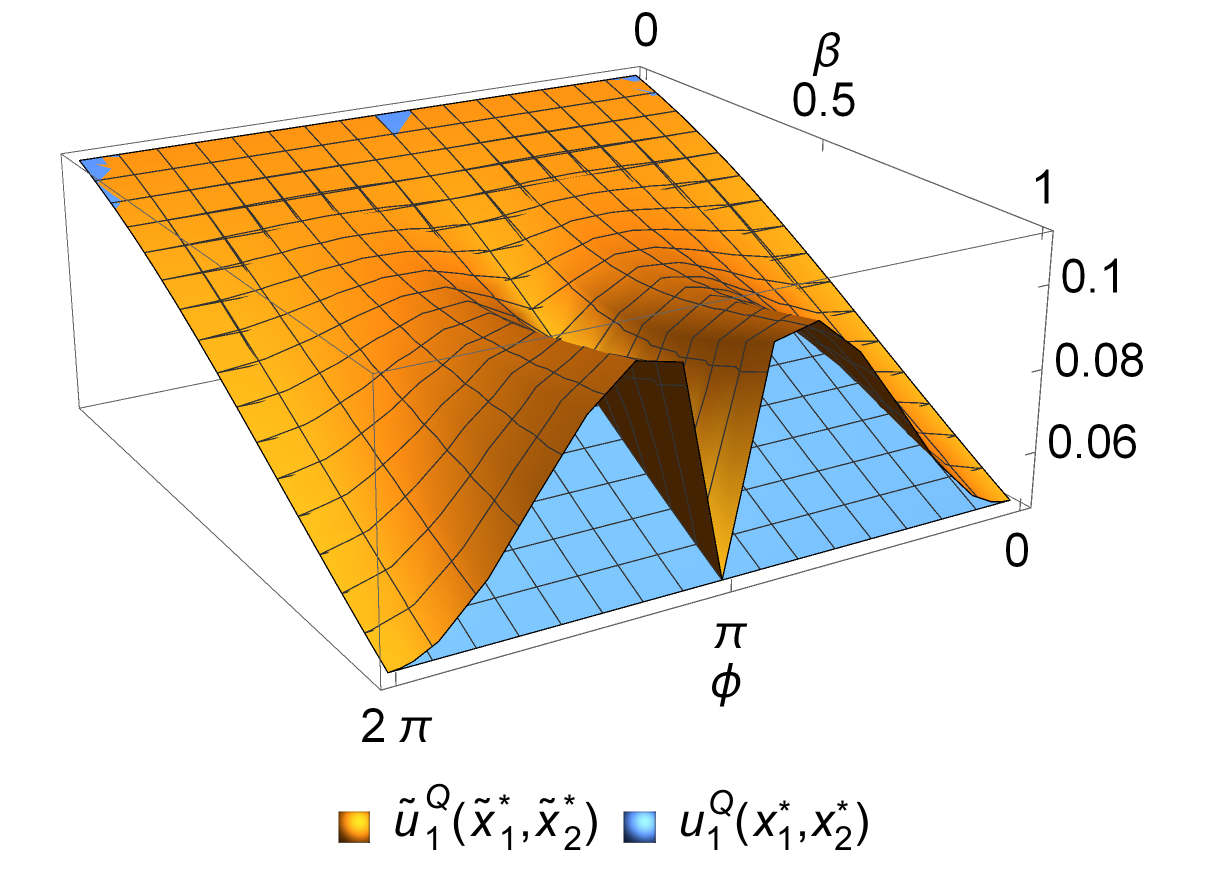}}\qquad
		\subfigure[$\theta=\frac{7\pi}{5}$]{\label{fig:edge-c}\includegraphics[width=0.3\textwidth]{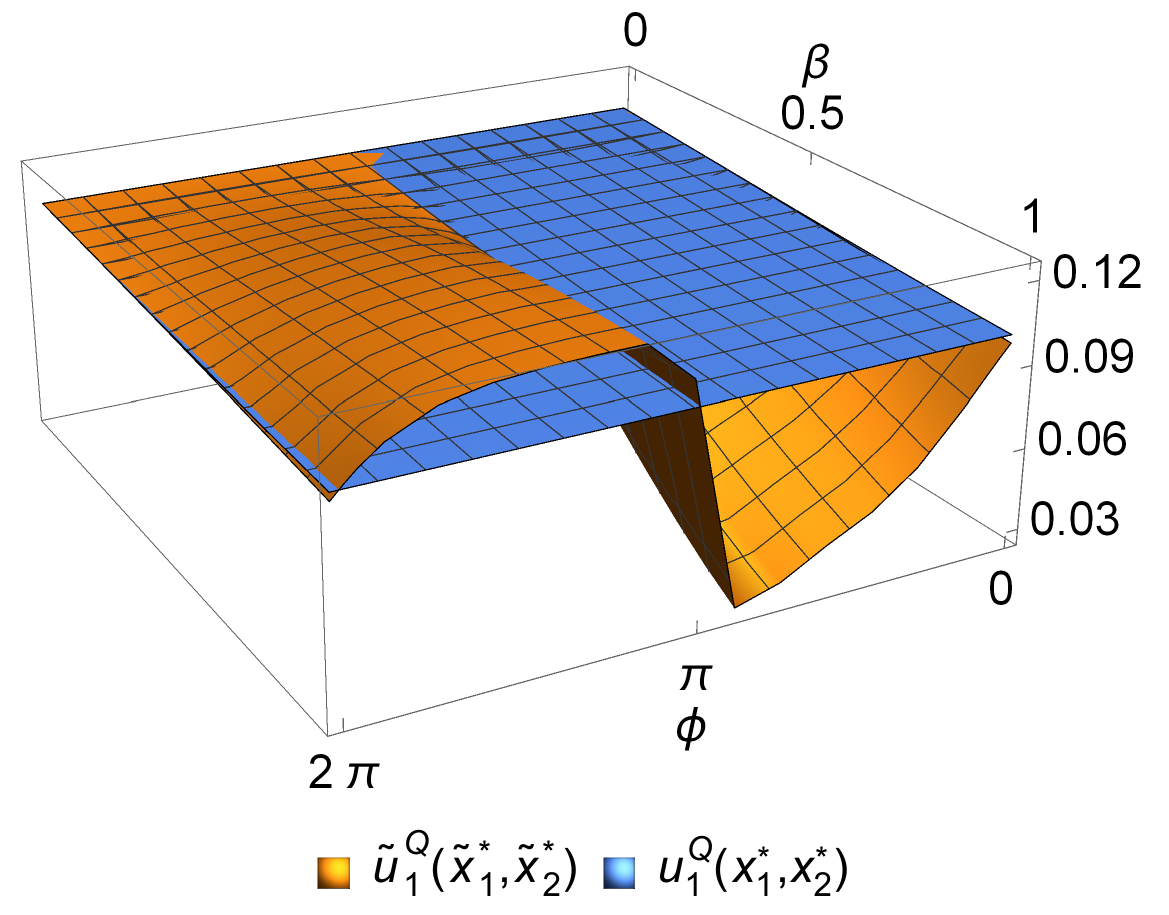}}\\
		
	\end{center}
	\caption{The payoffs $\tilde{u}_1^Q(\tilde{x}_1^*,\tilde{x}_2^*)$ and $u_1^Q(x_1^*,x_2^*)$ for $\alpha=\beta$ and  selected values of parameter $\theta$.} \label{F12}	
\end{figure}
\begin{figure}[h!]
	\begin{center}
		\includegraphics[width=0.6\textwidth]{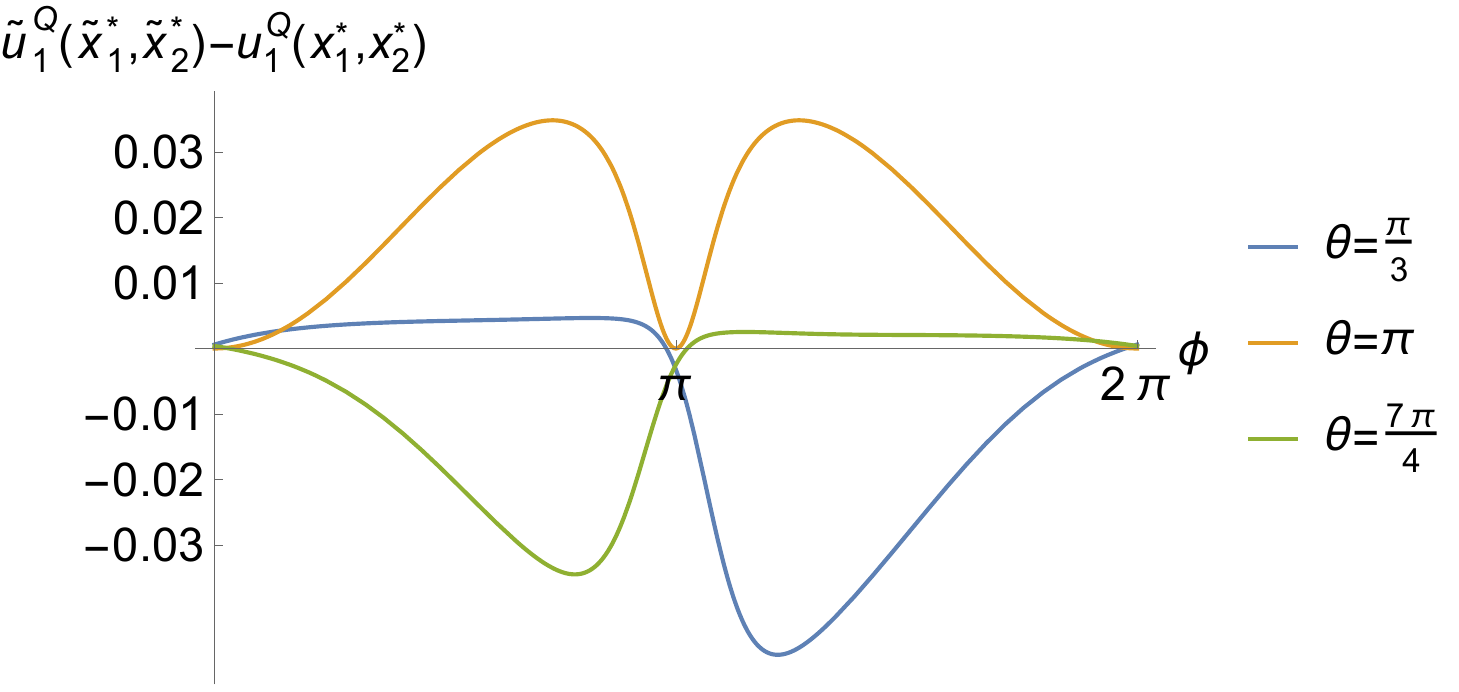}
	\end{center}
	\caption{The difference between payoffs  $\tilde{u}_1^Q(\tilde{x}_1^*,\tilde{x}_2^*)$ and $u_1^Q(x_1^*,x_2^*)$ for selected values of $\theta$ and $\alpha=\beta=0.8$.}\label{F13}
\end{figure}

Let us now analyze the properties of the $\tilde{u}_1^Q(\tilde{x}_1^*,\tilde{x}_2^*)$ and $u_1^Q(x_1^*,x_2^*)$, choosing the specific values of the parameter $\theta$. For $\theta=\pi$,  the payoffs for the game based on $\hat{J}_1(\delta,\xi)$ are always higher than those for the game based on $\hat{J}(\xi)$ (Figure \ref{fig12b}). It should be noted that  the payoff $u_1^Q(x_1^*,x_2^*)$ is independent of the $\phi$ parameter and is an increasing function of $\beta$ for $\theta\in[0,\frac{\pi}{2})\cup(\frac{3\pi}{2},2\pi)$ and a decreasing function for $\theta\in(\frac{\pi}{2},\frac{3\pi}{2})$. It can be deduced from Figures \ref{F12} and \ref{F13} that both phase parameters have a strong influence on the variation of the  function $\tilde{u}_1^Q(\tilde{x}_1^*,\tilde{x}_2^*)$. The entanglement entropy  $\tilde{S}_1$ is a function of the phase parameters through $\cos(\theta-\phi)$, so the graph of the function $\tilde{S}_1(\phi)$ at a fixed value of $\theta$ and $\beta$ is the same as graph  of $\tilde{S}_1(\theta)$ on Fig.~\ref{fig:edge-10b}. 
 
\begin{figure}[!h]
	\begin{center}
		\subfigure[$\theta-\phi=\frac{\pi}{6}$, $\phi=0.3*\frac{\pi}{3}$]{\label{14a}\includegraphics[width=0.3\textwidth]{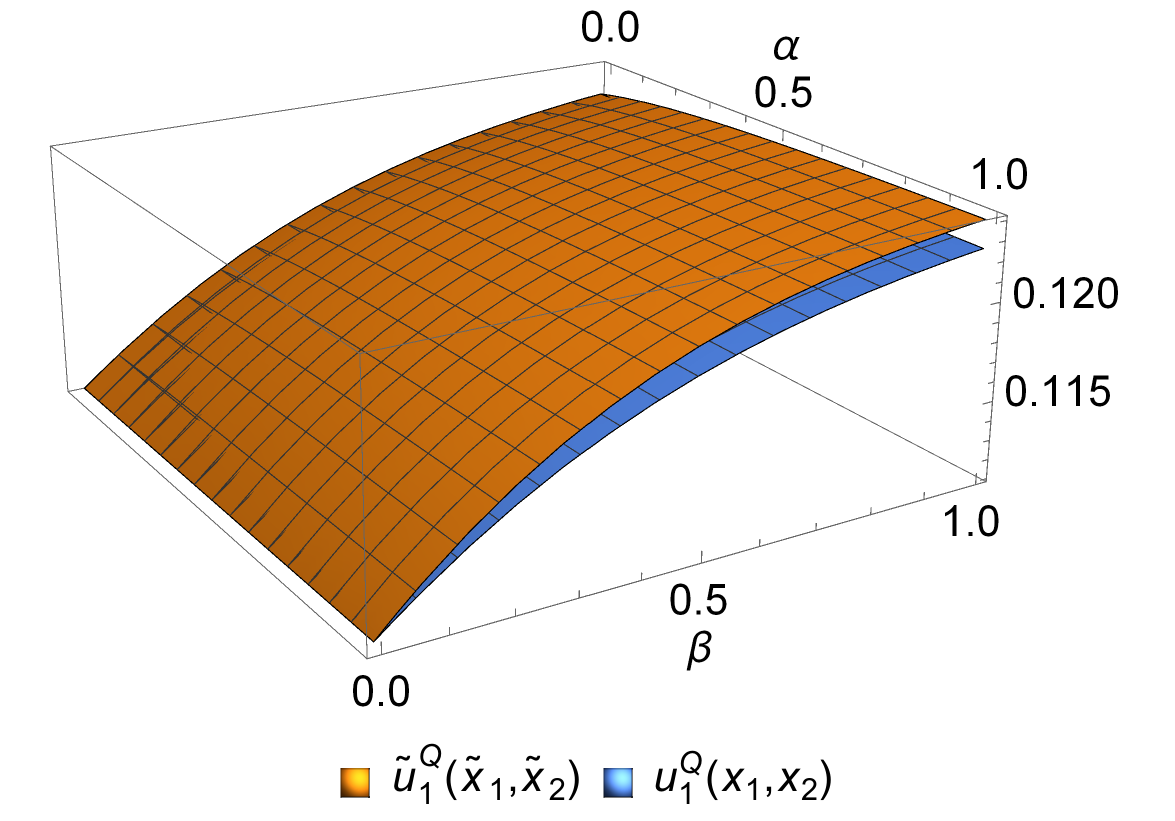}}\qquad
		\subfigure[$\theta-\phi=\frac{\pi}{6}$, $\phi=0.8*\frac{\pi}{3}$]{\label{14b}\includegraphics[width=0.3\textwidth]{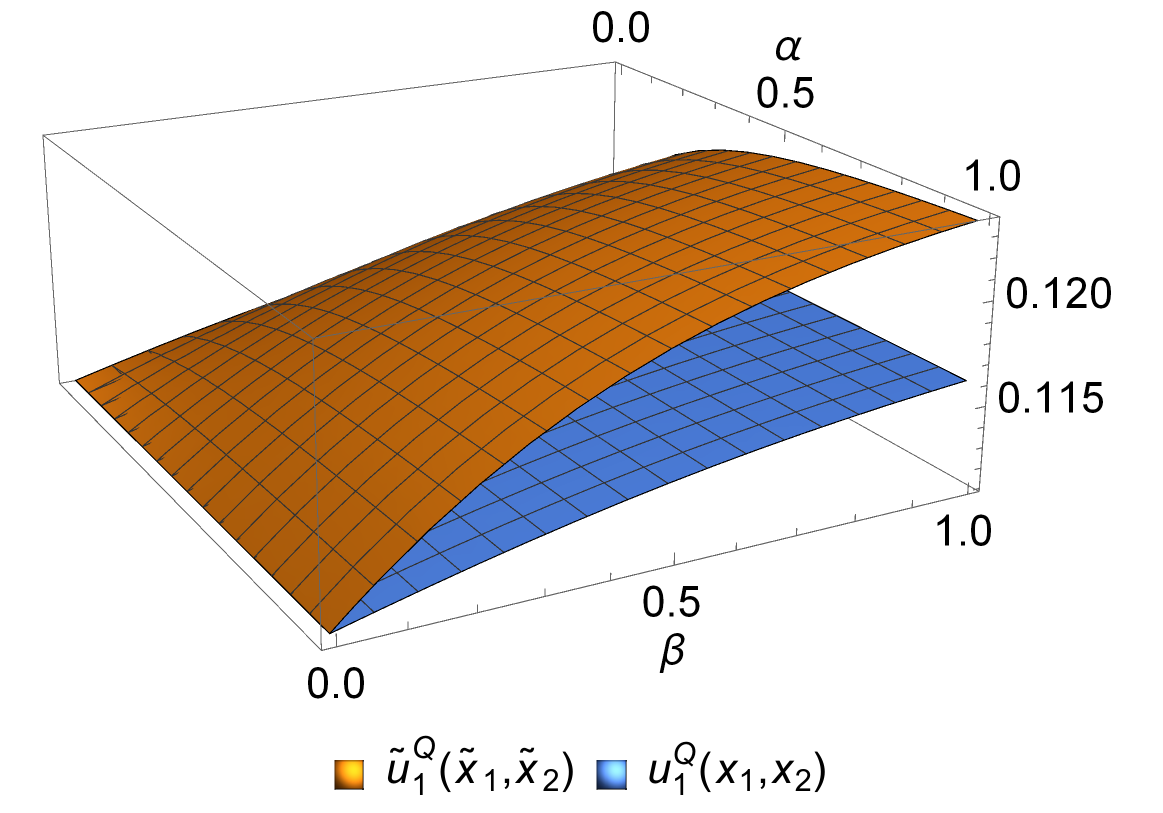}}\qquad
		\subfigure[$\theta-\phi=\frac{\pi}{6}$]{\label{14c}\includegraphics[width=0.3\textwidth]{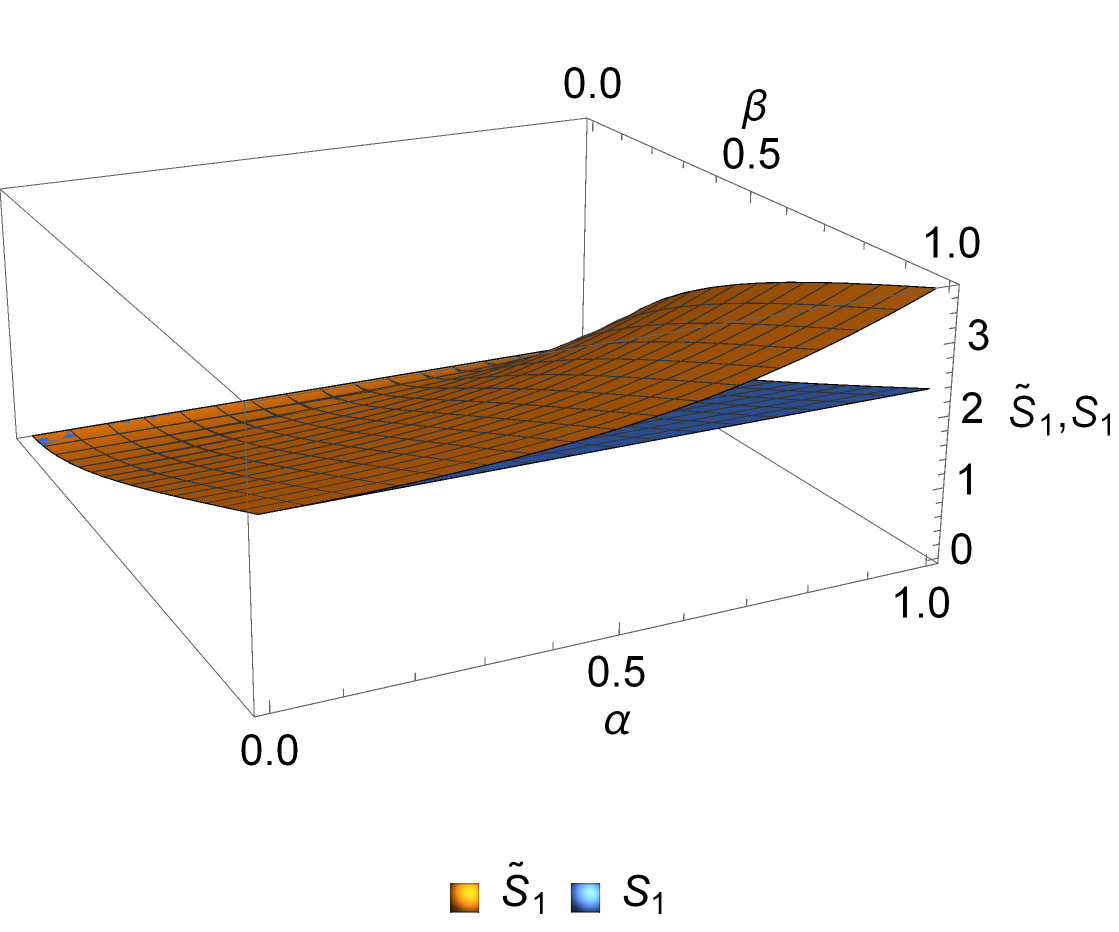}}\\
		\subfigure[$\theta-\phi=\frac{\pi}{4}$, $\phi=0.3*\frac{\pi}{4}$]{\label{14d}\includegraphics[width=0.3\textwidth]{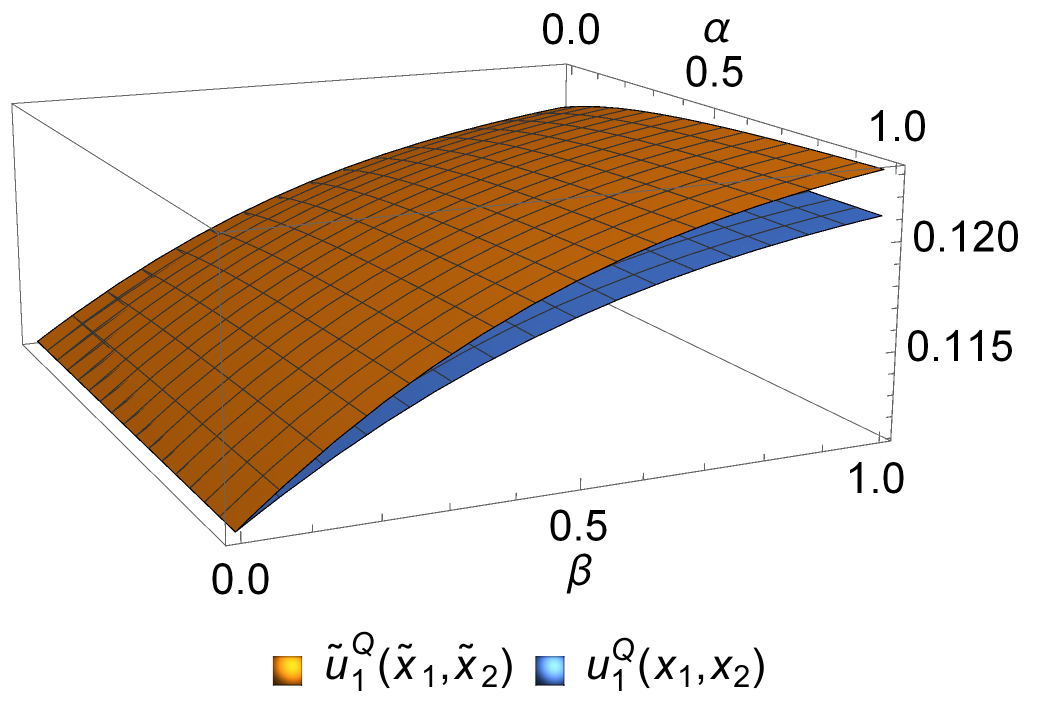}}\qquad
		\subfigure[$\theta-\phi=\frac{\pi}{4}$, $\phi=0.8*\frac{\pi}{4}$]{\label{14e}\includegraphics[width=0.3\textwidth]{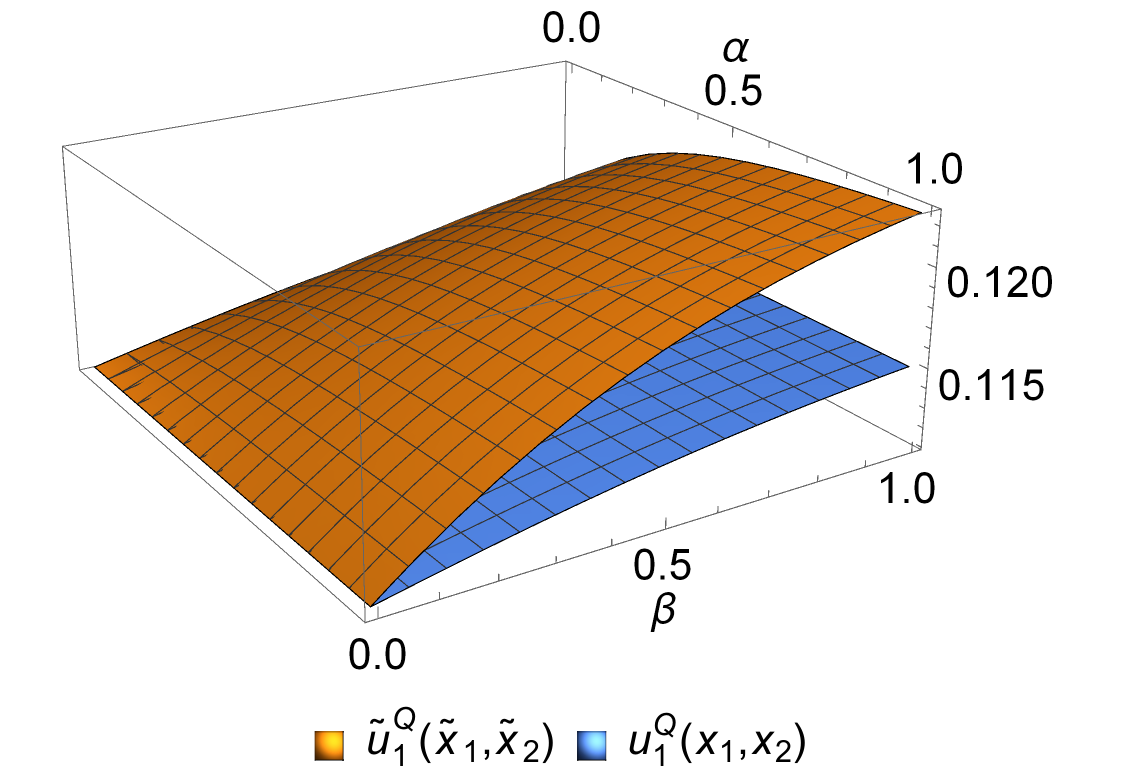}}\qquad	\subfigure[$\theta-\phi=\frac{\pi}{4}$]{\label{14f}\includegraphics[width=0.3\textwidth]{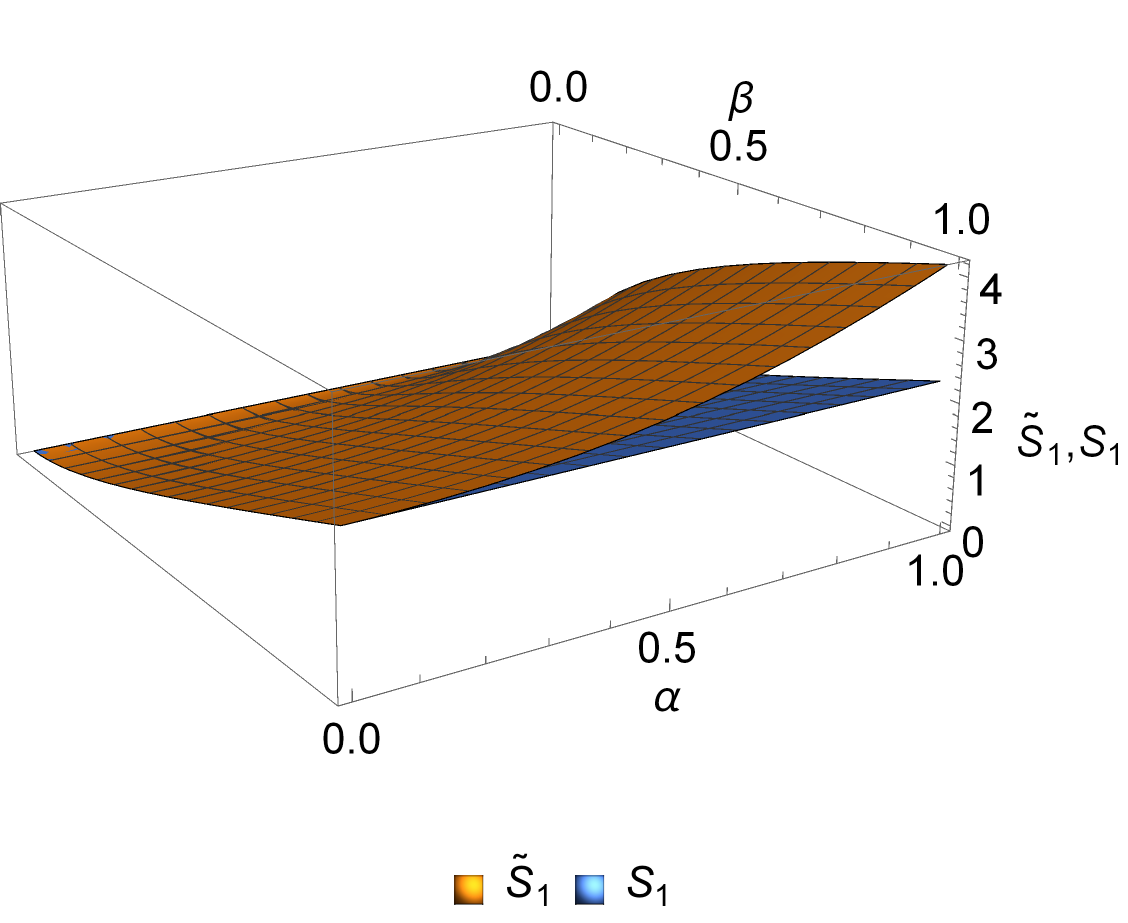}}\\
		\subfigure[$\theta-\phi=\frac{\pi}{3}$, $\phi=0.3*\frac{\pi}{6}$]{\label{14g}\includegraphics[width=0.3\textwidth]{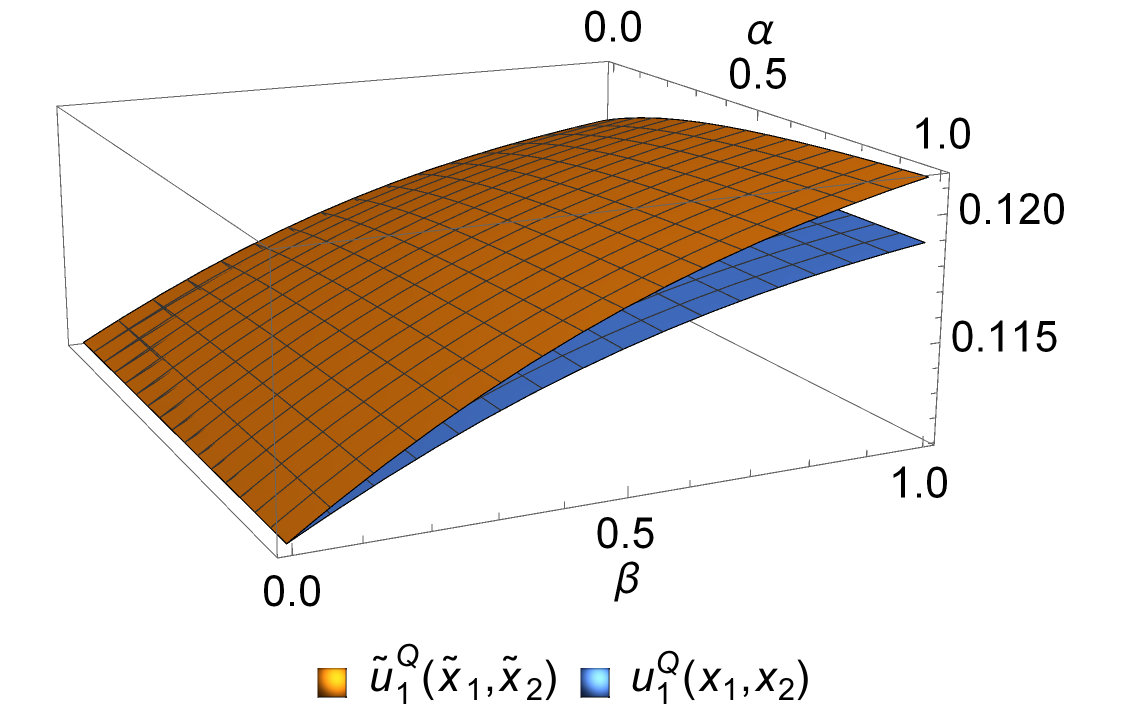}}\qquad
		\subfigure[$\theta-\phi=\frac{\pi}{3}$, $\phi=0.8*\frac{\pi}{6}$]{\label{14h}\includegraphics[width=0.3\textwidth]{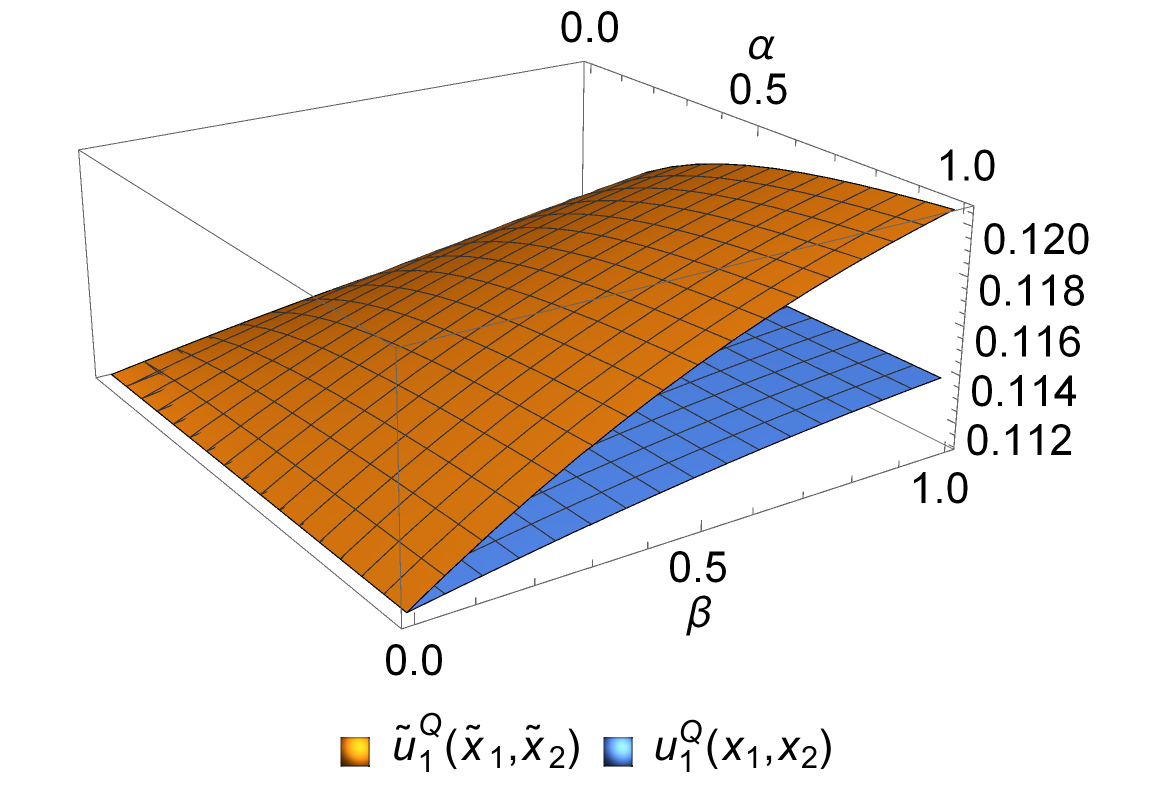}}\qquad	\subfigure[$\theta-\phi=\frac{\pi}{3}$]{\label{14i}\includegraphics[width=0.3\textwidth]{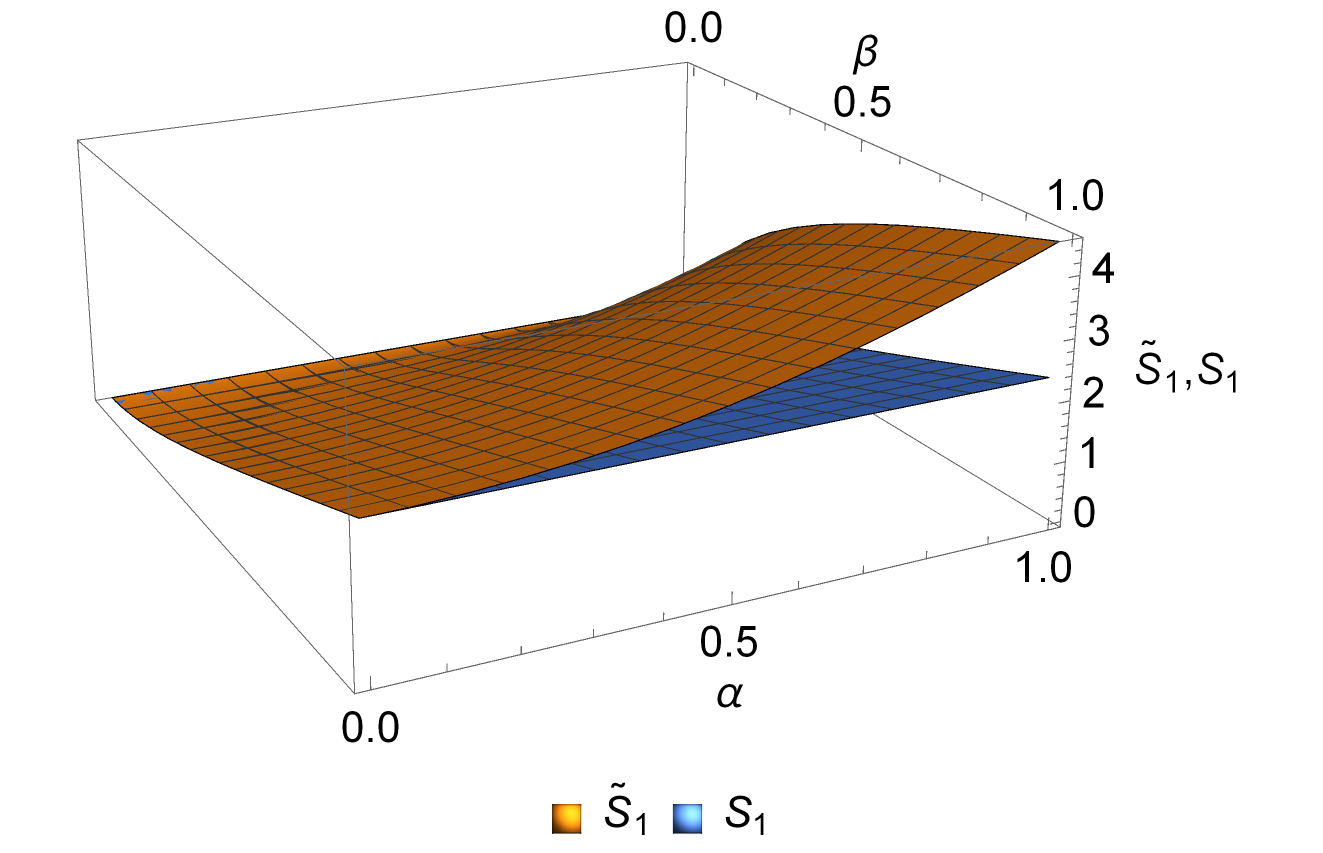}}		
	\end{center}
	\caption {The payoffs in Nash equilibrium of games based on the entangled operators $\hat{J}(\xi)$, $\hat{J}_1(\delta,\xi)$ and corresponding entanglement entropies for fixed values of $\theta$ and $\phi$ that satisfy the condition $\theta-\phi=\frac{\pi}{6},\frac{\pi}{4},\frac{\pi}{3}$.}\label{F14}	
\end{figure} 

\textbf{Case III.} 

Consider the case where $\theta-\phi=const.$ and $\alpha,\beta>0$, which is interesting because the entanglement entropy $\tilde{S}_1$ depends only on the  difference of phase parameters. Let us choose the range of the values of the parameter $\theta$ from 0 to $\frac{\pi}{2}$, for which the function  $u_1^Q(x_1^*,x_2^*)$ increases with the increasing squeezing parameter $\beta$.  When $\theta\in(0,\frac{\pi}{2})$, $\tilde{u}_1^Q(\tilde{x}_1^*,\tilde{x}_2^*)$ is also an increasing function of both squeezing parameters $\alpha$ and $\beta$.
As shown on Figures \ref{14c}, \ref{14f}, \ref{14i} the higher the values of the difference of phase parameters, the higher values reaches the measure of entanglement $\tilde{S}_1$ as the parameters $\alpha$ and $\beta$ increase; in all cases $\tilde{S}_1\geq S_1$. 

\begin{figure}
	\begin{center}
		\subfigure[ $\tilde{u}_1^Q(\tilde{x}_1^*,\tilde{x}_2^*)$ for $\alpha=0.5$ and fixed values of $\theta$ and $\phi$]{\label{15a}\includegraphics[width=0.3\textwidth]{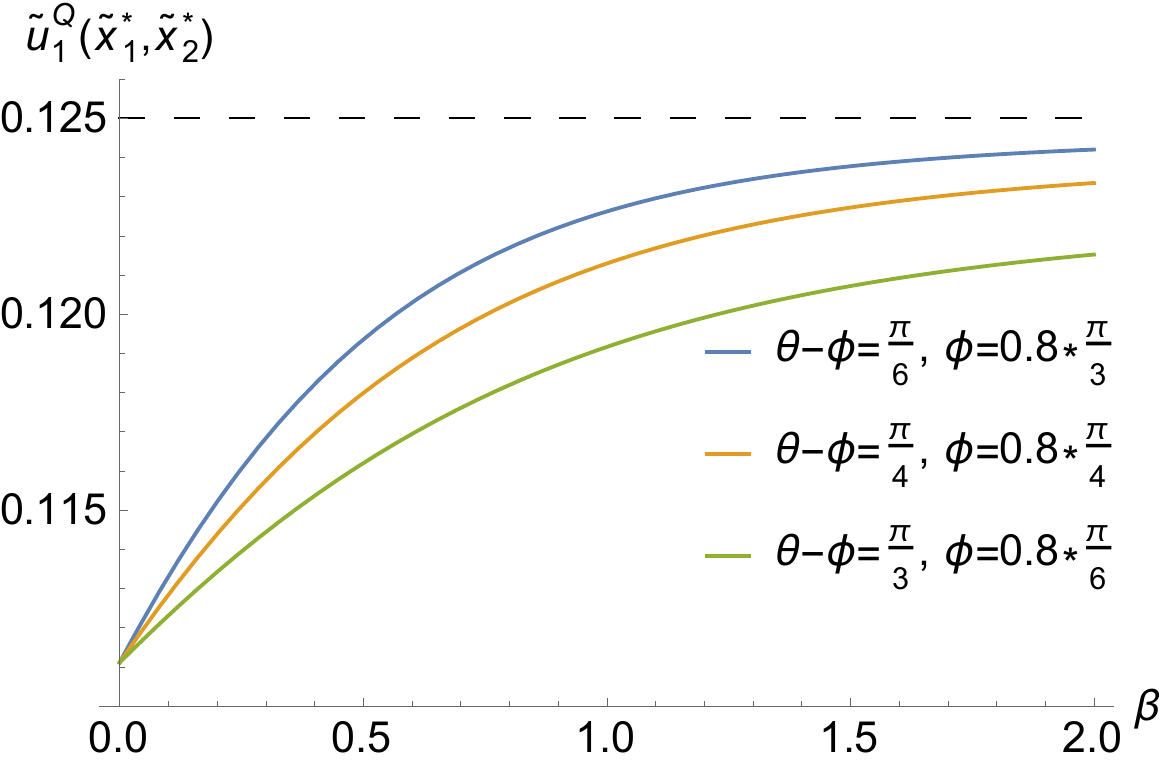}}\qquad
		\subfigure[$\tilde{u}_1^Q(\tilde{x}_1^*,\tilde{x}_2^*)$ for $\alpha=1$ and fixed values of $\theta$ and $\phi$]{\label{15b}\includegraphics[width=0.3\textwidth]{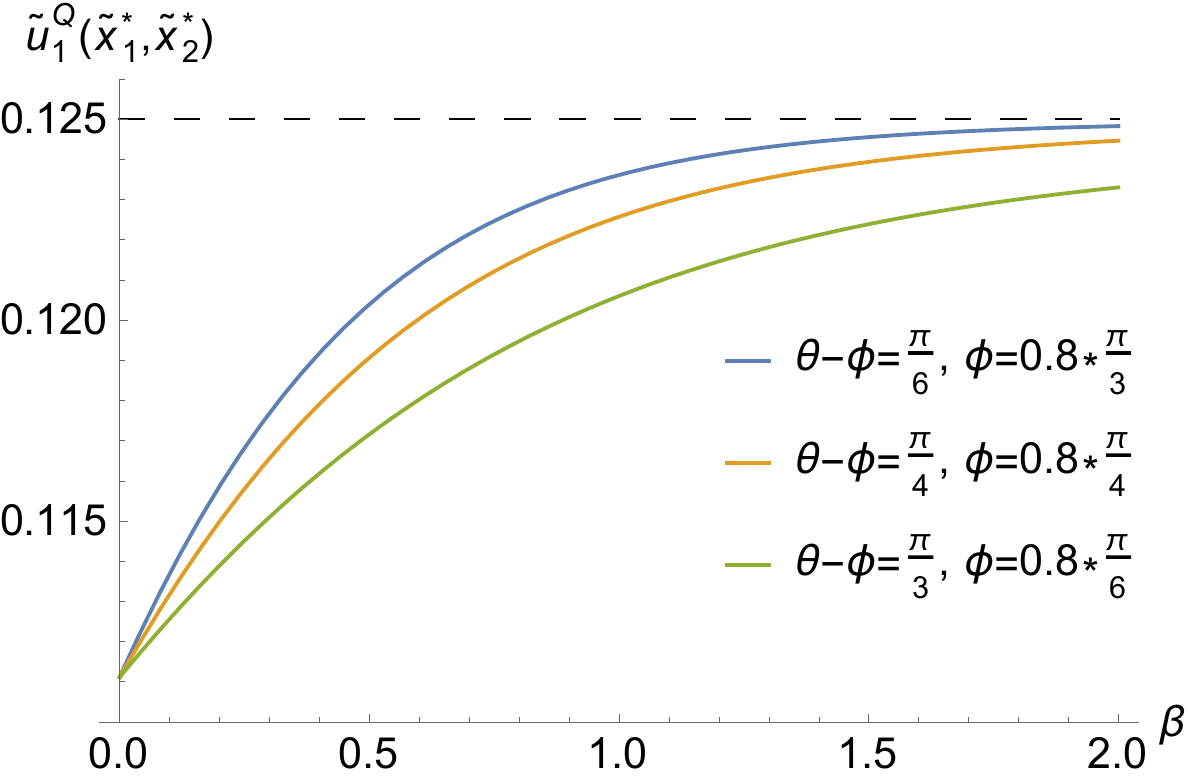}}\qquad
		\subfigure[$\tilde{S}_1$ for $\alpha=0.5$ and fixed values of $\theta-\phi$ ]{\label{15c}\includegraphics[width=0.3\textwidth]{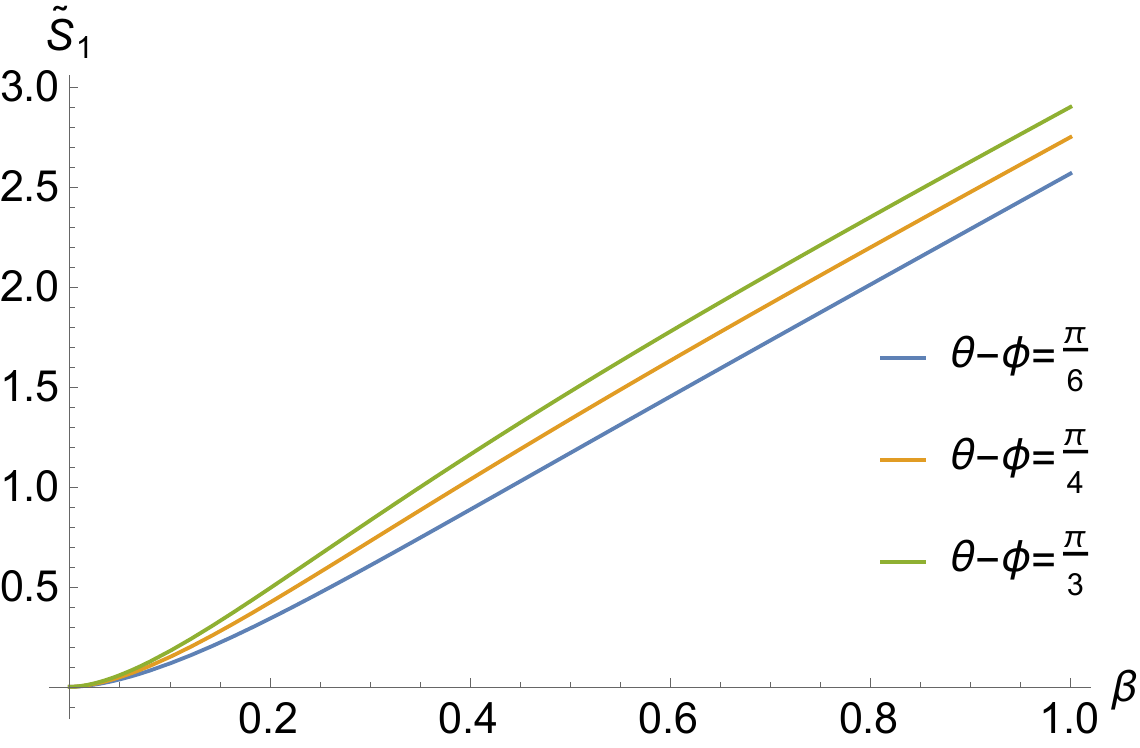}}		
	\end{center}
	\caption {The payoff $\tilde{u}_1^Q(\tilde{x}_1^*,\tilde{x}_2^*)$ and entanglement entropy $\tilde{S}_1$  for fixed values of $\alpha$, $\theta$ and $\phi$ (these are special cases of the graphs of the functions $\tilde{u}_1^Q(\tilde{x}_1^*,\tilde{x}_2^*)$ and $\tilde{S}_1$ in Fig.~\ref{F14}.}\label{F15}	
\end{figure}

	The measure $S_1$ depends only on the squeezing parameter $\beta$, therefore the graphs of $S_1$ in Figures \ref{14c}, \ref{14f} and \ref{14i} are the same. The difference between the graphs of $\tilde{S}_1$ in Figures \ref{14f} and \ref{14i} is hardly noticeable, but plotting the graphs of this function for a fixed value of the parameter $\alpha$ in Fig. \ref{15c} we see that for $\theta-\phi=\frac{\pi}{3}$ the measure of the degree of entanglement reaches higher values than for $\theta-\phi=\frac{\pi}{4}$.

Comparing Figure \ref{14a} with \ref{14b} (\ref{14d} with \ref{14e}, \ref{14g} with \ref{14h}), we see that the difference between the payoffs $\tilde{u}_1^Q(\tilde{x}_1^*,\tilde{x}_2^*)$ and $u_1^Q(x_1^*,x_2^*)$ increases as the parameter $\theta$ increases. The reason for this is that for $\theta\rightarrow\frac{\pi}{2}$  values of $u_1^Q(x_1^*,x_2^*)$ grow slower and slower as the functions of $\beta$  (see Fig.~\ref{F1}); the  $\tilde{u}_1^Q(\tilde{x}_1^*,\tilde{x}_2^*)$  also reaches lower values for a fixed parameter $\beta$ when the phase parameters are different from zero, but thanks to the parameter $\alpha$ the maximum possible payoff can be reached when $\alpha$ increases. To see this, we have plotted the graph of the $\tilde{u}_1^Q(\tilde{x}_1^*,\tilde{x}_2^*)$ for $\alpha=0.5, 1$ (Figs.~\ref{15a} and \ref{15b}) and the same fixed values of the parameters $\theta$ and $\phi$ chosen for the graphs in Figures \ref{14b}, \ref{14e} and \ref{14h}. Figures \ref{15a} and \ref{15c} have been plotted for the same values of the parameters $\alpha$ and $\theta-\phi$; we see that the effect of the phase parameters $\theta$, $\phi$ on the payoff function is quite different from that on the entanglement measure. However, the squeezing parameter $\alpha$ offsets the negative effect of the phase parameters on the payoff value (compare Fig.~\ref{15a} with \ref{15b}).

\section{Conclusions}
The quantization scheme of games containing a continuous set of strategies proposed by Li et al. made it possible to analyze the quantum counterpart of the duopoly models known in economics. The most well-known and simple model is Cournot's duopoly. In this case, the corresponding quantum model yields higher payoffs than the classical one. This phenomenon can be attributed to the entanglement of the initial state of the game; Li et al. showed that the higher the value of the squeezing parameter, the higher the payoffs of players in Nash equilibrium; the initial state of the game they considered depended on only one parameter - the squeezing parameter.

Due to the key role of the game degree of entanglement on its outcome, we decided to consider the quantum Cournot model based on the most general entanglement operator $\hat{J}_1(\delta,\xi)$ satisfying the assumptions concerning the symmetry with respect to the exchange of players and full representativeness of the classical game by its quantum counterpart and depending on two squeezing and two phase parameters. 

The main conclusion of the analysis is the observation that the relationship between the degree of entanglement of the initial state of the game and the payoff values in a Nash equilibrium is ambiguous.  Comparing the entanglement entropy values of the initial state of the game for fixed values of the parameters $\alpha$, $\beta$, $\theta$, $\phi$, one concludes that the degree of entanglement of state $\hat{J}_1(\delta,\xi)|00\big>$ is higher than that of state $\hat{J}(\xi)|00\big >$, while the payoffs of the players in games based on these states may have an inverse relationship (see Table \ref{tab1}).  

Phase parameters also have great impact on the outcome of the game. In the case of the game based on the entanglement operator $\hat{J}(\xi)$, the maximum possible outcome of the game with $\beta\rightarrow \infty$ is observed to decrease as the phase parameter $\theta$ approaches the value of $\frac{\pi}{2}$. When the game depends on two squeezing parameters, it is possible to reach the maximum payoff in Nash equilibrium when $\alpha\rightarrow\infty$ and the phase parameters satisfy the condition $\theta-\phi=const.$ (Fig.~\ref{F14}). As shown in Section \ref{sec3}, it is not possible to achieve an arbitrarily large payoff in a Nash equilibrium when $\alpha,\beta\rightarrow\infty$. The maximum payoff that a player can receive in a Nash equilibrium of the model under consideration is $\frac{k^2}{8}$.

\subsection*{Acknowledgment}
I am grateful to Prof. Piotr Kosiński and Prof. Krzysztof Andrzejewski for helpful discussion and useful remarks.

\begin{appendices}
	
	\section{}\label{AA1}
In paper \cite{LiQin}, the authors considered an entanglement operator of the game of the form 
\begin{equation}
\hat{J}(\gamma_1,\gamma_2,\gamma_{12})=e^{\gamma_{12}(a_1a_2-a_1^\dagger a_2^\dagger)}e^{\gamma_1(a_1^2-a_1^{\dagger\,2} )/2}e^{\gamma_2(a_2^2-a_2^{\dagger\, 2} )/2}\label{ap55}	
\end{equation}
where $\gamma_1,\,\gamma_2,\,\gamma_{12}\in R$ and $\gamma_{12}\geq 0$. This operator satisfies the condition that the classical game is to be fully represented by its quantum counterpart, but does not satisfy the symmetry condition with respect to the exchange of players. The entanglement operator defined by eq.~\eqref{a15} satisfies both of these conditions.

Using the position $\hat{X_j}=(a_j^\dagger+a_j)/\sqrt{2}$ and momentum $\hat{P_j}=i(a_j^\dagger-a_j)/\sqrt{2}$ operators the entanglement operator \eqref{ap55}  can be rewritten in the following form  
\begin{equation}
	\hat{J}(\gamma_1,\gamma_2,\gamma_{12})=e^{i\gamma_{12}(\hat{X}_1\hat{P}_2+\hat{X}_2\hat{P}_1)}e^{i\gamma_1(\hat{X}_1\hat{P}_1+\hat{P}_1\hat{X}_1)/2}e^{i \gamma_2(\hat{X}_2\hat{P}_2+\hat{P}_2\hat{X}_2 )/2}.\label{ap56}	
\end{equation} 
and is equivalent to the operator
\begin{equation}
		\hat{J}(\gamma_1,\gamma_2,\gamma_{12})=e^{iA_1\hat{X}_1\hat{P}_2+i A_2\hat{X}_2\hat{P}_1+iB_1(\hat{X}_1\hat{P}_1+\hat{P}_1\hat{X}_1)+iB_2(\hat{X}_2\hat{P}_2+\hat{P}_2\hat{X}_2 )+iC_1}\label{ap57}
	\end{equation}
where
\begin{displaymath}
	\begin{split}
A_1 &=\frac{r\sinh \gamma_{12}}{\sinh r}e^{-\frac{\gamma_1-\gamma_2}{2}},\\
A_2 &=\frac{r\sinh \gamma_{12}}{\sinh r}e^{\frac{\gamma_1-\gamma_2}{2}},\\
B_1 &=\frac{1}{4}\naw{\gamma_1+\gamma_2+\frac{2r\cosh\gamma_{12}}{\sinh r}\sinh\naw{\frac{\gamma_1-\gamma_2}{2}}},\\
B_2 &=\frac{1}{4}\naw{\gamma_1+\gamma_2-\frac{2r\cosh\gamma_{12}}{\sinh r}\sinh\naw{\frac{\gamma_1-\gamma_2}{2}}},\\
C_1&- \text{any real value},\\
r &=\arccosh \naw{\cosh\naw{\frac{\gamma_1-\gamma_2}{2}}\cosh\gamma_{12}}.		\end{split}
\end{displaymath}
The $iC_1$ factor contained in the operator $\hat{J}(\gamma_1,\gamma_2,\gamma_{12})$ \eqref{ap57} can be ignored because it has no effect on the final state of the game.	
	
	For $\gamma_1=\gamma_2$, the operator $\hat{J}(\gamma_1,\gamma_2,\gamma_{12})$ becomes symmetric due to the exchange of players and simplifies to a form
	\begin{equation}
	\hat{J}(\gamma_1,\gamma_1,\gamma_{12})=e^{\gamma_{12}(a_1a_2-a_1^\dagger a_2^\dagger)-\frac{\gamma_1}{2}(a_1^{\dagger\,2}+a_2^{\dagger\,2}-a_1^2-a_2^2)}
	\end{equation}
which is a special case of the operator $\hat{J}_1(\delta,\xi)$ defined by eq.~\eqref{a15} when $\theta=\phi=0$, $\beta=\gamma_{12}$ and $\delta=\frac{\gamma_1}{2}$.

	\section{}\label{AA2}
To find the expansion of $e^{M}$ where
\begin{equation}
 M=\left(\begin{array}{cccc}
	0 & 0 & 2\delta & \xi\\
	0 & 0 & \xi & 2\delta\\
	2\delta^* & \xi^* & 0 & 0\\
	\xi^* & 2\delta^* & 0 & 0
\end{array}\right)=\left(\begin{array}{cccc}
0 & 0 & 2\alpha e^{i\phi} & \beta e^{i\theta}\\
0 & 0 & \beta e^{i\theta} & 2\alpha e^{i\phi}\\
2\alpha e^{-i\phi} & \beta e^{-i\theta} & 0 & 0\\
\beta e^{-i\theta} & 2\alpha e^{-i\phi} & 0 & 0
\end{array}\right)
\end{equation}
one can diagonalize the matrix M by transformation
\begin{equation}
	M=SM_{diag}S^{-1}
\end{equation}
and then use the formula
\begin{equation}
	e^M=Se^{M_{diag}}S^{-1}.\label{A1}
\end{equation}
$M_{diag}$ is a diagonal matrix with elements on the main diagonal which are the eigenvalues of the matrix $M$ and the corresponding eigenvectors of $M$ are the consecutive columns of the matrix $S$. Finding the eigenvalues and eigenvectors of the matrix M, one can write 
\begin{equation}
	M_{diag}=\left( \begin{array}{cccc}
	-a	& 0 & 0 & 0 \\
	0	& a & 0 & 0 \\
	0	& 0 & -b & 0 \\
	0	& 0 & 0 & b
	\end{array}\right), \quad S=\left( \begin{array}{cccc}
\frac{e^{i(\theta+\phi)}a}{2e^{i\theta}\alpha-e^{i\phi}\beta}	&-\frac{e^{i(\theta+\phi)}a}{2e^{i\theta}\alpha-e^{i\phi}\beta}  &-\frac{e^{i(\theta+\phi)}b}{2e^{i\theta}\alpha+e^{i\phi}\beta}  &  \frac{e^{i(\theta+\phi)}b}{2e^{i\theta}\alpha+e^{i\phi}\beta}\\
-\frac{e^{i(\theta+\phi)}a}{2e^{i\theta}\alpha-e^{i\phi}\beta}	&\frac{e^{i(\theta+\phi)}a}{2e^{i\theta}\alpha-e^{i\phi}\beta}  &-\frac{e^{i(\theta+\phi)}b}{2e^{i\theta}\alpha+e^{i\phi}\beta}  &  \frac{e^{i(\theta+\phi)}b}{2e^{i\theta}\alpha+e^{i\phi}\beta}\\
-1	& -1 & 1 & 1 \\
1	& 1 & 1 & 1
\end{array} \right) 
\end{equation}
where $a=\sqrt{4\alpha^2+\beta^2-4\alpha\beta\cos(\theta-\phi)}$, $b=\sqrt{4\alpha^2+\beta^2+4\alpha\beta\cos(\theta-\phi)}$.

It is not difficult to find $e^{M_{diag}}$ 
\begin{equation}
	e^{M_{diag}}=\left( \begin{array}{cccc}
	\cosh a-\sinh a	& 0 & 0 & 0 \\
	0	& \cosh a+\sinh a & 0 & 0 \\
	0	& 0 & \cosh b-\sinh b & 0 \\
	0	& 0 & 0 & \cosh b+\sinh b
	\end{array}\right) .
\end{equation}
Using eq.~\eqref{A1} one can find  $e^M$; its final form is described by the eq.~\eqref{a26}.

\section{}\label{AA3}
The quantum payoff functions $\tilde{u}_{1,2}^Q(\tilde{x}_1,\tilde{x}_2)$ are obtained by substituting $\tilde{q}_1$ and $\tilde{q}_2$, described by the equation \eqref{ac29}, into the payoff functions defined by the equation \eqref{a4}. Their explicit forms are
\begin{small}
\begin{equation}
\begin{split}	\tilde{u}_1^Q(\tilde{x}_1,\tilde{x}_2)=&\frac{1}{8}\naw{2k-\naw{2\cosh b+\naw{\frac{1}{bf}+bf}\sinh b}(x_1+x_2)}\\
	&\cdot\naw{\naw{2\cosh a+\naw{\frac{1}{ad}+ad}\sinh a}(x_1-x_2)+\naw{2\cosh b+\naw{\frac{1}{bf}+bf}\sinh b}(x_1+x_2)},\\
\tilde{u}_2^Q(\tilde{x}_1,\tilde{x}_2)=&\frac{1}{8}\naw{2k-\naw{2\cosh b+\naw{\frac{1}{bf}+bf}\sinh b}(x_1+x_2)}\\
&\cdot\naw{\naw{2\cosh a+\naw{\frac{1}{ad}+ad}\sinh a}(x_2-x_1)+\naw{2\cosh b+\naw{\frac{1}{bf}+bf}\sinh b}(x_1+x_2)}.		
	\end{split}
\end{equation}\end{small}
Using eq.~\eqref{ab27} we obtain the following payoff functions 
\begin{footnotesize}\begin{equation}
	\begin{split}
&\tilde{u}_i^Q(\tilde{x}_1,\tilde{x}_2)=\\&=\frac{1}{2}\Bigg[k-\Bigg(\cosh\sqrt{4\alpha^2+\beta^2+4\alpha\beta\cos(\theta-\phi)}
		+\frac{(\beta\cos\theta+2\alpha\cos\phi)\sinh\sqrt{4\alpha^2+\beta^2+4\alpha\beta\cos(\theta-\phi)}}{\sqrt{4\alpha^2+\beta^2+4\alpha\beta\cos(\theta-\phi)}}\Bigg)(x_i+x_j)\Bigg]\\
&	\cdot\Bigg[\cosh\sqrt{4\alpha^2+\beta^2-4\alpha\beta\cos(\theta-\phi)}
		+\frac{(-\beta\cos\theta+2\alpha\cos\phi)\sinh\sqrt{4\alpha^2+\beta^2-4\alpha\beta\cos(\theta-\phi)}}{2\sqrt{4\alpha^2+\beta^2-4\alpha\beta\cos(\theta-\phi)}}\Bigg)(x_i-x_j)\\
		&+\Bigg(\cosh\sqrt{4\alpha^2+\beta^2+4\alpha\beta\cos(\theta-\phi)}
		+\frac{(\beta\cos\theta+2\alpha\cos\phi)\sinh\sqrt{4\alpha^2+\beta^2+4\alpha\beta\cos(\theta-\phi)}}{\sqrt{4\alpha^2+\beta^2+4\alpha\beta\cos(\theta-\phi)}}\Bigg)(x_i+x_j)\Bigg]
\end{split}
\end{equation}\end{footnotesize}
for $i,j=1,2$.
\end{appendices}


\begin{thebibliography}{99}
	\bibitem{EisertWL} Eisert J., Wilkens M., Lewenstein M.: Quantum Games and Quantum Strategies. \emph{Phys. Rev. Lett.} \textbf{83}, 3077-3080, (1999)
	\bibitem{EisertW} Eisert J., Wilkens M.: Quantum games. \emph{J. Mod. Opt.} \textbf{47}, 2543-2556, (2000)
	\bibitem{Meyer} Meyer D.: Quantum Strategies. \emph{Phys. Rev. Lett.} \textbf{82}, 1052-1055, (1999)
	\bibitem{Marinatto} Marinatto L., Weber T.: A quantum approach to static games of complete information. \emph{Phys. Lett A}, \textbf{272}, 291-303, (2000)
	\bibitem{Benjamin} Benjamin S.: Comment on "A quantum approach to static games of complete information". \emph{Phys. Lett.}, \textbf{A277}, 180-182, (2000)
	\bibitem{FlitneyA} Flitney A, Abbott D.: An introduction to quantum game theory. \emph{Fluct. Noise Lett.} \textbf{2}, R175-R187, (2000)
	\bibitem{BenjaminHay} Benjamin S., Hayden P.: Comment on “Quantum Games and Quantum Strategies". \emph{Phys. Rev. Lett.} \textbf{87(6)}, 069801, (2001)
	\bibitem{Iqbal} Iqbal A., Toor A.: Evolutionarily stable strategies in quantum games. \emph{Phys. Lett} \textbf{A280}, 249-256,  (2001)
	\bibitem{DuLi} Du J., Li H., Xu X.,  Zhou X., Han R.: Entanglement playing a dominating role in quantum games. \emph{Phys. Lett} \textbf{A289}, 9-15,  (2001)

	\bibitem{Enk}  van Enk S. J., Pike R.: Classical rules in quantum games. \emph{Phys. Rev} \textbf{A66}, 024306,  (2002)
	\bibitem{PiotrowskiS} Piotrowski E., Sladkowski J.:  An Invitation to Quantum Game Theory \emph{Int. Journ. Theor. Phys.} \textbf{42}, 1089-1099, (2003)
	
	\bibitem{Landsburg} Landsburg S.: Quantum Game Theory. \emph{Notices of the Am. Math. Soc.} \textbf{51}, 394-399,  (2004)
	\bibitem{NawazT1} Nawaz A., Toor A.: Generalized Quantization Scheme for Two-Person Non-Zero-Sum Games. \emph{Journ. Phys.} \textbf{A37}, 11457-11463, (2004)
	\bibitem{NawazT2} Nawaz A., Toor A.: Dilemma and quantum battle of sexes. \emph{Journ. Phys.} \textbf{A37}, 4437-4443,  (2004)

	\bibitem{FlitneyA2} Flitney A., Abbott D.: Quantum games with decoherence. \emph{Journ. Phys.} \textbf{A38}, 449-459,  (2005)
	\bibitem{Ichikawa} Ichikawa T., Tsutsui I.: Duality, phase structures, and dilemmas in symmetric quantum games. \emph{Ann. Phys.} \textbf{322}, 531-551, (2007)
	\bibitem{Cheon} Cheon T., Tsutsui I: Classical   and   quantum   contents   of   solvable   game   theory on Hilbert space. \emph{Phys. Lett.} \textbf{A348}, 147-152, (2006)
	\bibitem{Patel}  Patel N.: Quantum games: States of play. \emph{Nature} \textbf{445}, 144-146, (2007)
	\bibitem{Ichikawa1} Ichikawa T., Tsutsui I., Cheon T.: Quantum game theory based on the Schmidt decomposition. \emph{Journ. Phys. A: Math. and Theor.} \textbf{41}, 135303, (2008)
	\bibitem{FlitneyA3} Flitney A., Hollenberg L.: Nash equilibria in quantum games with generalized two-parameter strategies. \emph{Phys. Lett.} \textbf{A363}, 381-388,  (2007)
	\bibitem{Landsburg1} Landsburg S.: Nash equilibria in quantum games. \emph{Proc. Am. Math. Soc.} \textbf{139}, 4423-4434,  (2011)
	\bibitem{Landsburg2} Landsburg S.: Quantum Game Theory. \emph{Wiley Encyclopedia of operations Research and Management science}, Wiley and Sons, New York, (2011)
	\bibitem{Schneider2} Schneider D.: A periodic point-based method for the analysis of Nash equilibria in 2 × 2 symmetric quantum games. \emph{Journ. Phys.} \textbf{A44}, 095301, (2011)
	\bibitem{Schneider} Schneider D.: A new geometrical approach to Nash equilibria organization in Eisert's quantum games. \emph{Journ. Phys.} \textbf{A45}, 085303,  (2012)
	\bibitem{Avishai} Avishai Y.:Some Topics in Quantum Games. arXiv:1306.0284
	\bibitem{Bolonek}  Bolonek-Laso\'n K., Kosi\'nski P.: Some properties of the maximally entangled Eisert-Wilkens-Lewenstein game. \emph{Prog. Theor. Exp. Phys.} (7), 073A02, (2013)
	\bibitem{Ramzan1} Ramzan M.: Three-player quantum Kolkata restaurant problem under decoherence. \emph{ Quant. Inf. Process.} \textbf{12}, 577-586, (2013)
	\bibitem{Ramzan2} Ramzan M.,  Khan M. K.: Environment-assisted quantum Minority games. \emph{Fluctuation and Noise Letters} \textbf{12}, 1350025, (2013) 
	\bibitem{Nawaz4} Nawaz A.: The strategic form of quantum Prisoners' Dilemma. \emph{Chin. Phys. Lett.} \textbf{30(5)}, 050302, (2013)
	\bibitem{Frackiewicz} Frackiewicz P.: A comment on the generalization of the Marinatto-Weber quantum game scheme. \emph{Acta Phys. Polonica B} \textbf{44}, 29-33,  (2013)
	
	\bibitem{Nawaz6}  Nawaz A.: Prisoners' dilemma in the presence of collective dephasing. \emph{J. Phys. A: Math. Theor.} \textbf{45}, 195304, (2012)
	\bibitem{bolonek1} Bolonek-Laso\'n K., Kosi\'nski P.: On Nash equilibria in Eisert-Lewenstein-Wilkens game. \emph{Int. J. Quant. Inf.} \textbf{13}, 1550066, (2015)
	\bibitem{bolonek2} Bolonek-Laso\'n K.: Examining the effect of quantum strategies on symmetric conflicting interest games. \emph{Int. J. Quant. Inf.} \textbf{15} (2017), no. 5, 1750033
\bibitem{Dawkins} Dawkins R.: The Selfish Gene. Oxford University Press, Oxford, (1976)
\bibitem{Penrose}  Penrose R.: The Emperor's New Mind: Concerning Computers, Minds, and The Laws of Physics. Oxford University Press, Oxford, (1989)
		
	\bibitem{Li} Li H., Du J., Massar S.: Continuous-variable quantum games. \emph{Phys. Lett.} \textbf{A 306}, 73-78, (2002)
			\bibitem{varian} Varian H.R.: Intermediate Microeconomics, A Modern Approach. W.W. Norton \& Company, (2009)
		\bibitem{LoKiang0} Lo C.F., Kiang D.: Quantum Stackelberg duopoly. \emph{Phys. Lett. A} \textbf{318}, 333-336, (2003) 
		\bibitem{LoKiang3}	Lo C.F., Kiang D.: Quantum oligopoly. \emph{Europhys. Lett.} \textbf{64}(5), 592-598, (2003)
		\bibitem{khan} Khan S., Ramzan M., Khan M.K.: Quantum model of Bertrand Duopoly. \emph{Chin. Phys. Lett.} \textbf{27}(8), 080302 (2010) 
		\bibitem{rahaman} Rahaman R., Majumdar P., Basu B.: Quantum Cournot equilibrium for the Hotelling-Smithies model of product choice. \emph{J. Phys. A: Math. Theor.} \textbf{45}, 455301, (2012)
		\bibitem{FracSladk} Frackiewicz P., S\l adkowski J.: Quantum approach to Bertrand duopoly. 
		emph{Quant. Inf. Process.} \textbf{15}, 3637-3650, (2016)
		\bibitem{LoYeung} Lo C.F., Yeung C.F.: Quantum Stackelberg-Bertrand duopoly. \emph{Quant. Inf. Process.} \textbf{19}, 373, (2020)

	
	\bibitem{wang} Wang N., Yang Z.: Nonlinear quantum Cournot duopoly games. \emph{J. Phys. A: Math. Theor.} \textbf{55}, 425306, (2022)
	\bibitem{sekiguchi} Sekiguchi Y., Sakahara K., Sato T.: Uniqueness of Nash equilibria in a quantum Cournot duopoly game. \emph{Journ. Phys. A: Math. and Theor.} \textbf{43}(14), 145303, (2010)	
	\bibitem{wang1} Wang N., Yang Z.: Quantum mixed duopoly games with a nonlinear demand function. \emph{Quant. Inf. Process.} \textbf{22}, 139, (2023)
	\bibitem{frackiewiczbilski} Frackiewicz P., Bilski J.: Quantum games with unawareness with duopoly problems in view. \emph{Entropy} \textbf{21}, 1097, (2019)
	\bibitem{QinChen} Qin G., Chen X., Sun M., Du J.: Quantum Bertrand duopoly of incomplete information. \emph{J. Phys. A Math. Gen.} \textbf{38}, 4247, (2005)
	\bibitem{LoKiang1} Lo C.F., Kiang D.: Quantum Stackelberg duopoly with incomplete information. \emph{Phys. Lett.} \textbf{A 346}, 65-70, (2005)
	\bibitem{frackiewicz2} Frackiewicz P.: On subgame perfect equilibria in quantum Stackelberg duopoly with incomplete information. \emph{Phys. Lett.} \textbf{A 382}, 3463-3469, (2018)
	\bibitem{LoKiang2} Lo C.F., Kiang D.: Quantum Bertrand duopoly with differentiated products. \emph{Phys. Lett.} \textbf{A 321}, 94-98, (2004)
	\bibitem{Qin2} Qin G., Chen X., Sun M., Zhou X., Du J.: Appropriate quantization of asymmetric games with continuous strategies. \emph{Phys. Lett.}  \textbf{A 340}, 78-86,  (2005)
	\bibitem{WangXia} Wang X., Liu D., Zhang J-P.: Asymmetric model of the quantum Stackelberg Duopoly. \emph{Chin. Phys. Lett.} \textbf{30}(12), 120302, (2013)
		\bibitem{Zhong} Zhong Y., Shi L., Xu F.: Asymmetric quantum Stackelberg duopoly game based on isoelastic demand. \emph{Int. J. Theor. Phys.} \textbf{61}, 75, (2022)	
		\bibitem{LiQin} Li Y., Qin G., Zhou X., Du J.: The application of asymmetric entangled states in quantum games. \emph{Physics Letters} \textbf{A} 355, 447-451, (2006)
	\bibitem{gerry} Gerry C., Knight P.L., Introductory quantum optics, Cambridge University Press (2005)	
	\bibitem{plenio} Plenio M., Virmani S.: An introduction to entanglement measures. \emph{Quantum Inf. and Comp.} \textbf{7}(1), 1-51, (2007)
	\bibitem{adesso} Adesso G., Illuminati F.: Gaussian measures of entanglement versus negativities: Ordering of two-mode Gaussian state. \emph{Phys. Rev.} A \textbf{72}, 032334, (2005)
	\bibitem{Tommaso} Demarie T.F.: Pedagogical introduction to the entropy of entanglement for Gaussian states. \emph{Eur. J. Phys.} \textbf{39}, 035302, (2018)
	\bibitem{williamson}  Williamson J.: On the algebraic problem concerning the normal forms of linear dynamical systems. \emph{Am. J. Math.} \textbf{58}, 141-163,  (1936)
\end{thebibliography}
\end{document}